\documentstyle[12pt,epsfig,multicol]{article}
\newcommand{\cl}{\centerline}
\renewcommand{\theequation}{\arabic{equation}}
\newcommand\beq{\begin{equation}}
\newcommand\eeq{\end{equation}}
\newcommand\bea{\begin{eqnarray}}
\newcommand\eea{\end{eqnarray}}

\begin{document}

\input epsf

\begin{titlepage}
\setlength{\textwidth}{5.0in}
\setlength{\textheight}{7.5in}
\setlength{\parskip}{0.0in}
\setlength{\baselineskip}{18.2pt}
\hfill
{\tt SOGANG-HEP 284/01}
\vskip 0.3cm
\cl{\Large{{\bf Static properties of chiral models}}}\par
\cl{\Large{{\bf with SU(3) group structure}}}\par
\vskip 0.7cm
\begin{center}
{Soon-Tae Hong$^{1,2}$ and Young-Jai Park$^{1}$}\par
\end{center}
\begin{center}
{\small $^{1}$Department of Physics, Sogang University, Seoul 100-611, 
Korea}\par
{\small $^{2}$W.K. Kellogg Radiation Laboratory, California Institute of 
Technology, Pasadena, CA 91125, USA}\par
\end{center}
\vskip 0.3cm
\cl{\today}


\begin{quotation}
\vskip 0.5cm
We investigate the strangeness in the framework of chiral models, such as the 
Skyrmion, MIT bag, chiral bag and superqualiton models, with SU(3) flavor 
group structure.  We review the recent progresses in both the theoretical 
paradigm and experimental verification for strange hadron physics, and in 
particular the SAMPLE experiment results on the proton strange form factor.  
We study the color flavor locking phase in the color superconducting 
quark matter at high density, which might exist in the core of neutron stars, 
in the soliton-like superqualiton description.  We explain the difficulties 
encountered in the application of the standard Dirac quantization to the 
Skyrmion and superqualiton models and treat the geometrical constraints of 
these soliton models properly to yield the relevant mass spectrum including the 
Weyl ordering corrections and the BRST symmetry structures.            

\vskip 0.4cm
\noindent
PACS: 21.60.Fw, 12.39.Dc, 13.40.Gp, 14.20.-c, 11.10.Ef, 12.20.Ds\\
\noindent
Keywords: chiral models, Skyrmions, form factors, Dirac formalism, superqualiton
\noindent
\end{quotation}
\end{titlepage}

\newpage
\tableofcontents

\newpage
\section{Introduction}\label{intro}
\setcounter{equation}{0}
\renewcommand{\theequation}{\arabic{section}.\arabic{equation}}


Nowadays there has been significant discussion concerning the possibility of 
sizable strange quark matrix elements in the nucleon.  Especially the 
measurement of the spin structure function of the proton given by the European 
Muon Collaboration (EMC) experiments on 
deep inelastic muon scattering \cite{ashman88} has suggested a lingering 
question 
touched on by physicists that the effect of strange quarks on nucleon
structure is not small.  The EMC result has been interpreted as the 
possibility of a strange quark sea strongly polarized opposite to the proton 
spin.  Similarly such interpretation of the strangeness has been brought to 
other analyzes of low energy elastic neutrino-proton scattering\cite{kaplan88}.

Quite recently, the SAMPLE Collaboration\cite{sample01,sample00} reported the
experimental data of the proton strange form factor through parity violating
electron scattering~\cite{mck89,beise91,bob99aip}.  To be more precise, 
they measured the neutral weak form factors at a small momentum transfer 
$Q_{S}^2 = 0.1~{\rm (GeV/c)}^2$ 
to yield the proton strange magnetic form factor~\cite{sample01,sample00} 
$$
G_{M}^{s} (Q_{S}^2) = +0.14 \pm 0.29~{\rm (stat)} \pm 0.31~{\rm (sys)}.
$$

This result is contrary to the negative values of the proton strange form 
factor which result from most of the model 
calculations~\cite{jaffe89,musolf94,koepf92,holstein90,park91,phatak94,christov96,
hammer96,ito95,weigel95,leinweber96,geiger96,musolf96a,musolf96b,
meissner97} except those of Hong, Park and Min~\cite{hong97,hong99aip,hong932} 
based on the SU(3) chiral bag model 
(CBM)~\cite{gerry791,gerry792,gerry80,jaffe82ny,
jackson83,rho83,gerry84,park88,park89ccp,park90,hong92,hong931,hong94,
rho96cnd,rho97ccp,lee99,rho99aip,lee00} and the 
recent predictions of the chiral quark soliton model~\cite{kim298} and 
the heavy baryon chiral perturbation theory~\cite{meissner00,vankolck00}.  
Recently the anapole moment effects associated with the parity violating 
electron scattering  have been intensively studied to yield the more 
theoretical predictions~\cite{vankolck00,maekawa00,musolf00prd,musolf01nuc,bob01ph}.  
(For details see Ref.~\cite{bob01ph}.)  

In fact, if the strange quark content in the nucleon
is substantial then kaon condensation can be induced at a matter density lower
than that of chiral phase transition~\cite{kaplan86, lee95} affecting the 
scenarios for relativistic heavy-ion reactions~\cite{nelson87}, neutron star
cooling~\cite{brown88neu,brown92ku,brown94np,brown95ap,brown95be,brown96pr,
brown97np,brown97prl,brown98np,brown98prl} and so on. 

On the other hand, it is well known that baryons can be obtained from 
topological solutions, known as SU(2) Skyrmions, since the homotopy group 
$\Pi_{3}(SU(2))=Z$ admits 
fermions~\cite{adkins83,rho83,adkins841,zahed86,hong98}.  Using the collective 
coordinates of the isospin rotation of the Skyrmion, Witten and 
co-workers~\cite{adkins83} have performed semiclassical quantization having 
the static 
properties of baryons within 30$\%$ of the corresponding experimental data. 

Phenomenologically, the MIT bag model~\cite{chodos74,degrand75} firstly 
incorporated confinement and asymptotic freedom of QCD.  However, this model 
lacks chiral symmetry so that it cannot be directly applied to the nuclear 
interaction description.  Moreover, in order for the bag to be stable, a bag 
size should be approximately 1 fm, which is simply too big to naively exploit 
the MIT bag model in describing nuclear systems.  To overcome these difficulties, 
Brown and Rho proposed a "little bag"~\cite{gerry791,gerry792} where they 
implemented the spontaneously broken chiral symmetry and brought in Goldstine 
pion cloud to yield the pressure enough to squeeze the bag to a smaller size so 
that the bag can accommodate the nuclear physics of meson exchange 
interactions.  Here how to squeeze the bag without violating the uncertainty 
principle will be discussed later in accordance with the Cheshire cat 
principle~\cite{nadkarni85}.  On the other hand, the pion cloud was introduced 
outside the MIT bag to yield a "chiral bag"~\cite{chodos75} by imposing chiral 
invariant boundary conditions associated with the chiral invariance and 
confinement.  

As shown in the next chapter, based on an analogy to the monopole-isomultiplet 
system~\cite{jackiw76prl}, the baryon number was firstly 
noticed~\cite{vento80npa} to be fractionalized into the quark and pion phase 
contributions, and later established~\cite{rho83} for the special case of the 
"magic angle" of the pionic hedgehog field and then generalized for arbitrary 
chiral angle~\cite{goldstone83}.  Here one notes that, in the "cloud bag" 
model~\cite{thomas84np}, the hedgehog component of the pion field was ignored 
so that the baryon number could be lodged entirely inside the bag.  

The CBM, which is a hybrid of two different models: the MIT bag 
model at infinite bag radius on one hand and the SU(3) Skyrmion model at 
vanishing radius on the other hand, has enjoyed considerable success in 
predictions of the baryon static properties such as the EMC experiments and 
the magnetic moments of baryon octet and decuplet, as well as the strange form 
factors of baryons\cite{hong97} to confirm the SAMPLE Collaboration 
experiments.  After the discovery of the Cheshire cat 
principle~\cite{nadkarni85}, the CBM has been also regarded as a candidate 
which unifies the MIT bag and Skyrmion models and gives model independent 
relations insensitive to the bag radius.
 
On the other hand, Brown and co-workers~\cite{gerry80} calculated the pion 
cloud contributions to the baryon magnetic moments by using the SU(2) CBM as an 
effective nonrelativistic quark model (NRQM).  The Coleman-Glashow sum rules of 
the magnetic moments of the baryon octet were investigated in the SU(3) CBM so
that the bag was proposed as an effective NRQM with meson cloud {\it inside} 
and outside the bag surface \cite{hong931}.  The possibility of unification of 
the NRQM and Skyrmion and MIT bag models through the chiral bag, was proposed 
again for the baryon decuplet \cite{hong94}, as well as the baryon octet 
\cite{hong931}.

In the Skyrmion model \cite{skyrme61, zahed86}, many properties of baryon 
containing light u- and d-quarks have suggested that they can be described in
terms of solitons.  Provided the Wess-Zumino term \cite{wess71} is included in 
the nonlinear sigma model Lagrangian, the solitons have the correct quantum 
numbers to be QCD baryons \cite{witten83} with many predictions of their 
static properties \cite{adkins83}. Moreover the $N_{c}$ counting suggests that 
the baryons with a single heavy quark ($m_{q}\gg \Lambda_{QCD}$) can be 
described as solitons as baryon containing only light quarks.

Meanwhile, there has been considerable progress in understanding the properties 
of baryons containing a single heavy quarks \cite{callan85, scoccola88}.  
Callan and Klebanov (CK) \cite{callan85} suggested an interpretation of baryons
 containing heavy quarks as bound states of solitons of the pion chiral 
Lagrangian with mesons containing heavy quark.  In their formalism, the 
fluctuations in the strangeness direction are treated differently from those 
in the isospin directions \cite{callan85, scoccola88}.  Jenkins and Manohar 
\cite{jenkins93} recently reconsidered the model in terms of the heavy quark 
symmetry to conclude that a doublet of mesons containing the heavy quark can 
take place in the bound state if both the soliton and meson are taken as 
infinitely heavy.  On the other hand, in the scheme of the SU(3) cranking, 
Yabu and Ando \cite{yabu88} proposed the exact diagonalization of the symmetry 
breaking terms by introducing the higher irreducible representation (IR) 
mixing in the baryon wave function, which was later interpreted in terms of 
the multiquark structure \cite{kim89, lee89} in the baryon wave function.

On the other hand, the Dirac method \cite{dirac64} is a well known formalism 
to quantize physical systems with constraints.  The string theory is known to 
be restricted to obey the Virasoro conditions, and thus it is 
quantized~\cite{green87} by the Dirac method.  The Dirac quantization scheme 
has been also applied to the nuclear phenomenology~\cite{sulee88prd,sulee88plb}.  
In this method, the Poisson brackets in a second-class constraint system are converted 
into Dirac brackets to attain self-consistency.  The Dirac brackets, however, 
are generically field-dependent, nonlocal and contain problems related to 
ordering of field operators.  These features are unfavorable for finding 
canonically conjugate pairs.  However, if a first-class constraint system can 
be constructed, one can avoid introducing the Dirac brackets and can instead 
use Poisson brackets to arrive at the corresponding quantum commutators.

To overcome the above problems, Batalin, Fradkin, and Tyutin (BFT)
\cite{batalin86} developed a method which converts the second-class 
constraints into first-class ones by introducing auxiliary fields. Recently, 
this BFT formalism has been applied to several models of current interest 
\cite{banerjee93,wtkim94,kim98, ghosh94}, especially to the Skyrmion to 
obtain the modified mass spectrum of the baryons by including the Weyl 
ordering correction~\cite{hong991,hong001,hong00mod,hong01prd}.  

Furthermore, due to asymptotic freedom~\cite{gross73id,politzer73fx},
the stable state of matter at high density will be quark
matter~\cite{collins75ky}, which has been shown to exhibit
color superconductivity at low temperature~\cite{barrois77xd,bailin84bm}.  
The color superconducting quark matter~\cite{rapp98zu,alford98zt,evans99ek,
evans99nf,schafer99na,schafer99pb,schafer99ef,pisarski99nh,son99uk,
casalbuoni99wu,hong99dk,rho99kl,alford99mk,pisarski99av,schafer99jg,
hong00tn,rho00xf,manuel00,beane00,hong00ei,rapp00qa,hsu00mp,son00cm,
brown00aq,nowak00wa,rho00sh,park00rho,rajagopal00rs,
rajagopal00hep,kim00rho,hong00rho,hong01plb,gerry01rho,gerry01rhopr} might 
exist in the core of neutron
stars, since the Cooper-pair gap and the critical temperature
turn out to be quite large, of the order of $10\sim 100~{\rm
MeV}$, compared to the core temperature of the neutron star, 
which is estimated to be up to $\sim 0.7$ MeV~\cite{pines92}.
On the other hand, it is found that, when the density is large enough
for strange quark to participate in Cooper-pairing, not only
color symmetry but also chiral symmetry are spontaneously broken
due to so-called color-flavor locking (CFL)~\cite{alford99mk}: At
low temperature, Cooper pairs of quarks form to lock the color
and flavor indices as
\begin{eqnarray}
\left<{\psi_L}^a_{i\alpha}(\vec p){\psi_L}^b_{j\beta}(-\vec p)
\right>=-\left<{\psi_R}^a_{i\alpha}(\vec p)
{\psi_R}^b_{j\beta}(-\vec p)\right>
=\epsilon_{\alpha\beta}\epsilon^{abI}\epsilon_{ij I}\Delta (p_F),
\end{eqnarray}
where $a,b=1,2,3$ and $i,j=1,2,3$ are color and flavor indices,
respectively, and we ignore the small color sextet component in
the condensate. In this CFL phase, the particle spectrum can be 
precisely mapped into that of the hadronic phase at low density. 
Observing this map, Sch\"afer and Wilczek~\cite{schafer99ef,schafer99pb} 
have further conjectured that two phases are in fact continuously connected 
to each other.  The CFL phase at high density is complementary to the hadronic
phase at low density. This conjecture was subsequently
supported by showing that quarks in the CFL phase are realized as  
Skyrmions, called superqualitons, just like baryons are realized as 
Skyrmions in the hadronic phase~\cite{hong99dk,hong01plb}.


\section{Outline of the chiral models}
\setcounter{equation}{0}
\renewcommand{\theequation}{\arabic{section}.\arabic{equation}}



\subsection{Chiral symmetry and currents}
\setcounter{equation}{0}
\renewcommand{\theequation}{\arabic{section}.\arabic{equation}}

 
For a fundamental theory of hadron physics, we will consider in this work the 
chiral models such as Skyrmion, MIT bag and chiral bag models.  Especially, 
the CBM can be described as a topological extended object with hybrid phase 
structure: the quark fields surrounded by the meson cloud outside the bag.  In
the CBM, a surface coupling with the meson fields is introduced to restore the 
chiral invariance \cite{chodos75} which was broken in the MIT bag 
\cite{chodos74, degrand75}.  To discuss the symmetries of the CBM and to 
derive the vector and axial currents, which are crucial ingredients for the 
physical operators for the magnetic moments and EMC experiments, we introduce 
the realistic chiral bag Lagrangian
\beq
{\cal L}={\cal L}_{CS}+{\cal L}_{CSB}+{\cal L}_{FSB}
\label{cbmlagrangian}
\eeq
with the chiral symmetric (CS) part, chiral symmetry breaking (CSB) mass terms 
and SU(3) flavor symmetry breaking (FSB) pieces due to the corrections 
$m_{\pi}\neq m_{K}$ and $f_{\pi}\neq f_{K}$
\bea
{\cal L}_{CS}&=&\bar{\psi}i\gamma^{\mu}\partial_{\mu}\psi
-\frac{1}{2}\bar{\psi}U_{5}\psi\Delta_{B}\nonumber\\
& &+\left(-\frac{1}{4}f_{\pi}^{2}{\rm tr}(l_{\mu}l^{\mu})
       +\frac{1}{32e^{2}}{\rm tr}[l_{\mu},l_{\nu}]^{2}+{\cal L}_{WZW}\right)
        \bar{\Theta}_{B}\nonumber\\
{\cal L}_{CSB}&=& -\bar{\psi}M\psi\Theta_{B}
 +\frac{1}{4}f_{\pi}^{2}m_{\pi}^{2}{\rm tr}(U+U^{\dagger}
       -2)\bar{\Theta}_{B}\nonumber\\
{\cal L}_{FSB}&=&\frac{1}{6}(f_{\pi}^{2}m_{K}^{2}-f_{\pi}^{2}m_{\pi}^{2})
{\rm tr}((1-\sqrt{3}\lambda_{8})(U+U^{\dagger}-2))\bar{\Theta}_{B}
       \nonumber\\
& &-\frac{1}{12}f_{\pi}^{2}(\chi^{2}-1){\rm tr}((1-\sqrt{3}
    \lambda_{8})(Ul_{\mu}l^{\mu}+l_{\mu}l^{\mu}U^{\dagger}))\bar{\Theta}_{B}.
\label{lagcs}
\eea
Here the quark field $\psi$ has SU(3) flavor degrees of freedom and the chiral 
field $U=e^{i\lambda_{a}\pi_{a}/f_{\pi}} \in$ SU(3) is described by the 
pseudoscalar meson fields $\pi_{a}$ $(a=1,...,8)$\footnote{In this work we 
will use the convention that $a,b,c,...$ are the indices which run $1,2,...,8$ 
and $i,j,k,...$ for $1,2,3$ and $p,q,...$ for $4,5,6,7$. The Greek indices 
$\mu,\nu,...$ are used for the space-time with metric $g_{\mu\nu}=
{\rm diag}~(+,-,-,-)$.} 
and Gell-Mann matrices $\lambda_{a}$ with $\lambda_{a}\lambda_{b}=\frac{2}{3}
\delta_{ab}+(if_{abc}+d_{abc})\lambda_{c}$, and $l_{\mu}=U^{\dagger}\partial
_{\mu}U$.  In the numerical calculation in the CBM we will use the parameter 
fixing $e=4.75$, $f_{\pi}=93$ MeV and $f_{K}=114$ MeV. 

The interaction term crucial for the chiral symmetry restoration is given by 
\beq
U_{5}=\frac{1+\gamma_{5}}{2}U\frac{1+\gamma_{5}}{2}+
 \frac{1-\gamma_{5}}{2}U^{\dagger}\frac{1-\gamma_{5}}{2},
\eeq
and $\Delta_{B}=-n^{\mu}\partial_{\mu}\Theta_{B}$ where $\Theta_{B}$ is the bag
theta function with vanishing value (normalized to be unity) only inside the 
bag and $n^{\mu}$ is the outward normal unit four vector and the Skyrmion 
term is included to stabilize soliton solution of the meson phase Lagrangian in 
${\cal L}_{CS}$.  The WZW term, which will be discussed in terms of the topology 
in the next section, is described by the action
\beq
\Gamma_{WZW}=-\frac{iN_{c}}{240\pi^{2}}\int_{\bar{{\sf M}}}{\rm d}^{5}r
\epsilon^{\mu\nu\alpha\beta\gamma}{\rm tr}(l_{\mu}l_{\nu}l_{\alpha}l_{\beta}
l_{\gamma}), 
\label{wzwterm}
\eeq
where $N_{c}$ is the number of colors and the integral is done on the
five-dimensional manifold $\bar{{\sf M}}=\bar{V}\times S^{1}\times I$ with the
three-space volume $\bar{V}$ outside the bag, the compactified time $S^{1}$ 
and the unit interval $I$ needed for a local form of WZW term.  The chiral 
symmetry is explicitly broken by the quark mass term with 
$M={\rm diag}~(m_{u},m_{d},m_{s})$ and pion mass term, which is chosen such 
that it will vanish for $U=1$.

Now we want to construct Noether currents under the 
SU(3)$_{L}\times$SU(3)$_{R}$ local group transformation.  Under infinitesimal 
isospin transformation in the SU(3) flavor channel
\bea
\psi\rightarrow \psi^{\prime}&=&(1-i\epsilon_{a}\hat{Q}_{a})\psi,\nonumber\\
U\rightarrow U^{\prime}&=&(1-i\epsilon_{a}\hat{Q}_{a})U
(1+i\epsilon_{a}\hat{Q}_{a}),
\eea
where $\epsilon_{a}(x)$ the local angle parameters of the group transformation
and $\hat{Q}_{a}=\lambda_{a}/2$ are the SU(3) flavor charge operators given by 
the generators of the symmetry, the Noether theorem yields the flavor octet 
vector currents (FOVC) from the derivative terms in ${\cal L}_{CS}$ and ${\cal L}_{FSB}$
\bea
J_{V}^{\mu a}&=&\bar{\psi}\gamma^{\mu}\hat{Q}_{a}\psi\Theta_B
+\left(-\frac{i}{2}f_{\pi}^{2}{\rm tr}(\hat{Q}_{a}l^{\mu})
+\frac{i}{8e^2}{\rm tr}[\hat{Q}_{a},l^{\nu}][l^{\mu},l^{\nu}]+U\leftrightarrow
U^{\dagger}\right)\bar{\Theta}_B\nonumber\\
& &-\frac{i}{12}(f_{K}^{2}-f_{\pi}^{2}){\rm tr}((1-\sqrt{3}\lambda_{8})
(U\{\hat{Q}_{a},l^{\mu}\}+\{\hat{Q}_{a},l^{\mu}\}U^{\dagger})
+U\leftrightarrow U^{\dagger})\bar{\Theta}_B\nonumber\\
& &+\frac{N_{c}}{48\pi^{2}}\epsilon^{\mu\nu\alpha\beta}{\rm tr}(
\hat{Q}_{a}l_{\nu}l_{\alpha}l_{\beta}-
U\leftrightarrow U^{\dagger})\bar{\Theta}_B
\label{jvmua}
\end{eqnarray}
with $\epsilon^{0123}=1$.  Of course the $J_{V}^{\mu a}$ are conserved as 
expected in the chiral limit, but the mass terms in ${\cal L}_{CSB}$ and 
${\cal L}_{FSB}$ give rise to the nontrivial four-divergence
\bea
\partial_{\mu}J_{V}^{\mu a}&=&-\frac{i}{6}(f_{K}^{2}m_{K}^{2}
-f_{\pi}^{2}m_{\pi}^{2}){\rm tr}
((1-\sqrt{3}\lambda_{8})[\hat{Q}_{a},U+U^{\dagger}])\bar{\Theta}_B
\nonumber\\
& &+\frac{i}{12}(f_{K}^{2}-f_{\pi}^{2}){\rm tr}
((1-\sqrt{3}\lambda_{8})[\hat{Q}_{a},Ul_{\mu}l^{\mu}+l_{\mu}l^{\mu}
U^{\dagger}])\bar{\Theta}_B\nonumber\\
& &-i\bar{\psi}[\hat{Q}_{a},M]\psi\Theta_B.
\label{divvmua}
\eea
In addition one can see that the electromagnetic (EM) currents $J_{EM}^{\mu}$ 
can be easily constructed by replacing the SU(3) flavor charge operators 
$\hat{Q}_{a}$ with the EM charge operator $\hat{Q}_{EM}=\hat{Q}_{3}+\frac{1}
{\sqrt{3}}\hat{Q}_{8}$ in the FOVC (\ref{jvmua}) and that the four divergence 
(\ref{divvmua}) vanishes to yield the conserved EM currents.

Similarly under infinitesimal chiral transformation in the SU(3) flavor channel
\bea
\psi\rightarrow \psi^{\prime}&=&(1-i\epsilon_{a}\gamma_{5}\hat{Q}_{a})
\psi,\nonumber\\
U\rightarrow U^{\prime}&=&(1+i\epsilon_{a}\hat{Q}_{a})U
(1+i\epsilon_{a}\hat{Q}_{a}),
\eea
one obtains the flavor octet axial currents (FOAC)
\bea
J_{A}^{\mu a}&=&\bar{\psi}\gamma^{\mu}\gamma_{5}\hat{Q}_{a}\psi\Theta_B
+\left(-\frac{i}{2}f_{\pi}^{2}{\rm tr}(\hat{Q}_{a}l^{\mu})
+\frac{i}{8e^2}{\rm tr}[\hat{Q}_{a},l^{\nu}][l^{\mu},l^{\nu}]-U\leftrightarrow
U^{\dagger}\right)\bar{\Theta}_B\nonumber\\
& &-\frac{i}{12}(f_{K}^{2}-f_{\pi}^{2}){\rm tr}((1-\sqrt{3}\lambda_{8})
(U\{\hat{Q}_{a},l^{\mu}\}+\{\hat{Q}_{a},l^{\mu}\}U^{\dagger})
-U\leftrightarrow U^{\dagger})\bar{\Theta}_B\nonumber\\
& &+\frac{N_{c}}{48\pi^{2}}\epsilon^{\mu\nu\alpha\beta}{\rm tr}(
\hat{Q}_{a}l_{\nu}l_{\alpha}l_{\beta}
+U\leftrightarrow U^{\dagger})\bar{\Theta}_B.
\label{jamua}
\end{eqnarray}
Here one notes that the FOAC are conserved only in the chiral limit, but one 
has the nontrivial four-divergence from the mass terms of ${\cal L}_{CSB}$ 
and ${\cal L}_{FSB}$ 
\bea
\partial_{\mu}J_{A}^{\mu a}&=&\frac{i}{6}(f_{K}^{2}m_{K}^{2}
-f_{\pi}^{2}m_{\pi}^{2}){\rm tr}
((1-\sqrt{3}\lambda_{8})[\hat{Q}_{a},U-U^{\dagger}])\bar{\Theta}_B
\nonumber\\
& &-\frac{i}{12}(f_{K}^{2}-f_{\pi}^{2}){\rm tr}
((1-\sqrt{3}\lambda_{8})\{\hat{Q}_{a},Ul_{\mu}l^{\mu}-l_{\mu}l^{\mu}
U^{\dagger}\})\bar{\Theta}_B\nonumber\\
& &+i\bar{\psi}\gamma_{5}\{\hat{Q}_{a},M\}\psi\Theta_B.
\label{divamua}
\eea
In the meson phase currents of (\ref{jvmua}) and (\ref{jamua}), one should note that the 
terms with $U\leftrightarrow U^{\dagger}$ in the FOAC have the opposite sign 
of those in the FOVC.  Moreover the mesonic currents from the WZW term and the 
nontopological terms have also the sign difference in front of the term with 
$U\leftrightarrow U^{\dagger}$.

On the other hand, one can define the sixteen vector and axial vector charges 
\cite{gell62, adler68,lee72} of SU(3)$_{L}\times$SU(3)$_{R}$
\bea
\hat{Q}_{a}&=&\int {\rm d}^{3}x J_{V}^{0a}\nonumber\\
\hat{Q}_{a}^{5}&=&\int {\rm d}^{3}x J_{A}^{0a}
\label{qq}
\eea
where $J_{V}^{\mu a}$ and $J_{A}^{\mu a}$ are the octets of the FOVC and FOAC 
in (\ref{jvmua}) and (\ref{jamua}) respectively.  In the quantized theory discussed 
later these generators are the charge operators and satisfy their equal time 
commutator relations of the Lie algebra of SU(3)$_{L}\times$SU(3)$_{R}$
\bea
\left[\hat{Q}_{a},\hat{Q}_{b}\right]&=& if_{abc}\hat{Q}_{c}\nonumber\\
\left[\hat{Q}_{a},\hat{Q}_{b}^{5}\right]&=& if_{abc}\hat{Q}_{c}^{5}\nonumber\\
\left[\hat{Q}_{a}^{5},\hat{Q}_{b}^{5}\right]&=& if_{abc}\hat{Q}_{c}
\label{qqq}
\eea
and the chiral charges $\hat{Q}_{a}^{R}$ and $\hat{Q}_{a}^{L}$ defined as 
\beq
\hat{Q}_{a}^{R,L}=\frac{1}{2}(\hat{Q}_{a}\pm \hat{Q}_{a}^{5})
\label{qrl}
\eeq
form a disjoint Lie algebra of SU(3)s
\bea
\left[\hat{Q}_{a}^{R},\hat{Q}_{b}^{R}\right]&=& if_{abc}\hat{Q}_{c}^{R}
\nonumber\\
\left[\hat{Q}_{a}^{L},\hat{Q}_{b}^{L}\right]&=& if_{abc}\hat{Q}_{c}^{L}
\nonumber\\
\left[\hat{Q}_{a}^{R},\hat{Q}_{b}^{L}\right]&=&0 
\label{qqqrl}
\eea
from which the Adler-Weisberger sum rules \cite{adler65,weisberger65} can be 
obtained in terms of off-mass shell pion-nucleon cross sections.


\subsection{WZW action and baryon number}


More than thirty years ago Skyrme \cite{skyrme61} proposed a picture of the 
nucleon as a soliton in the otherwise uniform vacuum configuration of the 
nonlinear sigma model.  Quantizing the topologically twisted soliton, he 
suggested that the topological charge or winding number could be identified 
with baryon number $B$.  His conjecture for the definition of $B$ has been 
revived \cite{witten83,faddeev76} in terms of quantum chromodynamics (QCD).  
In particular Witten \cite{witten83} has established a unique relation between 
the topological charge and baryon number with the number of colors $N_{c}$ 
playing a crucial role.

In the large-$N_{c}$ limit of QCD \cite{thooft74}, meson interactions are 
described by the tree approximation to an effective local field theory of 
mesons, and baryons behave as if they were solitons \cite{witten79} so that 
the identification of the Skyrmion with a baryon can be consistent with QCD.

In this section we will briefly review and summarize the fermionization of the 
Skyrmion with the WZW action \cite{witten83} to obtain the baryon number in 
the CBM.

Now we consider the pure Skyrmion on a space-time manifold compactified to be 
$S^{4}=S^{3}\times S^{1}$ where $S^{3}$ and $S^{1}$ are compactified Euclidean 
three-space and time respectively.  The chiral field $U$ is then a mapping of 
$S^{4}$ into the SU(3) group manifold to yield the homotopy group 
$\pi_{4}($SU(3)$)=0$ so that the four-sphere in SU(3) defined by $U(x)$ is the 
boundary of a five dimensional manifold ${\sf M}=S^{3}\times S^{1}\times I$ 
with two dimensional disc $D=S^{1}\times I$ where $I$ is the unit interval.  
Here one notes that ${\sf M}$ is not unique so that the compactified space-time $S^{4}$ is 
also the boundary of another five-disc ${\sf M}^{\prime}$ with 
opposite orientation.

On the SU(3) manifold there is a unique fifth rank antisymmetric tensor 
$\omega_{\mu\nu\alpha\beta\gamma}$ invariant under 
SU(3)$_{L}\times$SU(3)$_{R}$, which enables us to define an action 
\beq
\Gamma_{{\sf M},{\sf M}^{\prime}}=\pm \int _{{\sf M},{\sf M}^{\prime}}
{\rm d}^{5}x \epsilon^{\mu\nu\alpha\beta\gamma}\omega_{\mu\nu\alpha\beta\gamma},
\label{omega}
\eeq
where the signs $\pm$ are due to the orientations of the five-discs 
${\sf M}$ and ${\sf M}^{\prime}$ respectively.  As in Dirac quantization for 
the monopole \cite{dirac31, jackiw84}, one should demand the uniqueness
condition in a Feynman path integral $e^{i\Gamma}{\sf M}=e^{i\Gamma}{\sf M}
^{\prime}$ to yield $\Gamma_{\sf M}-\Gamma_{{\sf M}^{\prime}}=\int_{{\sf M}
+{\sf M}^{\prime}}\omega=2\pi\times$integer for any five-sphere constructed 
from ${\sf M}+{\sf M}^{\prime}$ in the SU(3) group manifold.  Here one notes 
that every five-sphere in SU(3) is topologically a multiple of a basic 
five-sphere $S^{5}$ due to $\pi_{5}$(SU(3))=$Z$.  Normalizing $\omega$ on the 
basic five-sphere $S^{5}$ such that $\int_{S^{5}}\omega = 2\pi$ one can use in 
the quantum field theory the action of the form $n\Gamma$ where $n$ is an 
arbitrary integer.  On the other hand one can obtain the fifth rank 
antisymmetric tensor $\omega_{\mu\nu\alpha\beta\gamma}$ on the five-disc 
${\sf M}$ \cite{witten83}
\beq
\omega_{\mu\nu\alpha\beta\gamma}=-\frac{i}{240\pi^{2}}{\rm tr}(l_{\mu}l_{\nu}
l_{\alpha}l_{\beta}l_{\gamma}),
\label{omega12}
\eeq
which leads us to the condition that the $n\Gamma_{\sf M}$ is nothing but the 
WZW term in the pure Skyrmion model if $n=N_{c}$.  Here one notes in the weak 
field approximation that the right hand side of (\ref{omega12}) can be reduced 
into a total divergence so that by Stokes's theorem $\int_{\sf M}\omega$ can 
be rewritten as an integral over the boundary of ${\sf M}$, namely 
compactified space-time $S^{5}$.  In the CBM the five-disc ${\sf M}=S^{3}
\times S^{1}\times I$ is modified into $\bar{{\sf M}}=\bar{V}\times S^{1}
\times I$ where $\bar{V}=S^{3}-V$ with $V$ being the three-space volume inside 
the bag.  On the modified five-manifold one can construct the WZW term 
(\ref{wzwterm}) in the CBM.

Also it is shown \cite{witten83} that the above action $\Gamma_{\sf M}$ is a 
homotopy invariant under SU(2) mappings with the homotopy group $\pi_{4}$(SU(2))=$Z_{2}$ 
and for a $2\pi$ adiabatic rotation of a soliton the action gains the 
value $\Gamma_{\sf M}=\pi$ corresponding to the nontrivial homotopy class in 
$\pi_{4}$(SU(2)) so that one can obtain an extra phase $e^{in\pi}=(-1)^{n}$ in 
the amplitude, with respect to a soliton at rest with $\Gamma_{\sf M}=0$ 
belonging to the trivial homotopy class.  Here the factor $(-1)^{n}$ indicates 
that the soliton is a fermion (boson) for odd (even) $n$.  On the other hand, 
one remembers that a baryon constructed with $n$ quarks is a fermion (boson) 
if $n$ is odd (even).  With the WZW term with three flavor $N_{c}=3$, one then 
concludes that the Skyrmion can be fermionized.  Here one notes that the 
nontrivial homotopy class in $\pi_{4}$(SU(2)) can be depicted \cite{witten83} 
by the creation and annihilation mechanism of a Skyrmion-anti Skyrmion pair in 
the vacuum through the channel of $2\pi$ rotation of the Skyrmion and it 
corresponds to quantization of the Skyrmion as a fermion.  Such a mechanism 
has also been used \cite{jaroszewicz85} in the (2+1) dimensional nonlinear 
sigma model to discuss the Hopf topological invariant and linking number
\cite{wilczek83}.

In fact, since the (2+1) dimensional O(3) nonlinear sigma model (NLSM) was 
first discussed by Belavin and Polyakov~\cite{polyakov75}, there have 
been lots of attempts to improve this soliton model associated 
with the homotopy group $\pi_{2}(S^{2})=Z$.  In particular, the 
configuration space in the O(3) NLSM is infinitely connected to yield 
the fractional spin statistics, which was first shown by Wilczek and 
Zee~\cite{wilczek82,wilczek83} via the additional Hopf term.  Moreover the O(3) 
NLSM with the Hopf term was canonically quantized~\cite{bowick86} and 
the $CP^{1}$ model with the Hopf 
term~\cite{pan88,kovner89,semenoff92,ban94,oh98plb}, which can be related with 
the O(3) NLSM via the Hopf map projection from $S^{3}$ to 
$S^{2}$, was also canonically quantized later~\cite{kovner89}.  In fact, the $CP^{1}$ model has 
better features than the O(3) NLSM, in the sense that the action of the 
$CP^{1}$ model with the Hopf invariant has a desirable manifest locality, 
since the Hopf term has a local integral representation in 
terms of the physical fields of the $CP^{1}$ model~\cite{wilczek83}.  
Furthermore, this manifest locality in time is crucial for a consistent 
canonical quantization~\cite{hong00cph}.  Recently, the geometrical constraints in the O(3) NLSM and $CP^{1}$ model are systematically analyzed to yield the 
first class Hamiltonian and the corresponding BRST invariant effective 
Lagrangian~\cite{hong99o3,hong00cp,hong00cph}.  Meanwhile, the $CP^{N}$ model 
was studied~\cite{bhlee00cpn} on the noncommutative geometry~\cite{seiberg99}, 
which was quite recently analyzed in the framework of the improved Dirac 
quantization scheme~\cite{hong00brane}.  

Now using the Noether theorem as in the previous section one can obtain the 
conserved flavor singlet vector currents (FSVC) $J_{V}^{\mu}$ which can be 
practically derived by simple replacement of $\hat{Q}_{a}$ with 1 in the 
FOVC (\ref{jvmua}).  If one defines the baryon number of a quark to be 
$1/N_{c}$ so that a baryon constructed from $N_{c}$ quarks has baryon number
one, then the baryon number current $B^{\mu}$ can be shown to be 
$(1/N_{c})J_{V}^{\mu}$, namely
\beq
B^{\mu}=\frac{1}{N_{c}}\bar{\psi}\gamma^{\mu}\psi \Theta_{B}+\frac{1}{24\pi
^{2}}\epsilon^{\mu\nu\alpha\beta}{\rm tr}(l_{\nu}l_{\alpha}l_{\beta})\bar
{\Theta}_{B},
\label{bmu}
\eeq
and the baryon number of the chiral bag is given by
\beq
B=\int {\rm d}^{3}x B^{0}=\int_{B}{\rm d}^{3}x \frac{1}{N_{c}}\psi^{\dagger}
\psi+\int_{\bar{V}}{\rm d}^{3}x \frac{1}{24\pi^{2}}\epsilon_{ijk}
{\rm tr}(l_{i}l_{j}l_{k}),
\label{bb0}
\eeq
which will be discussed in terms of the hedgehog solution ansatz in the next 
section.


\subsection{Hedgehog solution}


Since the Euler equation for the meson fields in the nonlinear sigma model 
was analytically investigated \cite{chodos75} to obtain a specific classical 
solution for the meson fields whose isospin index points radially $\pi_{i}
(\vec{r})/f_{\pi}=\hat{r}^{i}\theta (r)$, the so-called hedgehog solution, 
this spherically symmetric classical solution has been commonly used as a 
prototype ansatz in the literature of the Skyrmion related hadron physics.

In this section we will consider the classical configuration in the meson and 
quark phases to review and summarize briefly the baryon number fractionization 
\cite{rho83, goldstone83} in the CBM.

Assuming maximal symmetry in the meson phase of the chiral bag, we describe the hedgehog 
solution $U_{0}$ embedded in the SU(2) isospin subgroup of SU(3)
\beq
U_{0}=
\left(
\begin{array}{cc}
e^{i\vec{\tau}\cdot\hat{r}\theta (r)} & 0\\
0 & 1\\
\end{array}
\right)
\label{hedgehog}
\eeq
where $\tau_{i}$ $(i=1,2,3)$ are the Pauli matrices, $\hat{x}=\vec{x}/r$ and 
$\theta (r)$ is the chiral angle determined by minimizing the static mass $M$ 
of the chiral bag and constrained by the boundary condition at the bag surface.

In the CBM Lagrangian (\ref{cbmlagrangian}), due to the symmetry breaking mass 
terms, the static mass has an additional pion mass term~\cite{adkins841,zahed86,nam90} as below
\bea
M&=&\frac{2\pi f_{\pi}}{e}\int_{ef_{\pi}R}^{\infty}{\rm d}z~z^{2}\left(\left(
\frac{d\theta}{dz}\right)^{2}+\left(2+2\left(\frac{d\theta}{dz}\right)^{2}
+\frac{\sin^{2}\theta}{z^{2}}\right)\frac{\sin^{2}\theta}{z^{2}}
\right.
\nonumber\\
& &\left.+2\mu_{\pi}^{2}
(1-\cos\theta)\right)
\label{stmass}
\eea
with the dimensionless quantities $z=ef_{\pi}r$ and $\mu_{\pi}=m_{\pi}/
ef_{\pi}$.  Minimizing the above static mass $M$, one obtain the equation of 
motion for the chiral angle $\theta$ outside the bag
\bea
& &(z^{2}+2\sin^{2}\theta)\frac{d^{2}\theta}{dz^{2}}+2z\frac{d\theta}{dz}
+\left(\left(\frac{d\theta}{dz}\right)^{2}-1-\frac{\sin^{2}\theta}{z^{2}}
\right)\sin 2\theta 
\nonumber\\
& &-\mu_{\pi}^{2}z^{2}\sin \theta =0
\label{eom}
\eea
which yields the static Skyrmion chiral angle defining a stationary point of 
the chiral bag action.  

Together with the boundary term $-\frac{1}{2}\bar{\psi}U_{5}\psi \Delta_{B}$, 
the static mass $M$ also yields the boundary condition for the chiral angle at 
$z=ef_{\pi}R$
\beq
\left(1+\frac{2\sin^{2}\theta}{z^{2}}\right)\frac{d\theta}{dz}=\frac{1}{2ef_{\pi}^{3}}
\bar{\psi}i\gamma_{5}\vec{\tau}\cdot\hat{r}e^{i\gamma_{5}\vec{\tau}\cdot
\hat{r}\theta}\psi
\label{bcpsi}
\eeq
which allows the flow of currents in the two phase via the bag boundary.  Here 
one notes that the baryon number (\ref{bb0}) obtained from the topological 
WZW term and quark fields remains constant \cite{rho83, goldstone83} 
regardless of the bag radius through the continuity of the current at the bag 
boundary, even though one has the additional pion mass term in the static 
mass $M$.

On the other hand, in the chiral symmetric limit, the conventional variation 
scheme with respect to the quark fields yields the Dirac equation inside the 
bag and the boundary condition on the bag surface
\bea
i\gamma^{\mu}\partial_{\mu}\psi&=&0,~~~~r<R,\label{bccb}\\
i\gamma^{\mu}n_{\mu}\psi&=&U_{5}\psi,~~r=R
\label{bccbm}
\eea
where the missing quark masses will be discussed later after the collective 
coordinate quantization is performed.

Due to the coupling of spin and isospin in the boundary condition 
(\ref{bccbm}), the u and d quarks can be coalesced \cite{mulders84} into 
hedgehog (h) quark states, eigenstates of grand spin $\vec{K}=\vec{I}+\vec{J}$, 
not of the isospin $\vec{I}$ and the spin $\vec{J}=\vec{L}+\vec{S}$ 
separately, while the s quark is decoupled from the hedgehog quark states.  
The h quark state is then specified by a set of quantum numbers $(K,m_{K},P,m)$ 
where $K(K+1)$ and $m_{K}$\footnote{Here we have used the same symbol $m_{K}$ 
for the quantum number and the kaon mass.  However, a reader can easily recognize the 
meaning of the symbol from the context.} are the eigenvalues of the squared 
operator $\vec{K}^{2}$ and $K_{3}$ the third component of $\vec{K}$, and $P$ 
and $m$ are the parity and radial excitation quantum numbers, respectively.  
Similarly the s quark states are labeled by another set $(j,m_{j},P,n)$ 
with $j(j+1)$, $m_{j}$ and $n$, the eigenvalues of $\vec{J}^{2}$, $J_{3}$ and 
radial quantum number.

Now the quark field $\psi$ can be expanded in terms of the wave functions of 
the hedgehog and strange quark states
\bea
\psi(\vec{r},t)&=&\sum_{n}\psi_{n}^{h}(\vec{r})e^{-i\varepsilon_{n}t}a_{n}
+\psi_{n}^{h*}(\vec{r})e^{i\varepsilon_{n}t}b_{n}^{\dagger}\nonumber\\
& &+\psi_{n}^{s}(\vec{r})e^{-i\omega_{n}t}c_{n}
+\psi_{n}^{s*}(\vec{r})e^{i\omega_{n}t}d_{n}^{\dagger}
\label{quarkfield}
\eea
where the hedgehog quark states are expressed by the spatial wave functions 
$\psi_{n}^{h}(\vec{r})$ with grand spin quantum numbers, whose explicit forms 
will be given in the Appendix B, and the annihilation operator $a_{n}$ ($b_{n}
^{\dagger}$) for the positive (negative) energy fulfills the usual 
anticommutator rules and also defines the vacuum $a_{n}|0>=b_{n}|0>=0$, and 
the strange quark states are analogously described.  Here we do not bother to 
include the color index explicitly since every particle is a color singlet.  
The energy spectrum of the hedgehog quark states \cite{mulders84} is 
subordinate to $\theta (R)$, the chiral angle at bag surface, while the 
strange quark states remain intact regardless of the chiral angle.

Finally in the framework of the previous literatures \cite{rho83, goldstone83}
we reconsider the baryon number (\ref{bb0}) in the hedgehog ansatz to see that 
the total baryon number is still an integer in the CBM.  Using the hedgehog 
solution (\ref{hedgehog}) in the meson piece in (\ref{bb0}) one can obtain 
the fractional baryon number in terms of the chiral angle at the bag surface 
$\theta=\theta (R)\in [-\pi,0]$ \cite{goldstone83}
\beq
B_{m}=-\frac{1}{2\pi}\chi_{E}(\theta-\sin\theta\cos\theta )
\label{bm}
\eeq
where $\chi_{E}$ is the Euler characteristic, which has an inter two in the 
spherical bag surface.

In general the Euler characteristic of a compact surface is the topological 
invariant defined by the integer $v-e+f$ \cite{oneill66} with $v$, $e$ and $f$ 
the numbers of vertices, edges and faces in a decomposition of the surface so 
that one can easily see $\chi_{E}$(sphere) = 2 and $\chi_{E}$(torus) = 0, for 
instance.  Also it is interesting to see that adding a handle ${\sf H}$, or a 
torus with the interior of one face removed, to a compact surface ${\sf S}$ 
reduces its Euler characteristic by two, since to obtain the coalesced surface 
${\sf S^{\prime}}$ one needs the surgery of removing the interior of a face of 
${\sf S}$ so that ${\sf S^{\prime}}$ has two faces less than ${\sf S}$ and 
${\sf H}$ combined.  For a coalesced surface with $h$ handles, one has the 
generalized identity $\chi_{E}({\sf S^{\prime}})=\chi_{E}({\sf S})-2h$ 
\cite{oneill66}.

On the other hand, it has been noted \cite{rho83} in the CBM that the quark 
phase spectrum is asymmetric about zero energy to yield the nonvanishing 
vacuum contribution to the baryon number
\beq
B_{0}=-\frac{1}{2}\lim_{s\rightarrow 0}\sum_{n}{\rm sgn}(E_{n})e^{-s|E_{n}|}
\label{b0lim}
\eeq
where the sum runs over all positive and negative energy eigenstates and the 
symmetrized operator $\frac{1}{2}[\psi^{\dagger},\psi]$ is used in the quark 
part of (\ref{bb0}).  Here one notes that the regularized factor is closely 
related \cite{goldstone83} to the eta invariant of Atiyah et al. 
\cite{atiyah75},
\beq
\eta (s)=\frac{1}{2}\lim_{s\rightarrow 0}\sum_{n}{\rm sgn}(E_{n})|E_{n}|^{-s}
\label{atiyah}
\eeq
which has been also discussed in connection with the phase factor of the path 
integral in quantum field theory associated with the Jones polynomial and 
knot theory~\cite{witten89}, and recently has been exploited in investigation 
of the semiclassical partition functions and the Jacobi fields in the 
framework of the Morse theory of differential geometry~\cite{hong01mor}.

Except at the magic angle $\theta=-\pi/2$, where the baryon number 
is shared equally with both quark and meson phases and $B_{0}$ jumps by unity 
due to the Dirac sea~\cite{rho83}, the chiral angle dependence of the quark vacuum baryon 
number $dB_{0}/d\theta =\frac{1}{2}\lim_{s\rightarrow 0}\sum_{n}s(dE_{n}/d\theta)
e^{-s|E_{n}|}$ is given in terms of the integration of the Gaussian curvature 
$\kappa$ on the bag surface ${\sf S}$ \cite{goldstone83}
\beq
\frac{dB_{0}}{d\theta}=\frac{1}{2\pi^{2}}\sin^{2}\theta\int_{\sf S}{\rm d}^{2}x
\kappa
\label{curvature}
\eeq
where a multiple-reflection expansion of the Euclidean Green's function, as 
well as the Dirac equation (\ref{bccb}) and the boundary condition
(\ref{bccbm}), has been used \cite{goldstone83}. 
  
Using the Gauss-Bonnet theorem \cite{kobayashi69} one can rewrite 
(\ref{curvature}) in terms of the Euler characteristic 
$dB_{0}/d\theta = (\chi/\pi)\sin^{2}\theta$ 
to yield the total quark phase baryon number
\beq
B_{q}=1+\frac{1}{2\pi}\chi_{E}(\theta-\sin\theta\cos\theta)
\label{bqgb}
\eeq
where, in addition to the $\theta$-dependent vacuum contribution $B_{0}$, one 
has the unity factor contributed by the $N_{c}$ degenerate valence quarks to 
fill the $K^{P}=0^{+}$ h-quark and $j^{P}=\frac{1}{2}^{+}$ s-quark eigenstates.  
In the $K^{P}=0^{+}$ level we can define the static hedgehog ground state 
$|H>_{0}$: $a_{v}^{\dagger}|0>$ ($a_{v}^{\dagger}$ being the valence quark 
creation operator with the quantum number $K^{P}=0^{+}$) for $-\pi/2\leq \theta
\leq 0$ and $|0>$ in $-\pi \leq\theta<-\pi/2$, since the quarks in the 
positive energy level are the valence quarks while those in the negative energy 
level can be considered to sink into the vacuum.

Here one notes that in the MIT bag limit at $\theta=0$, where there are no 
vacuum and meson contributions, only the $N_{c}$ degenerate valence quarks 
yield the baryon number.  Also for $-\pi/2\leq\theta<0$ the valence quarks 
and $\theta$-dependent vacuum contribute to $B_{q}$ while for $-\pi\leq\theta
<-\pi/2$ only the quark vacuum does in the static hedgehog ground state.


\subsection{Collective coordinate quantization}


Until now we have considered the baryon quantum number in the classical 
static hedgehog solution in the meson phase of the CBM. As in the Skyrmion 
model \cite{adkins83}, the other quantum numbers such as spin, isospin and 
hypercharge can be obtained in the CBM by quantizing the zero modes associated 
with the slow collective rotation
\beq
U_{0}\rightarrow AU_{0}A^{\dagger},~~~~\psi\rightarrow A\psi
\label{collective}
\eeq
on the SU(3)$_{F}$ group manifold where $A(t)\in$ SU(3)$_{F}$ is the time dependent 
collective variable restrained by the WZW constraint.

In the $\lambda$ dimensional IR, the baryon is then described by a wave function of the form
\beq
|B\rangle^{\lambda}=\Phi_{B}^{\lambda}(A)\otimes|{\rm intrinsic}\rangle
\label{blambda}
\eeq
where $\Phi_{B}^{\lambda}(A)$ is the baryon dependent collective coordinate 
wave function 
\beq
\Phi_{B}^{\lambda}(A)=\sqrt{{\rm dim}(\lambda)}D_{ab}^{\lambda}(A)
\label{phib}
\eeq
with the quantum numbers $a=(Y,I,I_{3})$ ($Y$; hypercharge, $I$; isospin) and 
$b=(Y_{R};J,J_{3})$ ($Y_{R}$; right hypercharge, $J$; spin) in the Wigner $D$-
matrix.  In the 8-dimensional adjoint representation the matrix is given by 
$D_{ab}^{8}(A)=\frac{1}{2}{\rm tr}(A^{\dagger}\lambda_{a}A\lambda_{b})$.  On the other hand, 
the intrinsic state degenerate to all the baryons is described by 
the classical meson configuration approximated by a rotated hedgehog solution 
$AU_{0}A^{\dagger}$ and a rotated hedgehog ground state discussed later.

With the introduction of the collective rotation, the Dirac equation 
(\ref{bccb}) is modified and the boundary condition (\ref{bccbm}) is rewritten 
in the hedgehog ansatz as below \cite{park88}
\bea
\left(i\gamma^{\mu}\partial_{\mu}+\frac{1}{2}\dot{q}_{a}\gamma^{0}
\lambda_{a}\right)\psi&=&0,~~~~r<R\label{bccbh}\\
\left(i\hat{r}\cdot\vec{\gamma}+e^{i\gamma_{5}\lambda_{i}\hat{r}_{i}
\theta}\right)\psi&=& 0~~~~r=R
\label{bccbmh}
\eea
where we have used the collective coordinates $q_{a}$ defined by $A^{\dagger}
\dot{A}=-\frac{i}{2}\lambda_{a}\dot{q}_{a}$.  The collective rotation of the 
chiral bag induces \cite{park88} the particle-hole excitations which will be 
treated perturbatively in this work to yield the correction to the wave 
functions $\psi_{n}^{h}(\vec{r})$ and $\psi_{n}^{s}(\vec{r})$ in 
(\ref{quarkfield})
\bea
\psi_{n}^{h}&=&\psi_{n}^{0h}(\vec{r})+\frac{1}{2}\dot{q}_{i}
\sum_{m\neq n}\frac{_{h}\langle m|\lambda_{i}|n\rangle_{h}}
{\varepsilon_{m}-\varepsilon_{n}}\psi_{m}^{0h}(\vec{r})\nonumber\\
& &+\frac{1}{2}\dot{q}_{p}\sum_{m}\frac{_{s}\langle m|\lambda_{p}|n\rangle_{h}}
{\omega_{m}-\varepsilon_{n}}\psi_{m}^{0s}(\vec{r})
\nonumber\\
\psi_{n}^{s}&=&\psi_{n}^{0s}(\vec{r})+\frac{1}{2}\dot{q}_{p}
\sum_{m\neq n}\frac{_{h}\langle m|\lambda_{p}|n\rangle_{s}}
{\varepsilon_{m}-\omega_{n}}
\psi_{m}^{0h}(\vec{r})
\label{psins}
\eea
where the matrix elements with the unperturbed states $\psi_{n}^{0h}(\vec{r})$ 
and/or $\psi_{n}^{0s}(\vec{r})$ are defined as the following Dirac notations
\bea
_{h}\langle m|\lambda_{i}|n\rangle_{h}&=&\int_{V}{\rm d}^{3}x\psi_{m}
^{\dagger 0h}(\vec{r})\lambda_{i}\psi_{n}^{0h}(\vec{r})\nonumber\\
_{s}\langle m|\lambda_{p}|n\rangle_{h}&=&\int_{V}{\rm d}^{3}x\psi_{m}
^{\dagger 0s}(\vec{r})\lambda_{p}\psi_{n}^{0h}(\vec{r}).
\label{unperstate}
\eea
Here one notes that since $\lambda_{8}\dot{q}_{8}$ related to the WZW term plays the 
role of a constraint, it does not appear explicitly in the above quark 
wave functions.

With the collective rotation, the Fock space should then be modified for 
$N_{c}$ quarks to fill up the new single states (\ref{psins}) with the minimum 
energy so that the rotated hedgehog ground state has a form analogous to the 
cranking formula in nuclear physics \cite{inglis54}
\bea
|H\rangle&=&\left(1+\frac{1}{2}\sum_{m,n}\left(\dot{q}_{i}\frac{
_{h}\langle m|\lambda_{i}|n\rangle_{h}}{\varepsilon_{m}-\varepsilon_{n}}
a_{m}^{\dagger}b_{n}^{\dagger}
+\dot{q}_{p}\frac{_{h}\langle m|\lambda_{p}|n\rangle_{s}}
{\varepsilon_{m}-\omega_{n}}a_{m}^{\dagger}d_{n}^{\dagger}\right.\right.
\nonumber\\
& &+\left.\left.\dot{q}_{p}\frac{_{s}\langle m|\lambda_{p}|n\rangle_{h}}
{\omega_{m}-\varepsilon_{n}}c_{m}^{\dagger}b_{n}^{\dagger}\right)\right.
\nonumber\\
& &\left.+\frac{1}{2}\sum_{m}\left(\dot{q}_{i}\frac{
_{h}\langle m|\lambda_{i}|v\rangle}{\varepsilon_{m}-\varepsilon_{v}}
a_{m}^{\dagger}a_{v}
+\dot{q}_{p}\frac{_{s}\langle m|\lambda_{p}|v\rangle}
{\omega_{m}-\varepsilon_{v}}c_{m}^{\dagger}a_{v}\right)\right)|H\rangle_{0}
\label{nuclear}
\eea
where $|v\rangle$ stands for the valence quark state for 
$-\pi/2\leq\theta\leq 0$.

To obtain the chiral bag Hamiltonian in the chiral symmetric limit (see 
Section 2.2 for the symmetry breaking case) we can construct the canonical 
momenta $\Pi_{a}$ conjugate to the collective variables $q_{a}$
\beq
\Pi_{a}={\cal I}_{1}\dot{q}_{i}\delta_{ia}+{\cal I}_{2}\dot{q}_{p}\delta_{pa}
+\frac{\sqrt{3}}{2}B\delta_{8a}.
\label{pia}
\eeq
Here we have used the parameter fixing $N_{c}=3$ and the identity 
$B_{q}+B_{m}=1$ discussed before where $B_{m}$ comes from the WZW term and 
$B_{q}$ is calculated from the equation 
$\langle H|\int_{V}{\rm d}^{3}x\psi^{\dagger}(\lambda_{8}/2)|H\rangle
=\frac{\sqrt{3}}{2}B_{q}$.

The moments of inertia ${\cal I}_{1}$ and ${\cal I}_{2}$ are explicitly 
given by sum of two contributions from the quark and meson phases as below
\bea
{\cal I}_{1}&=&\frac{3}{2}\sum_{m,n} 
\frac{|_{h}\langle m|\lambda_{3}|n\rangle_{h}|^{2}}
{\varepsilon_{m}-\varepsilon_{n}}
+\frac{3}{2}\sum_{m} 
\frac{|_{h}\langle m|\lambda_{3}|v\rangle|^{2}}
{\varepsilon_{m}-\varepsilon_{v}}
\nonumber\\
& &+\frac{8\pi}{3e^{3}f_{\pi}}\int_{ef_{\pi}R}^{\infty}{\rm d}z~z^{2}\sin^{2}
\theta \left(1+\left(\frac{d\theta}{dz}\right)^{2}+\frac{\sin^{2}\theta}{z^{2}}
\right)
\nonumber\\
{\cal I}_{2}&=&\frac{3}{2}\sum_{m,n} 
\left(\frac{|_{h}\langle m|\lambda_{4}|n\rangle_{s}|^{2}}
{\varepsilon_{m}-\omega_{n}}
+\frac{|_{s}\langle m|\lambda_{4}|n\rangle_{h}|^{2}}
{\omega_{m}-\varepsilon_{n}}\right)
+\frac{3}{2}\sum_{m} 
\frac{|_{s}\langle m|\lambda_{4}|v\rangle|^{2}}
{\omega_{m}-\varepsilon_{v}}
\nonumber\\
& &+\frac{2\pi}{e^{3}f_{\pi}}\int_{ef_{\pi}R}^{\infty}{\rm d}z~z^{2}
(1-\cos\theta)\left(1+\left(\frac{d\theta}{dz}\right)^{2}
+\frac{2\sin^{2}\theta}{z^{2}}
\right)
\label{i1i2}
\eea
where we have used the symmetry properties of the matrix elements to employ 
only $\lambda_{3}$ and $\lambda_{4}$.

The chiral bag Hamiltonian is then given by
\beq
H_{0}=M+\frac12  \left(\frac{1}{{\cal I}_1}-\frac{1}{{\cal I}_2}\right) 
\hat{J}^2+\frac{1}{2{\cal I}_2} \left(\hat{C}_2^{2}
-\frac{3}{4}\hat{Y}_{R}^{2}\right)
\label{chiralham}
\eeq
where $M$ is the static mass (\ref{stmass}) and $\hat{J}^{2}$ and 
$\hat{C}_{2}^{2}$ are the Casimir operators in the SU(2) and SU(3) groups, 
respectively, and $\hat{Y}_{R}$ is the right hypercharge operator to yield the 
WZW constraint $\hat{Y}_{R}|{\rm phys}\rangle=+1|{\rm phys}\rangle$ for any 
physical state $|{\rm phys}\rangle$.


\subsection{Cheshire cat principle}


As we have seen in the previous sections the CBM can be considered as a 
hybrid or combination of two different models: the MIT bag model at infinite 
bag radius on one hand and Skyrmion model at vanishing radius on the other
hand.  Of course the meson phase Lagrangian in (\ref{cbmlagrangian}) can be 
generalized by a more complicated version including vector meson fields such 
as $\rho$ and $\omega$ \cite{meissner86}.

In the hybrid model there has been considerable discussion concerning the 
conjecture that the bag itself has only notational but no physical 
significance, the so called Cheshire cat principle 
(CCP)~\cite{nadkarni85,gerry84,rho96cnd,rho97ccp,rho99aip}.\footnote{
Based on phenomenology, a similar idea of the CCP was proposed by Brown and 
co-workers, simultaneously and independently of Ref.~\cite{nadkarni85}.}  
The jargon Cheshire cat originates from the quotation in the fable "Alice in 
Wonderland" \cite{caroll65}: "Well, I've often seen a cat without a grin," 
thought Alice, "but a grin without a cat!  It is the most curious thing, 
I ever saw in my life!"  According to the Cheshire cat viewpoint, the bag 
wall (Cheshire cat) tends to fade away, when examined closely, leaving 
behind the bag boundary conditions translating the fermionic and bosonic 
descriptions into one another (the grin of the Cheshire cat)~\cite{nadkarni85}.

In (1+1) dimensions where exact bosonization and fermionization relations are 
known \cite{coleman75}, Nadkarni and co-workers proposed the Cheshire cat model 
where the CCP is exactly obeyed so that physics is invariant under changes in 
bag shape and/or size~\cite{nadkarni85}.  Namely, in a simple model with a free 
massless fermion inside the bag and the equivalent free massless boson outside 
the bag, the bag boundary conditions are shown, via bosonization relations, to 
yield a clue to the CCP: shifting the bag wall has no physical effect.

Now, we briefly recapitulate the CCP in the (1+1) dimensional CBM, by 
introducing a massless free single-flavored fermionic quark $\psi$ confined to 
a region of volume $V$ (inside) and a massless free bosonic meson $\phi$ 
located in a region $\bar{V}$ (outside).  Here we assume these two fields are 
coupled to each other via the surface $\partial V$.  Now we consider the 
following action $S$ which is invariant under global chiral rotations and 
parity\footnote{Here we have used the metric $g_{\mu\nu}={\rm diag} (1,-1)$ 
and the Weyl representation for the gamma matrices, 
$\gamma_{0}=\gamma^{0}=\sigma_{1}$, 
$\gamma_{1}=-\gamma^{1}=-i\sigma_{2}$, $\gamma_{5}=\gamma^{5}=\sigma_{3}$ with 
Pauli matrices $\sigma_{i}$.} 
\bea
S&=&S_{V}+S_{\bar{V}}+S_{\partial V},
\label{sss}\\
S_{V}&=&\int_{V}{\rm d}^{2}x~\bar{\psi}i\gamma^{\mu}\partial_{\mu}\psi+\cdots,
\label{sv11}\\
S_{\bar{V}}&=&\int_{\bar{V}}{\rm d}^{2}x~\frac{1}{2}(\partial_{\mu}\phi)^{2}
+\cdots,
\label{svbar11}\\
S_{\partial V}&=&\int_{\partial V}{\rm d}\Sigma^{\mu}\frac{1}{2}n_{\mu}
\bar{\psi}e^{i\gamma_{5}\phi/f}\psi,
\label{spv11}
\eea
where the ellipsis stands for other terms such as interactions, masses and so 
on.  Here we have assumed that chiral symmetry holds on the boundary even if 
as in nature it is broken both inside and outside due to mass terms, and that 
the boundary term does not break the discrete symmetries $P$, $C$ and $T$.  In 
the boundary action (\ref{spv11}), $f=1/\sqrt{4\pi}$ is the $\phi$ meson decay 
constant and ${\rm d}\Sigma^{\mu}$ is an area element with the normal vector 
$n^{\mu}$, namely, $n^{2}=-1$ and picked outward-normal.   

From the action (\ref{sss}), one can obtain the classical equations of motion
\bea
i\gamma^{\mu}\partial_{\mu}\psi&=&0,
\label{quark11}\\
\partial^{\mu}\partial_{\mu}\phi&=&0,
\label{boson11}
\eea
and the boundary conditions associated with the MIT confinement condition
\bea
in^{\mu}\gamma_{\mu}\psi&=&-e^{i\gamma_{5}\phi/f}\psi,
\label{bdry111}\\
n^{\mu}\partial_{\mu}\phi&=&\frac{1}{2f}\bar{\psi}n^{\mu}
\gamma_{\mu}\gamma_{5}\psi.
\label{bdry112}
\eea
Here one can have the conserved vector current 
$j_{\mu}=\bar{\psi}\frac{1}{2}\gamma_{\mu}\psi$ with $\partial^{\mu}j_{\mu}=0$ 
or $\bar{\psi}\frac{1}{2}n^{\mu}\gamma_{\mu}\psi=0$ at the surface from Eq. 
(\ref{bdry111}), and the conserved axial vector current 
$j_{\mu}^{5}=\bar{\psi}\frac{1}{2}\gamma_{\mu}\gamma_{5}\psi$ with 
$\partial^{\mu}j_{\mu}^{5}=0$ from Eq. (\ref{bdry112}).  Note that at quantum 
level the vector current is not conserved due to quantum anomaly, contrast to 
the usual open space case where anomaly is in the axial current.  

For simplicity we assume that the quark is confined to the space 
$-\infty \leq r\leq R$ with a boundary at $r=R$.  The vector current $j_{\mu}$ 
is then conserved inside the bag
\beq
\partial^{\mu}j_{\mu}=0,
\label{consjmu11}
\eeq
to, after integration, yield the time-rate change of the fermion (quark) number    
\beq
\frac{{\rm d}B}{{\rm d}t}=2\int_{-\infty}^{R}{\rm d}r~\partial_{0}j_{0}
=2\int_{-\infty}^{R}{\rm d}r~\partial_{1}j_{1}=2j_{1}(R),
\label{qdot11}
\eeq
so that one can obtain on the boundary 
\beq
\frac{{\rm d}B}{{\rm d}t}=\bar{\psi}n^{\mu}\gamma_{\mu}\psi,
\label{barpsi11}
\eeq
which vanishes classically as mentioned above.  However, at quantum level the above quantity 
is not well-defined locally in time since $\psi^{\dagger}(t)\psi(t+\epsilon)$ is singular as 
$\epsilon\rightarrow 0$ due to vacuum fluctuation.  Now we regulate this bilinear operator by 
exploiting the following point-splitting ansatz at $r=R$
\beq
j_{1}=\bar{\psi}(t)\frac{1}{2}\gamma_{1}\psi(t+\epsilon)
=-\frac{i}{4f}\epsilon \dot{\phi}(t)\psi^{\dagger}(t)\psi(t+\epsilon)
=\frac{1}{4\pi f}\epsilon \dot{\phi}(t)+O(\epsilon),
\label{j1oepsilon}
\eeq    
where we have used the boundary condition (\ref{bdry111}), the commutation relation 
$[\phi(t),\phi(t+\epsilon)]=i~{\rm sgn}\epsilon$ and $\psi^{\dagger}(t)\psi(t+\epsilon)
=\frac{i}{\pi\epsilon}+{\rm regular~terms}$~\cite{rho96cnd}.  The quarks can then 
flow in or out if the meson fields change in time.  In order to understand the leakage of 
the quarks from the bag, we consider the surface tangent 
$t^{\mu}=\epsilon^{\mu\nu}n_{\nu}$ to obtain at $r=R$
\beq
t^{\mu}\partial_{\mu}\phi=-\frac{1}{2f}\bar{\psi}n^{\mu}\gamma_{\mu}
\psi=\frac{1}{2f}\bar{\psi}t^{\mu}\gamma_{\mu}\gamma_{5}\psi,
\label{tpartialphi}
\eeq  
where we have used the relation $\bar{\psi}\gamma_{\mu}\gamma_{5}\psi
=\epsilon_{\mu\nu}\bar{\psi}\gamma^{\nu}\psi$ valid in (1+1) dimensions.  
Combination of Eqs. (\ref{bdry112}) and (\ref{tpartialphi}) yields the 
bosonization relation at the boundary $r=R$ and time $t$
\beq
\partial_{\mu}\phi=\frac{1}{2f}\bar{\psi}\gamma_{\mu}\gamma_{5}\psi,
\label{bosoniz}
\eeq
which is a unique feature of (1+1) dimensional fields~\cite{coleman75}.  
Moreover, the quark field can be written in terms of the meson field as follows
\beq
\psi(x)={\rm exp}\left(-\frac{i}{2f}\int_{x_{0}}^{x}{\rm d}z~\left[\pi (z)+\gamma_{5}
\frac{{\rm d}\phi}{{\rm d}z}\right]\right)\psi (x_{0}),
\eeq
where $\pi(z)$ is the momentum field conjugate to $\phi (z)$.  Here one notes 
that the nonvanishing vector current (\ref{j1oepsilon}) is not conserved due 
to quantum effects to yield the vector anomaly as shown in Eq. (\ref{qdot11}), 
and that the amount of fermion number $\Delta t \dot{\phi}/\pi f$ is pushed 
into the Dirac sea through the bag boundary to yield the following fractional 
fermion numbers $B_{V}$ inside the bag and $B_{\bar{V}}$ outside the bag, 
respectively
\beq
B_{V}=1-\frac{\theta}{\pi},~~~B_{\bar{V}}=\frac{\theta}{\pi}
\label{bvbbarv11}
\eeq
with $\theta=\phi(R)/f$.  Note that due to the identity $B_{V}+B_{\bar{V}}=1$, 
the total fermion number $B$ is invariant under such changes of the bag 
location and/or size so that one can conclude that the CCP in (1+1) dimensions 
is realized.  Until now we have considered the colorless fermions without 
introducing a gauge field $A^{\mu}$.  If one includes the additional gauge 
degrees of freedom inside the bag, one can have another type of anomaly, 
so-called color anomaly~\cite{rho91color,rho92color}, which also appears in 
the realistic (3+1) dimensional CBM.  (For more details see 
Ref.~\cite{rho96cnd}.)  
    
\begin{figure}
\centerline{\psfig{figure=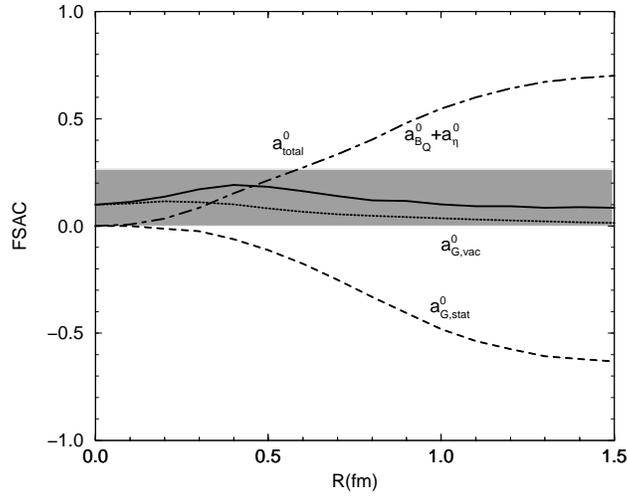,height=3.0in}}
\caption {The flavor singlet axial current of the proton as a function of 
bag radius: (a) the quark and $\eta$ meson contribution 
$a_{B_{Q}}^{0}+a_{\eta}^{0}$, (b) the gluon contributions $a_{G,static}^{0}$ 
and $a_{G,vac}^{0}$ from static gluon due to quark source and gluon vacuum, 
respectively, (c) the total contribution $a_{total}^{0}$.  
The shaded area stands for the range admitted by experiments.}
\label{fsac}
\end{figure}

Now, we would like to briefly comment on the case of the CCP in (3+1) 
dimensions.  One remembers in (\ref{bm}) and (\ref{bqgb}) that the fractional 
baryon numbers $B_{m}$ and $B_{q}$ are described in terms of the Euler 
characteristic and chiral angle, which depend on the bag shape and size, 
respectively, so that one can enjoy the freedom to fix the fractional baryon 
numbers in both phases by adjusting these bag parameters.  Moreover, due to 
the identity $B_{m}+B_{q}=1$, the total baryon number $B$ is invariant under 
such changes of the bag shape and/or size so that one can conclude that 
the CCP in the CBM is realized at least in the physical quantity $B$ in (3+1) 
dimensions as in the above case of (1+1) dimensions.  This fact supports the 
CCP in the (3+1) dimensional CBM even though there is still no rigorous 
verification for this principle in other physical quantities evaluated in the 
CBM.  For instance, one can see the approximate CCP in the flavor 
singlet axial current evaluated in the (3+1) dimensional CBM, as shown in 
Figure~\ref{fsac}.  (For more details see Ref.~\cite{lee99}.)  In the 
following sections, we will see that the CBM can be regarded as a candidate 
unifying the MIT bag and Skyrmion models since the other physical quantities 
are also insensitive enough to suggest the CCP.

\section{Baryon octet magnetic moments}
\setcounter{equation}{0}
\renewcommand{\theequation}{\arabic{section}.\arabic{equation}}


\subsection{Coleman-Glashow sum rules}


Since Coleman and Glashow \cite{coleman61} predicted the magnetic moments 
of the baryon octet about forty years ago, there has been a lot of progress 
in both the theoretical paradigm and experimental verification for the 
baryon magnetic moments.

In this section, we will investigate the explicit Coleman-Glashow 
sum rules and spin symmetries of the magnetic moments of the baryon octet 
in the adjoint representation of the SU(3) flavor group by assuming that the 
chiral bag has the SU(3) flavor symmetry with $m_{u}=m_{d}=m_{s}$, $m_{\pi}
=m_{K}$ and $f_{\pi}=f_{K}$.  Even though the quark and pion masses in 
${\cal L}_{CSB}$ in (\ref{lagcs}) break both the SU(3)$_{L}\times$SU(3)$_{R}$
and the diagonal SU(3) symmetry so that chiral symmetry cannot be conserved, 
these terms without derivatives yield no explicit contribution to the EM 
currents $J_{EM}^{\mu}$ obtainable from (\ref{jvmua}), and at least in the 
adjoint representation of the SU(3) group the EM currents are conserved and 
of the same form as the chiral limit result $J_{EM,CS}^{\mu}$ to preserve 
the U-spin symmetry.

The higher representation mixing in the baryon wave functions, induced by the 
different pseudoscalar meson masses and decay constants outside and different 
quark masses inside the bag, will be discussed in the next section in terms of 
the multiquark structure scheme where the chiral bag has additional meson 
contribution from the $\bar{\rm q}$q content inside the bag.

In the collective quantization scheme of the CBM which was discussed in the 
previous section, the EM currents yield the magnetic moment operators of the 
same form as the chiral symmetric limit consequence $\hat{\mu}_{CS}^{i}=
\hat{\mu}_{CS}^{i(3)}+\frac{1}{\sqrt{3}}\hat{\mu}_{CS}^{i(8)}$ where
\beq
\hat{\mu}_{CS}^{i(a)} = - {\cal N}D_{ai}^{8}
      - {\cal N}^{\prime}d_{ipq}D_{ap}^{8}\hat{T}_{q}^{R}
      + \frac{N_c}{2\sqrt3}{\cal M} D_{a8}^{8}\hat{J}_{i}.
\label{mucsia}
\eeq
Here $\hat{J}_{i}=-\hat{T}_{i}^{R}$ are the SU(2) spin operators, and 
$\hat{T}_{i}^{R}$ and $\hat{T}_{p}^{R}$ are the right SU(3) isospin operators 
along the isospin and strangeness directions respectively, and the inertia 
parameters are of complicated forms given by
\bea
{\cal N}&=& \frac{3}{2}\sum_{m}{\rm sgn}(\varepsilon_{m}){}_{h}\langle m|
\mu_{3}^{(3)}|m\rangle_{h}-3\langle v|\mu_{3}^{(3)}|v\rangle
\nonumber\\
& &+\frac{4\pi}{3e^{3}f_{\pi}}\int_{ef_{\pi}R}^{\infty}{\rm d}z z^{2}
\sin^{2}\theta\left[1+(\frac{d\theta}{dz})^{2}+\frac{\sin^{2}\theta}{z^{2}}\right]
\nonumber\\
{\cal N}^{\prime}&=&-\frac{1}{{\cal I}_{2}}\frac{3}{2}\sum_{m,n} 
\left[\frac{_{h}\langle m|\lambda_{4}|n\rangle_{s} 
{}_{s}\langle n|\mu_{3}^{(4)}|m\rangle_{h}}
{\varepsilon_{m}-\omega_{n}}
+\frac{_{s}\langle m|\lambda_{4}|n\rangle_{h}
{}_{h}\langle n|\mu_{3}^{(4)}|m\rangle_{s}}
{\omega_{m}-\varepsilon_{n}}\right]
\nonumber\\
& &-\frac{1}{{\cal I}_{2}}\frac{3}{2}\sum_{m} 
\frac{\langle v|\lambda_{4}|m\rangle_{s} 
{}_{s}\langle m|\mu_{3}^{(4)}|v\rangle}
{\omega_{m}-\varepsilon_{v}}
+\frac{1}{3\pi e^{2}f_{\pi}^{2}}\int_{ef_{\pi}R}^{\infty}{\rm d}z~z^{2}\sin^{2}
\theta \frac{d\theta}{dz}
\nonumber\\
{\cal M}&=&-\frac{1}{{\cal I}_{1}}\frac{3}{2}\sum_{m,n} 
\frac{{}_{h}\langle m|\lambda_{3}|n\rangle_{h} 
{}_{h}\langle n|\mu_{3}^{(0)}|m\rangle_{h}}
{\varepsilon_{m}-\varepsilon_{n}}
-\frac{1}{{\cal I}_{1}}\frac{3}{2}\sum_{m} 
\frac{\langle v|\lambda_{3}|m\rangle_{h} 
{}_{h}\langle m|\mu_{3}^{(0)}|v\rangle}
{\varepsilon_{m}-\varepsilon_{v}}
\nonumber\\
& &+\frac{1}{{\cal I}_{1}}
\frac{1}{3\pi e^{2}f_{\pi}^{2}}\int_{ef_{\pi}R}^{\infty}{\rm d}z~z^{2}\sin^{2}
\theta \frac{d\theta}{dz}
\label{parameters}
\eea
with $\mu_{3}^{(3)}=\frac{1}{4}\cdot\frac{1}{3}\vec{\lambda}\cdot\vec{V}$, 
$\mu_{3}^{(4)}=\frac{1}{4}V_{3}$ and $\mu_{3}^{(0)}=\frac{1}{4}
\frac{2}{3}V_{3}$ where $V_{i}=\epsilon_{ijk}x_{j}\gamma^{0}\gamma^{k}$ 
and the hermitian conjugate matrix elements are understood in the quark 
phase parts of ${\cal N}^{\prime}$ and ${\cal M}$.  The numerical values 
\cite{park90, hong932} of these inertia parameters are summarized in 
Table~\ref{inertia} and their quark phase inertia parameters are discussed in 
Ref.~\cite{park90} and Appendix B.  Here one notes that ${\cal M}$ and 
${\cal N}^{\prime}$ originate from the topological WZW term along the 
isospin and strangeness directions, respectively.
\begin{table}[t]
\caption{The inertia parameters as a function of the bag radius $R$
with $f_{\pi}=93MeV$, $f_{K}=114MeV$ and $e=4.75$.}
\begin{center}
\begin{tabular}[t]{ccccccc}
\hline
$R$ &${\cal M}$ &${\cal N}$ &${\cal N}^{\prime}$
&${\cal P}$ &${\cal Q}$ &$\omega$\\
\hline
    0.00  &0.671  &5.028  &0.908  &0.762  &0.986  &5.372\\
    0.10  &0.671  &5.088  &0.835  &0.772  &1.000  &6.008\\
    0.20  &0.669  &5.371  &0.791  &0.822  &1.062  &7.144\\
    0.30  &0.660  &5.660  &0.752  &0.886  &1.125  &8.290\\
    0.40  &0.647  &5.697  &0.699  &0.944  &1.159  &8.991\\
    0.50  &0.643  &5.834  &0.615  &1.022  &1.205 &10.133\\
    0.60  &0.656  &6.000  &0.519  &1.112  &1.265 &11.875\\
    0.70  &0.693  &6.128  &0.424  &1.184  &1.305 &14.022\\
    0.80  &0.768  &6.167  &0.335  &1.212  &1.302 &16.550\\
    0.90  &0.886  &6.130  &0.266  &1.185  &1.249 &19.280\\
    1.00  &1.042  &6.056  &0.222  &1.114  &1.156 &21.987\\
\hline
\end{tabular}\\
\end{center}
\label{inertia}
\end{table}
With respect to the octet baryon wave function $\Phi_{B}^{\lambda}$ 
discussed in (\ref{phib}), the spectrum of the magnetic moment operator 
$\hat{\mu}^{i}$ in the adjoint representation of the SU(3) flavor symmetric 
limit has the following U-spin symmetric Coleman-Glashow sum rules 
\cite{coleman61, okubo62, adkins842} due to the degenerate d- and s-flavor 
charges in the SU(3) EM charge operator $\hat{Q}_{EM}$ in the EM currents
\bea
\mu_{\Sigma^{+}}&=&\mu_{p}=\frac{1}{10}{\cal M}+\frac{4}{15}
({\cal N}+\frac{1}{2}{\cal N}^{\prime})
\nonumber\\
\mu_{\Xi^{0}}&=&\mu_{n}=\frac{1}{20}{\cal M}-\frac{1}{5}({\cal N}+\frac{1}{2}
{\cal N}^{\prime})
\nonumber\\
\mu_{\Xi^-}&=&\mu_{\Sigma^{-}}=-\frac{3}{20}{\cal M}-\frac{1}{15}
({\cal N}+\frac12 {\cal N}^\prime)
\nonumber\\
\mu_{\Sigma^{0}}&=&-\mu_{\Lambda}=-\frac{1}{40}{\cal M}+\frac{1}{10}
({\cal N}+\frac12 {\cal N}^\prime).
\label{magmomcs}
\eea
Here one should note that the U-spin symmetry originates from the SU(3) group 
theoretical fact that the matrix elements of the magnetic moment  operators 
in (\ref{mucsia}) in the adjoint representation, such as $\langle 8|D_{38}^{8}
+\frac{1}{\sqrt{3}}D_{88}^{8}|8\rangle$, have degenerate values for the U-spin 
multiplets ($p$, $\Sigma^{+}$), ($n$, $\Xi^{0}$) and ($\Xi^{-}$, $\Sigma^{-}$) 
with the same electric charges.

In (\ref{magmomcs}) one can easily see that $\mu_{\Sigma}(I_{3})=\mu_{\Sigma
^{0}}+I_{3}\Delta_{\mu_{\Sigma}}$ where $\Delta_{\mu_{\Sigma}}=\frac{1}{8}
{\cal M}+\frac{1}{6}({\cal N}+\frac12{\cal N}^{\prime})$ so that the summation 
$\mu_{\Sigma^{+}}+\mu_{\Sigma^{-}}$ is independent of the third component of 
the isospin $I_{3}$ so that one can obtain the other Coleman-Glashow sum 
rules \cite{coleman61, okubo62, adkins842, okubo63}
\beq
\mu_{\Sigma^{0}}=\frac12(\mu_{\Sigma^{+}}+\mu_{\Sigma^{-}}).
\label{musigma0}
\eeq
Since there is no SU(3) singlet contribution to the magnetic moment, the 
summation of the magnetic moments over the octet baryon vanishes to yield 
the identity \cite{adkins842}
\beq
\sum_{B\in {\rm octet}}\mu_{B}=0.
\label{sumboctet}
\eeq

Introducing in the meson pieces of the CBM Lagrangian (\ref{cbmlagrangian}) 
the minimal photon coupling to the derivative terms, $\partial_{\mu}U
\rightarrow \nabla_{\mu}U=\partial_{\mu}U+ie A_{\mu}[\hat{Q}_{EM},U]$ with 
the SU(3) EM charge operator $\hat{Q}_{EM}$ one obtains the $\Lambda\Sigma
^{0}$ transition matrix element for the decay $\Sigma^{0}\rightarrow \Lambda
+\gamma$
\beq
\frac{1}{\sqrt{3}}\mu_{\Lambda\Sigma^{0}}=-\frac{1}{40}{\cal M}+\frac{1}{10}
({\cal M}+\frac12{\cal N}^{\prime})
\label{mulambdasigma}
\eeq
which, in incorporating an SU(3) singlet contribution of the photon, satisfies 
the modified Coleman-Glashow sum rules \cite{okubo63, gupta75}
\bea
\frac{1}{\sqrt{3}}\mu_{\Lambda\Sigma^{0}}&=&\mu_{\Lambda}-\mu_{n}
\nonumber\\
\frac{6}{\sqrt{3}}\mu_{\Lambda\Sigma^{0}}&=&\mu_{\Sigma^{0}}
-2\mu_{\Xi^{0}}+3\mu_{\Lambda}-2\mu_{n}.
\label{muls02}
\eea

It is also interesting to note that the hyperon and transition magnetic moments in the SU(3) 
flavor symmetric limit can be expressed in terms of the nucleon 
magnetic moments only \cite{coleman61, okubo62, lipkin84}
\bea
& &\mu_{\Lambda}=\frac12 \mu_{n}\nonumber\\
& &\mu_{\Xi^{-}}=-(\mu_{p}+\mu_{n})\nonumber\\
& &\mu_{\Sigma^{+}}-\mu_{\Sigma^{-}}+\mu_{\Xi^{0}}-\mu_{\Xi^{-}}=3(\mu_{p}
+\mu_{n})\nonumber\\
& &\frac{1}{\sqrt{3}}\mu_{\Lambda\Sigma^{0}}=-\frac12\mu_{n}.
\label{musums}
\eea
Here one should note that the transition magnetic moment possesses an arbitrary 
global phase factor in itself, while the other octet magnetic moments have a 
definite overall sign.  In (\ref{musums}) we have used the phase convention 
of Ref.~\cite{dydak77}, which is consistent with de Swart convention 
\cite{deswart63} of the SU(3) isoscalar factors used in the CBM.


\subsection{Strangeness in Yabu-Ando scheme}


In the previous section we have considered the CBM in the adjoint representation 
with the SU(3) flavor symmetry, where the U-spin symmetry is conserved even 
though we have the chiral symmetry breaking mass terms.  Now we include the 
SU(3) flavor symmetry breaking terms ${\cal L}_{FSB}$ in (\ref{lagcs}) to yield 
the magnetic moment operators $\hat{\mu}_{FSB}^{i(a)}$ of (\ref{mufsbia}) 
induced by the symmetry breaking kinetic terms.  However, the symmetry is also 
broken nonperturbatively by the mass terms via the higher dimensional IR 
channels where the CBM can be treated in the Yabu-Ando scheme \cite{yabu88} to 
yield the multiquark structure with the meson cloud ${\it inside}$ the bag.  
The quantum mechanical perturbative scheme to the symmetry breaking effects 
in the multiquark structure will be discussed in terms of the V-spin symmetry 
in the next section.

Assuming that the CBM includes the kinetic term in ${\cal L}_{FSB}$ in the 
collective quantization, the Noether scheme gives rise to the U-spin symmetry 
breaking conserved EM currents $J_{EM,FSB}^{\mu}$ so that $J_{EM}^{\mu}=
J_{EM,CS}^{\mu}+J_{EM,FSB}^{\mu}$.  With the spinning CBM ansatz the EM 
currents yield the magnetic moment operators $\hat{\mu}^{i}=\hat{\mu}^{i(3)}
+\frac{1}{\sqrt{3}}\hat{\mu}^{i(8)}$ where $\hat{\mu}^{i(a)}=\hat{\mu}_{CS}
^{i(a)}+\hat{\mu}_{FSB}^{i(a)}$.  Here $\hat{\mu}_{CS}^{i(a)}$ is given in 
(\ref{mucsia}) and $\hat{\mu}_{FSB}^{i(a)}$ is described as below
\beq
\hat{\mu}^{i(a)}_{FSB}=
      - {\cal P} D_{ai}^{8} (1-D_{88}^{8})
      + \frac{\sqrt{3}}{2}{\cal Q}d_{ipq}D_{ap}^{8}D_{8q}^{8}
\label{mufsbia}
\eeq
where ${\cal P}$ and ${\cal Q}$ are the inertia parameters along the 
isospin and strangeness directions obtained from the mesonic Lagrangian 
${\cal L}_{FSB}$ 
\bea
{\cal P}&=&\frac{8\pi}{9e^{3}f_{\pi}^{3}}(f_{K}^{2}-f_{\pi}^{2})
\int_{ef_{\pi}R}^{\infty}{\rm d}z z^{2}\sin^{2}\theta\cos\theta\nonumber\\
{\cal Q}&=&\frac{8\pi}{9e^{3}f_{\pi}^{3}}(f_{K}^{2}-f_{\pi}^{2})
\int_{ef_{\pi}R}^{\infty}{\rm d}z z^{2}\sin^{2}\theta\nonumber\\
m{\cal I}_{2}&=&\frac{8\pi}{3e^{3}f_{\pi}^{3}}(f_{K}^{2}m_{K}^{2}
-f_{\pi}^{2}m_{\pi}^{2})\int_{ef_{\pi}R}^{\infty}{\rm d}z z^{2}(1-\cos\theta)
\nonumber\\
& &+\frac{4\pi}{3ef_{\pi}}(f_{K}^{2}-f_{\pi}^{2})\int_{ef_{\pi}R}^{\infty}
{\rm d}z z^{2}\left(\left(\frac{{\rm d}\theta}{{\rm d}z}\right)^{2}
+\frac{2\sin^{2}\theta}{z^{2}}\right)\cos\theta\nonumber\\
& &+\frac{1}{3}m_{s}N_{c}\sum_{n}{}_{h}\langle n|\gamma^{0}|n\rangle_{h}
\label{inertiapq}
\eea
whose numerical values are shown in Table 1.

Breaking up the tensor product of the Wigner $D$ functions into a sum of the 
single $D$ functions \cite{deswart63},
\beq
D_{a_{1}b_{1}}^{8}D_{a_{2}b_{2}}^{8}=\sum_{a,b,\lambda,\gamma}
\left(
\begin{array}{ccc}
8 & 8 & \lambda_{\gamma}\\
a_{1} & a_{2} & a
\end{array}
\right)
\left(
\begin{array}{ccc}
8 & 8 & \lambda_{\gamma}\\
b_{1} & b_{2} & b 
\end{array}
\right)
D_{ab}^{\lambda},
\label{ddsum}
\eeq
one can rewrite the isovector and isoscalar parts of the operator 
$\hat{\mu}_{FSB}^{i(a)}$ as
\bea
\hat{\mu}^{i(3)}_{FSB}&=&{\cal P}(-\frac{4}{5}D_{3i}^{8}+\frac{1}{4}(
D_{3i}^{10}+D_{3i}^{\bar{10}})+\frac{3}{10}D_{3i}^{27})
+{\cal Q}(\frac{3}{10}D_{3i}^{8}-\frac{3}{10}D_{3i}^{27})
\nonumber \\
\hat{\mu}^{i(8)}_{FSB}&=&{\cal P}(-\frac{6}{5}D_{8i}^{8}+\frac{9}{20}
D_{8i}^{27})+{\cal Q}(-\frac{3}{10}D_{8i}^{8}-\frac{9}{20}D_{8i}^{27}).
\label{muop38}
\eea
Here the ${\bf 10}$ and $\bar{\bf 10}$ IRs, which are absent in 
the isoscalar channel due to their nonvanishing hypercharge, come out together 
to conserve the hermitian property of the operator in the isovector channel, 
while the singlet operator constructed in the singlet IR ${\bf 1}$ cannot allow the 
quantum number ($Y_{R}$;$J$,-$J_{3}$)=(0;1,0) \cite{adkins842} so that the 
operator does not occur in either channel.

Using the octet baryon wave function (\ref{phib}) for the matrix elements of 
the full magnetic moment operator $\hat{\mu}^{i}$, one can obtain the 
hyperfine structure in the adjoint representation
\bea
\mu_{p}&=&\frac{1}{10}{\cal M}+\frac{4}{15}({\cal N}+\frac{1}{2}
{\cal N}^{\prime})+\frac{8}{45}{\cal P}-\frac{2}{45}{\cal Q}\nonumber\\
\mu_{n}&=&\frac{1}{20}{\cal M}-\frac{1}{5}({\cal N}+\frac{1}{2}
{\cal N}^{\prime})-\frac{1}{9}{\cal P}+\frac{7}{90}{\cal Q}\nonumber\\
\mu_{\Lambda}&=&\frac{1}{40}{\cal M}-\frac{1}{10}({\cal N}+\frac12
{\cal N}^\prime)-\frac{1}{10}{\cal P}-\frac{1}{20}{\cal Q}\nonumber\\
\mu_{\Xi^0}&=&\frac{1}{20}{\cal M}-\frac{1}{5}({\cal N}+\frac12
{\cal N}^\prime)-\frac{11}{45}{\cal P}-\frac{1}{45}{\cal Q}\nonumber\\
\mu_{\Xi^-}&=&-\frac{3}{20}{\cal M}-\frac{1}{15}({\cal N}+\frac12
{\cal N}^\prime)-\frac{4}{45}{\cal P}-\frac{2}{45}{\cal Q}\nonumber\\
\mu_{\Sigma^+}&=&\frac{1}{10}{\cal M}+\frac{4}{15}({\cal N}+\frac12
{\cal N}^\prime)+\frac{13}{45}{\cal P}-\frac{1}{45}{\cal Q}\nonumber\\
\mu_{\Sigma^0}&=&-\frac{1}{40}{\cal M}+\frac{1}{10}({\cal N}+\frac12
{\cal N}^\prime)+\frac{11}{90}{\cal P}+\frac{1}{36}{\cal Q}
\nonumber\\
\mu_{\Sigma^-}&=&-\frac{3}{20}{\cal M}-\frac{1}{15}({\cal N}+\frac12
{\cal N}^\prime)-\frac{2}{45}{\cal P}+\frac{7}{90}{\cal Q}.
\label{octetmus}
\eea
Here one notes that the Coleman-Glashow sum rules (\ref{musigma0}) and 
(\ref{sumboctet}) are still valid while the other relations (\ref{magmomcs}) 
and(\ref{musums}) are no longer retained due to the SU(3) flavor symmetry 
breaking effects of $m_{u}=m_{d}\neq m_{s}$, $m_{\pi}\neq m_{K}$ and 
$f_{\pi}\neq f_{K}$ through the inertia parameters ${\cal P}$ and ${\cal Q}$.

By substituting the EM charge operator $\hat{Q}_{EM}$ with the q-flavor EM 
charge operator $\hat{Q}_{q}$, one can obtain the q-flavor currents 
$J_{EM}^{\mu (q)}=J_{EM,CS}^{\mu (q)}+J_{EM,FSB}^{\mu (q)}$ in the SU(3) flavor 
symmetry broken case to yield the EM currents with three flavor pieces 
$J_{EM}^{\mu}=J_{EM}^{\mu (u)}+J_{EM}^{\mu (d)}+J_{EM}^{\mu (s)}$.  Here one 
notes that by defining the flavor projection operators
\bea
\hat{P}_{u}&=&\frac{1}{3}+\frac12 \lambda_{3}+\frac{1}{2\sqrt{3}}\lambda_{8},
\nonumber\\
\hat{P}_{d}&=&\frac{1}{3}-\frac12 \lambda_{3}+\frac{1}{2\sqrt{3}}\lambda_{8},
\nonumber\\
\hat{P}_{s}&=&\frac{1}{3}-\frac{1}{2\sqrt{3}}\lambda_{8},
\label{projectionop}
\eea
satisfying $\hat{P}_{q}^{2}=\hat{P}_{q}$ and $\sum_{q}\hat{P}_{q}=1$, one can 
easily construct the q-flavor EM charge operators $\hat{Q}_{q}=\hat{Q}_{EM}
\hat{P}_{q}=Q_{q}\hat{P}_{q}$.

As in the previous section, one can then find the magnetic moment operator in 
the u-flavor channel
\bea
\hat{\mu}^{i(u)}&=&{\cal M}\frac{2N_c}{9}(1+\frac{\sqrt{3}}{2}D_{38}^{8}
+\frac12 D_{88}^{8})\hat{J}_{i}
- {\cal N}\frac{2}{3}(D_{3i}^{8}
+\frac{1}{\sqrt{3}}D_{8i}^{8})
\nonumber\\
& &-{\cal N}^{\prime}\frac{2}{3}d_{ipq}
(D_{3p}^{8}+\frac{1}{\sqrt{3}}D_{8p}^{8})\hat{T}_{q}^{R}
-{\cal P}\frac{2}{3}(D_{3i}^{8}+\frac{1}{\sqrt{3}}D_{8i}^{8})
(1-D_{88}^{8})
\nonumber\\
& &+{\cal Q}\frac{1}{\sqrt{3}}d_{ipq}
(D_{3p}^{8}+\frac{1}{\sqrt{3}}D_{8p}^{8})
    D_{8q}^{8}
\label{opmuiu}
\eea
to yield the u-components of the baryon octet magnetic moments in the adjoint 
representation
\bea
\mu_{p}^{(u)}&=&\frac{2}{510}{\cal M}+\frac{8}{45}({\cal N}+\frac{1}{2}
{\cal N}^{\prime})+\frac{16}{135}{\cal P}-\frac{4}{135}{\cal Q}\nonumber\\
\mu_{n}^{(u)}&=&\frac{11}{30}{\cal M}-\frac{2}{15}({\cal N}+\frac{1}{2}
{\cal N}^{\prime})-\frac{2}{27}{\cal P}+\frac{7}{135}{\cal Q}\nonumber\\
\mu_{\Lambda}^{(u)}&=&\frac{7}{20}{\cal M}-\frac{1}{15}({\cal N}+\frac{1}{2}
{\cal N}^{\prime})-\frac{1}{15}{\cal P}-\frac{1}{30}{\cal Q}\nonumber\\
\mu_{\Xi^{0}}^{(u)}&=&\frac{11}{30}{\cal M}-\frac{2}{15}({\cal N}+\frac{1}{2}
{\cal N}^{\prime})-\frac{22}{135}{\cal P}-\frac{2}{135}{\cal Q}\nonumber\\
\mu_{\Xi^{-}}^{(u)}&=&\frac{7}{30}{\cal M}-\frac{1}{45}({\cal N}+\frac{1}{2}
{\cal N}^{\prime})-\frac{8}{135}{\cal P}-\frac{4}{135}{\cal Q}\nonumber\\
\mu_{\Sigma^{+}}^{(u)}&=&\frac{2}{5}{\cal M}+\frac{8}{45}({\cal N}+\frac{1}{2}
{\cal N}^{\prime})+\frac{26}{135}{\cal P}-\frac{2}{135}{\cal Q}\nonumber\\
\mu_{\Sigma^{0}}^{(u)}&=&\frac{19}{60}{\cal M}+\frac{1}{15}({\cal N}+\frac{1}{2}
{\cal N}^{\prime})+\frac{11}{135}{\cal P}+\frac{1}{54}{\cal Q}\nonumber\\
\mu_{\Sigma^{-}}^{(u)}&=&\frac{7}{30}{\cal M}-\frac{2}{45}({\cal N}+\frac{1}{2}
{\cal N}^{\prime})-\frac{4}{135}{\cal P}+\frac{7}{135}{\cal Q}.
\label{muus}
\eea
Similarly one can construct the s-flavor magnetic moment operator
\bea
\hat{\mu}^{i(s)} &=&-{\cal M}\frac{N_c}{9}(1-D_{88}^{8})\hat{J}_{i}
    -{\cal N}\frac{2}{3\sqrt{3}}D_{8i}^{8}
    -{\cal N}^{\prime}\frac{2}{3\sqrt{3}}d_{ipq}D_{8p}^{8}\hat{T}_{q}^{R}
\nonumber\\
& &-{\cal P}\frac{2}{3\sqrt{3}}D_{8i}^{8}(1-D_{88}^{8})
   +{\cal Q}\frac{1}{3}d_{ipq}D_{8p}^{8}D_{8q}^{8}
\label{opmuis}
\eea
to obtain the baryon octet magnetic moments in the s-flavor channel 
\bea
\mu_{N}^{(s)}&=&-\frac{7}{60}{\cal M}+\frac{1}{45}({\cal N}+\frac{1}{2}
{\cal N}^{\prime})+\frac{1}{45}{\cal P}+\frac{1}{90}{\cal Q}\nonumber\\
\mu_{\Lambda}^{(s)}&=&-\frac{3}{20}{\cal M}-\frac{1}{15}({\cal N}+\frac{1}{2}
{\cal N}^{\prime})-\frac{1}{15}{\cal P}-\frac{1}{30}{\cal Q}\nonumber\\
\mu_{\Xi}^{(s)}&=&-\frac{11}{530}{\cal M}-\frac{42}{415}({\cal N}+\frac{1}{2}
{\cal N}^{\prime})-\frac{122}{9135}{\cal P}-\frac{12}{4135}{\cal Q}\nonumber\\
\mu_{\Sigma}^{(s)}&=&-\frac{11}{60}{\cal M}+\frac{1}{15}({\cal N}+\frac{1}{2}
{\cal N}^{\prime})+\frac{11}{135}{\cal P}+\frac{1}{54}{\cal Q}.
\label{muss}
\eea
Here one notes that all the baryon magnetic moments satisfy the 
model-independent relations in the u- and d-channels and the I-spin symmetry 
in the s-flavor channel where the isomultiplets have the same strangeness 
number
\bea
\mu_{B}^{(d)}&=&\frac{Q^{d}}{Q^{u}}\mu_{B}^{(u)}
\label{udsym}\\
\mu_{B}^{(s)}&=&\mu_{\bar{B}}^{(s)}.
\label{udssymmetry}
\eea
Here $\bar{B}$ is the isospin conjugate baryon in the isomultiplets of the 
baryon.

For the $\Lambda \Sigma^{0}$ transition, one can obtain the u- and d-flavor 
components given by the different pattern
\beq
\frac{1}{\sqrt{3}}\mu_{\Lambda\Sigma^{0}}^{(u)}=\frac{2}{\sqrt{3}}\mu
_{\Lambda\Sigma^{0}}^{(d)}=-\frac{1}{60}{\cal M}+\frac{1}{15}({\cal N}
+\frac{1}{2}{\cal N}^{\prime})+\frac{8}{135}{\cal P}-\frac{7}{270}{\cal Q},
\label{lambdasigma2}
\eeq
and the vanishing s-flavor component.

Until now we have considered the explicit SU(3) flavor symmetry breaking 
effects in the magnetic moment operators $\hat{\mu}^{i}$ of the CBM in the 
adjoint representation, where the mass terms in ${\cal L}_{CSB}$ and 
${\cal L}_{FSB}$ cannot contribute to $\hat{\mu}^{i}$ due to the absence of 
the derivative term.  Treating the mass terms as the representation dependent 
fraction in the Hamiltonian approach, one can see that the term with 
$D_{88}^{8}$ induces the representation mixing effects in the baryon wave 
functions.  In order to investigate explicitly the mixing effects in the 
Yabu-Ando scheme, we quantize the collective variables $A(t)$ so that we can 
obtain the Hamiltonian of the form
\beq
H=M+\frac12  \left(\frac{1}{{\cal I}_1}-\frac{1}{{\cal I}_2}\right) \hat{J}^2
   +\frac{1}{2{\cal I}_2}\left(h_{SB}-\frac{3}{4}\hat{Y}_{R}^{2}\right)
\label{yabuham}
\eeq
where ${\cal I}_{1}$ and ${\cal I}_{2}$ are the moments of inertia of the CBM 
along the isospin and the strangeness directions respectively and their 
explicit expressions are given in (\ref{i1i2}).

Here one remembers that the static mass $M$ obtainable from (\ref{stmass}) 
satisfies the equation of motion for the chiral angle (\ref{eom}).  The pion 
mass in (\ref{eom}) also yields deviation from the chiral limit chiral angle 
for a fixed bag radius so that the numerical results in the massive CBM can 
be worsened when one uses the experimental decay constant.

In order to obtain the numerical results in Table~\ref{inertia}, we use the 
massless 
chiral angle and the experimental data $f_{\pi}=93$ MeV, $f_{K}=114$ MeV and 
$e=4.75$ since $(m_{u}+m_{d})/m_{s}\approx m_{\pi}^{2}/m_{K}^{2}\approx 0.1$, 
so that we can neglect the light quark and pion masses.  This approximation 
would not be contradictory to our main purpose to investigate the massive 
kaon contributions to the baryon magnetic moments.

On the other hand, the chiral and SU(3) flavor symmetry breaking induces the 
representation dependent part\footnote{To be consistent with the massless 
chiral angle approximation, we also neglect the u- and d-quark contributions, 
$\frac{2}{3}\omega_{u,d}(1\pm \frac{\sqrt{3}}{2}D_{38}^{8}+\frac12 D_{88}^{8}
)$ with $\omega_{u,d}={\cal I}_{2}m_{u,d}N_{c}\sum_{n}\langle n|\gamma^{0}
|n\rangle$, which can break the I-spin symmetry through $D_{38}^{8}$.}  
\beq
h_{SB}=\hat{C}_{2}^{2}+\frac{2}{3}\omega (1-D_{88}^{8}),
\label{hsb}
\eeq
where $\hat{C}_{2}^{2}$ is the Casimir operator in the SU(3)$_{L}$ group and 
the symmetry breaking strength is given by
\bea
\omega&=&\frac{8\pi}{e^{3}f_{\pi}^{3}}{\cal I}_{2}(f_{K}^{2}m_{K}^{2}
-f_{\pi}^{2}m_{\pi}^{2})\int_{ef_{\pi}R}^{\infty}{\rm d}z z^{2}(1-\cos\theta)
+{\cal I}_{2}m_{s}N_{c}\sum_{n}\langle n|\gamma^{0}|n\rangle_{h}
\nonumber\\
& &+\frac{4\pi}{ef_{\pi}}{\cal I}_{2}(f_{K}^{2}-f_{\pi}^{2})
\int_{ef_{\pi}R}^{\infty}{\rm d}z z^{2} \left(\left(\frac{d\theta}{dz}
\right)^{2}+\frac{2\sin^{2}\theta}{z^{2}} \right)\cos\theta
\label{omegapara}
\eea
with the numerical values in Table~\ref{inertia}.  Of course one can easily see that, in 
the vanishing $\omega$ limit, the Hamiltonian (\ref{yabuham}) approaches to 
the previous one $H_{0}$ in (\ref{chiralham}) with the SU(3) flavor symmetry.

Now one can directly diagonalize the Hamiltonian $h_{SB}$ in the 
eigenvalue equation $h_{SB}|B\rangle=\varepsilon_{SB}|B\rangle$ of the 
Yabu-Ando scheme \cite{yabu88} with the eigenstate denoted by 
$|B\rangle=\sum_{\lambda}C_{\lambda}^{B}|B\rangle^{\lambda}$ where $C_{\lambda}^{B}$ 
is the representation mixing coefficient and $|B\rangle$ are octet baryon 
wave function in the $\lambda$ dimensional IR discussed in (\ref{blambda}). 
\begin{table}[t]
\caption{The baryon octet magnetic moments in the U-spin symmetry broken 
case in the Yabu-Ando scheme of the CBM, compared with the SU(2) CBM and 
naive NRQM predictions and the experimental data}
\begin{center}
\begin{tabular}{crrrrrrrrr}
\hline
R &$\mu_{p}$ &$\mu_{n}$ &$\mu_{\Lambda}$ &$\mu_{\Xi^{0}}$  &$\mu_{\Xi^{-}}$ 
  &$\mu_{\Sigma^{+}}$ &$\mu_{\Sigma^{0}}$ &$\mu_{\Sigma^{-}}$ 
  &$\mu_{\Lambda\Sigma^{0}}$\\
\hline
0.00 &1.69 &$-1.28$ &$-0.56$  &$-1.22$ &$-0.47$ &1.73 &0.69 &$-0.36$ &1.19\\ 
0.10 &1.71 &$-1.30$ &$-0.56$  &$-1.22$ &$-0.46$ &1.73 &0.69 &$-0.36$ &1.20\\
0.20 &1.80 &$-1.39$ &$-0.56$  &$-1.27$ &$-0.45$ &1.80 &0.72 &$-0.36$ &1.28\\
0.30 &1.89 &$-1.48$ &$-0.58$  &$-1.32$ &$-0.45$ &1.86 &0.75 &$-0.36$ &1.36\\
0.40 &1.91 &$-1.50$ &$-0.57$  &$-1.33$ &$-0.44$ &1.87 &0.76 &$-0.35$ &1.38\\
0.50 &1.96 &$-1.54$ &$-0.57$  &$-1.36$ &$-0.42$ &1.89 &0.77 &$-0.35$ &1.43\\
0.60 &2.02 &$-1.59$ &$-0.57$  &$-1.38$ &$-0.40$ &1.90 &0.78 &$-0.34$ &1.48\\
0.70 &2.07 &$-1.62$ &$-0.55$  &$-1.39$ &$-0.37$ &1.89 &0.78 &$-0.34$ &1.52\\
0.80 &2.10 &$-1.62$ &$-0.53$  &$-1.37$ &$-0.34$ &1.85 &0.76 &$-0.34$ &1.51\\
0.90 &2.10 &$-1.59$ &$-0.49$  &$-1.33$ &$-0.31$ &1.79 &0.73 &$-0.34$ &1.48\\
1.00 &2.10 &$-1.55$ &$-0.45$  &$-1.26$ &$-0.28$ &1.73 &0.69 &$-0.35$ &1.26\\
${\rm SU(2)}$ &2.27  &$-1.35$  &$-0.61$  &$-1.33$  &$-0.60$ &2.28 &0.82 
     &$-0.64$ &1.26\\  
${\rm naive}$ &2.79  &$-1.86$  &$-0.61$  &$-1.43$  &$-0.50$ &2.68 &0.82 
     &$-1.04$ &1.61\\
${\rm exp}$ &2.79  &$-1.91$  &$-0.61$  &$-1.25$ &$-0.65$ &2.46 &$-$
     &$-1.16$ &1.61\\
\hline
\end{tabular}
\end{center}
\label{octet}
\end{table}

The possible SU(3) representations of the minimal multiquark Fock space 
qqq+qqq$\bar{\rm q}$q are restricted by the Clebsch-Gordan series 
${\bf 8}\oplus\bar{\bf 10}\oplus{\bf 27}$\footnote{Because of the baryon 
constraint $Y_{R}=1$ originated from the WZW term, the spin-$\frac12$ decuplet 
baryons to ${\bf 10}\oplus{\bf 27}\oplus{\bf 35}$.  In the qqq$\bar{\rm q}$q 
multiquark structure the Clebsch-Gordan decomposition of the tensor product of 
the two IR's is given by $({\bf 3}\otimes{\bf 3}\otimes{\bf 3})\otimes
(\bar{\bf 3}\otimes{\bf 3})=({\bf 1}\oplus{\bf 8}^{2}\oplus{\bf 10})\otimes
({\bf 1}\oplus{\bf 8})={\bf 1}^{3}\oplus{\bf 8}^{8}\oplus{\bf 10}^{4}\oplus
\bar{\bf 10}^{2}\oplus{\bf 27}^{3}\oplus{\bf 35}$ where the superscript stands 
for the number of different IR's with the same dimension.} 
in the baryon octet with $Y_{R}=1$ and $J=\frac12$, so that the representation 
mixing coefficients can be evaluated by solving the eigenvalue equation of the 
3$\times$3 Hamiltonian matrix in (\ref{yabuham}).

Since in the multiquark scheme of the CBM the baryon wave functions act 
nonperturbatively on the magnetic moment operators with the quark and meson 
phase contributions in their inertia parameters, one could have the meson 
cloud content $\bar{\rm q}$q inside the bag via the channel of 
qqq$\bar{\rm q}$q multiquark Fock space.  Here in order to construct the 
pseudoscalar mesons inside the bag, the $\bar{\rm q}$q contents refer to all 
the appropriate flavor combinations.

In the SU(3) flavor sector of the CBM, the mechanism explaining the meson 
cloud inside the bag surface seems \cite{hong931} closely related to the 
pseudoscalar composite operators $\bar{\psi}i\gamma_{5}\lambda_{a}\psi\sim
\pi_{a}$ ($a=1,...,8$) since the pseudoscalar quark bilinears transform like 
$({\bf 3},\bar{\bf 3})\oplus(\bar{\bf 3},{\bf 3})$, while in the U(1) flavor 
sector the mechanism is supposed \cite{hong931} to be described with the 
anomalous gluon effect in the quark-antiquark annihilation channel 
\cite{derujula75}.  In the SU(3) CBM with the minimal multiquark Fock space, the meson cloud 
content $\bar{\rm q}$q inside the bag surface can be then phenomenologically 
illustrated \cite{hong931} by sum of two topologically different Feynman 
diagrams.  One notes here that, in the multiquark scheme
of the SU(3) CBM, the baryon magnetic moments have two-body operator effect 
as well as one-body self interaction in the sense of quasi-particle model in 
the many body problem.  The gluons are supposed to mediate the pseudoscalar 
$\eta_{0}$ meson cloud via the $\bar{\rm q}$q pair creation and annihilation 
process.

As shown in Table~\ref{octet}, the U-spin symmetry breaking effect, 
through the explicit operator $\hat{\mu}_{FSB}^{i(a)}$ and the Yabu-Ando 
scheme in the multiquark structure, improves the fit to most of the baryon 
octet magnetic moments.  However if the experimental data \cite{datagroup9x} 
is correct, the fit to the $\mu_{\Sigma^{-}}$ seems a little bit worsened.  
Here one should note that $\mu_{\Lambda}$ seems to be well predicted in the CBM 
as in the naive NRQM since $\mu_{\Lambda}$ could be mainly determined from the 
strange quark and kaon whose masses are kept in our massless profile 
approximation.  From the numerical values in Table~\ref{octet}, one can see 
that the SU(3) 
CBM could be regarded to be a good candidate of the unification of the bag 
and Skyrmion models with predictions almost independent of the bag radius.  
For the $\Sigma^{0}\rightarrow \Lambda+\gamma$ transition matrix element, we 
obtain the numerical prediction of the CBM $\mu_{\Lambda\Sigma^{0}}=1.19-1.53$ 
comparable to the experimental data $\mu_{\Lambda\Sigma^{0}}^{exp}=1.61$ 
\cite{datagroup9x}.

In the q-flavor channels, the I-spin symmetry and model-independent relations 
(\ref{udssymmetry}) hold in the multiquark scheme since the Hamiltonian 
$h_{SB}$ has the eigenstates degenerate with the isomultiplets in our 
approximation, where the I-spin symmetry breaking light quark masses are 
neglected.  


\section{Baryon decuplet magnetic moments}
\setcounter{equation}{0}
\renewcommand{\theequation}{\arabic{section}.\arabic{equation}}


\subsection{Model-independent sum rules}


In the previous section we have calculated the magnetic moments of baryon 
octet in the SU(3) flavor case \cite{hong931}, where the Coleman-Glashow 
sum rules~\cite{coleman75} including the U-spin symmetry hold up to the SU(3) 
flavor symmetric limit of the adjoint representation to suggest the 
possibility of a unification of the SU(3) CBM and the naive NRQM.  The 
measurements of the magnetic moments of the decuplet baryons were reported for 
$\mu_{\Delta^{++}}$~\cite{bosshard91} and $\mu_{\Omega^{-}}$~\cite{diehl91} 
to yield a new avenue for understanding hadronic structure.

In this section we will calculate the magnetic moments of the baryon 
decuplet~\cite{hong94} to compare with the known experimental data, to make new 
predictions in the CBM for the unknown experiments and to derive the 
model-independent sum rules which will be used later to generalize the CBM 
conjecture~\cite{hong931} for the baryon decuplet.

In order to estimate the magnetic moments of the decuplet baryons in the 
U-spin broken symmetry case, we have at first derived the explicit magnetic 
moment operators $\hat{\mu}_{FSB}^{i(a)}$ from the flavor symmetry 
breaking Lagrangian ${\cal L}_{FSB}$ in the adjoint representation where 
$\hat{\mu}_{CSB}^{i(a)}$ vanishes.  In the SU(3) cranking scheme described 
in the previous sections, the magnetic moment operators $\hat{\mu}^{i}$ are 
then given by (\ref{mucsia}) and (\ref{mufsbia}) and the tensor product of 
the Wigner $D$ functions in $\hat{\mu}_{FSB}^{i(a)}$ can be decomposed into a 
sum of the single $D$ functions to yield the isovector and isoscalar parts 
as below
\bea
\hat{\mu}^{i(3)}_{FSB}&=&{\cal P}(-\frac{4}{5}D_{3i}^{8}+\frac{3}{10}
D_{3i}^{27})+{\cal Q}(\frac{3}{10}D_{3i}^{8}-\frac{3}{10}D_{3i}^{27})
\nonumber \\
\hat{\mu}^{i(8)}_{FSB}&=&{\cal P}(-\frac{6}{5}D_{8i}^{8}+\frac{9}{20}
D_{8i}^{27})+{\cal Q}(-\frac{3}{10}D_{8i}^{8}-\frac{9}{20}D_{8i}^{27}).
\label{mu3810}
\end{eqnarray}
Here one notes that, to conserve the hermitian property of the magnetic moment
operator, ${\bf 10}$ and $\bar{\bf 10}$ IRs appear together in the isovector 
channel of the baryon octet as discussed in the previous section while the 
${\bf 1}$, ${\bf 10}$ and $\bar{{\bf 10}}$ IRs do not take place in the 
decuplet baryons.

With respect to the decuplet baryon wave function $\Phi_{B}^{\lambda}$ in 
(\ref{phib}) the magnetic moment operator $\hat{\mu}^{i}$ has the
spectrum for the decuplet in the adjoint representation 
\begin{eqnarray}
\mu_{\Delta^{++}}&=&\frac{1}{8}{\cal M}+\frac{1}{2}({\cal N}-\frac{1} {2%
\sqrt{3}}{\cal N}^{\prime})+\frac{3}{7}{\cal P}-\frac{3}{56}{\cal Q} 
\nonumber \\
\mu_{\Delta^{+}}&=&\frac{1}{16}{\cal M}+\frac{1}{4}({\cal N}-\frac{1} {2%
\sqrt{3}}{\cal N}^{\prime})+\frac{5}{21}{\cal P}+\frac{1}{84}{\cal Q} 
\nonumber \\
\mu_{\Delta^{0}}&=&\frac{1}{21}{\cal P}+\frac{13}{168}{\cal Q}  \nonumber \\
\mu_{\Delta^{-}}&=&-\frac{1}{16}{\cal M}-\frac{1}{4}({\cal N}-\frac{1} {2%
\sqrt{3}}{\cal N}^{\prime})-\frac{1}{7}{\cal P}+\frac{1}{7}{\cal Q} 
\nonumber \\
\mu_{\Sigma^{*+}}&=&\frac{1}{16}{\cal M}+\frac{1}{4}({\cal N}-\frac{1} {2%
\sqrt{3}}{\cal N}^{\prime})+\frac{19}{84}{\cal P}-\frac{17}{168}{\cal Q} 
\nonumber \\
\mu_{\Sigma^{*0}}&=&\frac{1}{84}{\cal P}-\frac{1}{84}{\cal Q}  \nonumber \\
\mu_{\Sigma^{*-}}&=&-\frac{1}{16}{\cal M}-\frac{1}{4}({\cal N}-\frac{1} {2%
\sqrt{3}}{\cal N}^{\prime})-\frac{17}{84}{\cal P}+\frac{13}{168}{\cal Q} 
\nonumber \\
\mu_{\Xi^{*0}}&=&-\frac{1}{42}{\cal P}-\frac{17}{168}{\cal Q}  \nonumber \\
\mu_{\Xi^{*-}}&=&-\frac{1}{16}{\cal M}-\frac{1}{4}({\cal N}-\frac{1} {2\sqrt{%
3}}{\cal N}^{\prime})-\frac{11}{42}{\cal P}+\frac{1}{84}{\cal Q}  \nonumber
\\
\mu_{\Omega^{-}}&=&-\frac{1}{16}{\cal M}-\frac{1}{4}({\cal N}-\frac{1} {2%
\sqrt{3}}{\cal N}^{\prime})-\frac{9}{28}{\cal P}-\frac{3}{56}{\cal Q}.
\label{dec}
\end{eqnarray}

In the SU(3) flavor symmetric limit with the chiral symmetry breaking masses 
$m_{u}=m_{d}=m_{s}$, $m_{K}=m_{\pi}$ and decay constants $f_{K}=f_{\pi}$, the 
magnetic moments of the decuplet baryons are simply given by \cite{beg64}
\begin{equation}
\mu_{B}=Q_{EM}(\frac{1}{16}{\cal M}+\frac{1}{4}({\cal N}-\frac{1}{2\sqrt{3}}
{\cal N}^{\prime}))  
\label{mubqem}
\eeq
where $Q_{EM}$ is the EM charge.  Here one remembers that for the case of the 
CBM in the adjoint representation, the prediction of the baryon magnetic 
moments with the chiral symmetry is the same as that with the SU(3) flavor 
symmetry since the mass-dependent term in ${\cal L}_{CSB}$ and 
${\cal L}_{FSB}$ do not yield any contribution to $J^{\mu}_{FSB}$ so that 
there is no terms with ${\cal P}$ and ${\cal Q}$ in (\ref{dec}).

Due to the degenerate d- and s-flavor charges in the SU(3) EM charge
operator $\hat{Q}_{EM}$, the CBM possesses the generalized U-spin symmetry 
relations in the baryon decuplet magnetic moments, similar to those in the 
octet baryons (\ref{magmomcs}),
\begin{eqnarray}
\mu_{\Delta^{-}}&=&\mu_{\Sigma^{*-}}=\mu_{\Xi^{*-}}=\mu_{\Omega^{-}}
\nonumber \\
\mu_{\Delta^{0}}&=&\mu_{\Sigma^{*0}}=\mu_{\Xi^{*0}}  \nonumber \\
\mu_{\Delta^{+}}&=&\mu_{\Sigma^{*+}}  
\label{uspin10}
\end{eqnarray}
which will be shown to be shared with the naive NRQM, to support the effective 
NRQM conjecture of the CBM.

Since the SU(3) FSB quark masses do not affect the magnetic moments of the
baryon decuplet in the ${\it adjoint}$ representation of the CBM, in the more
general SU(3) flavor symmetry broken case with $m_{u}=m_{d}\neq m_{s}$, $%
m_{\pi}\neq m_{K}$ and $f_{\pi}\neq f_{K}$, the decuplet baryon magnetic
moments with ${\cal P}$ and ${\cal Q}$ satisfy the other sum rules 
\cite{hong94}
\begin{eqnarray}
\mu_{\Sigma^{*0}}&=&\frac{1}{2}(\mu_{\Sigma^{*+}}+\mu_{\Sigma^{*-}})
\label{sumrule1} \\
\mu_{\Delta^{-}}+\mu_{\Delta^{++}}&=&\mu_{\Delta^{0}}+\mu_{\Delta^{+}}
\label{sumrule2} \\
\sum_{B\in {\rm decuplet}}\mu_{B}&=&0.  \label{cg10}
\end{eqnarray}
Here one notes that the $\Sigma^{*}$ hyperons satisfy the identity $%
\mu_{\Sigma^{*}}(I_{3})=\mu_{\Sigma^{*0}}+I_{3}\Delta\mu_{\Sigma^{*}}$,
where $\Delta\mu_{\Sigma^{*}}=\frac{1}{16}{\cal M}+\frac{1}{4}({\cal N}-%
\frac{1}{2 \sqrt{3}}{\cal N}^{\prime})+\frac{3}{14}{\cal P}-\frac{5}{56}%
{\cal Q}$, such that $\mu_{\Sigma^{*+}}+\mu_{\Sigma^{*-}}$ is independent of
$I_{3}$ as in (\ref{sumrule1}). For the $\Delta$ baryons one can formulate the
relation $\mu_{\Delta}(I_{3})=\mu_{\Delta}^{0}+I_{3}\Delta\mu_{\Delta}$ with
$\mu_{\Delta}^{0}=\frac{1}{32}{\cal M}+\frac{1}{8}({\cal N}-\frac{1}{2\sqrt{3%
}} {\cal N}^{\prime})+\frac{1}{7}{\cal P}+\frac{5}{112}{\cal Q}$ and $%
\Delta\mu_{\Delta}=\frac{1}{16}{\cal M}+\frac{1}{4}({\cal N}-\frac{1} {2%
\sqrt{3}}{\cal N}^{\prime})+\frac{4}{21}{\cal P}-\frac{11}{168}{\cal Q}$, so
that $\Delta$ baryons can be easily seen to fulfill the sum rule 
(\ref{sumrule2}).  Also the summation of the magnetic moments over all the 
decuplet baryons vanish to yield the model independent relation (\ref{cg10}). 
Also the summation of the magnetic moments over all the decuplet baryons 
vanishes to yield the model independent relation, namely the third sum rule in 
(\ref{cg10}), since there is no SU(3) singlet contribution to the magnetic 
moments as in the baryon octet magnetic moments.

In the SU(3) flavor symmetry broken case, by using the projection operators in 
(\ref{projectionop}) we can decompose the EM currents into three flavor pieces 
to obtain the baryon decuplet magnetic moments in the u-flavor channels of the 
adjoint representation
\bea
\mu_{\Delta^{++}}^{(u)}&=&\frac{5}{12}{\cal M}+\frac{1}{3}({\cal N}-\frac{1}
{2\sqrt{3}}{\cal N}^{\prime})+\frac{2}{7}{\cal P}-\frac{1}{28}{\cal Q} 
\nonumber \\
\mu_{\Delta^{+}}^{(u)}&=&\frac{3}{8}{\cal M}+\frac{1}{6}({\cal N}-\frac{1} 
{2\sqrt{3}}{\cal N}^{\prime})+\frac{10}{63}{\cal P}+\frac{1}{126}{\cal Q} 
\nonumber \\
\mu_{\Delta^{0}}^{(u)}&=&\frac{1}{3}{\cal M}+\frac{2}{63}{\cal P}
+\frac{13}{252}{\cal Q}  \nonumber \\
\mu_{\Delta^{-}}^{(u)}&=&\frac{7}{24}{\cal M}-\frac{1}{6}({\cal N}-\frac{1}
{2\sqrt{3}}{\cal N}^{\prime})-\frac{2}{21}{\cal P}+\frac{2}{21}{\cal Q} 
\nonumber \\
\mu_{\Sigma^{*+}}^{(u)}&=&\frac{3}{8}{\cal M}+\frac{1}{6}({\cal N}-\frac{1}
{2\sqrt{3}}{\cal N}^{\prime})+\frac{19}{126}{\cal P}-\frac{17}{252}{\cal Q} 
\nonumber \\
\mu_{\Sigma^{*0}}^{(u)}&=&\frac{1}{3}{\cal M}+\frac{1}{126}{\cal P}
-\frac{1}{126}{\cal Q}  \nonumber \\
\mu_{\Sigma^{*-}}^{(u)}&=&\frac{7}{24}{\cal M}-\frac{1}{6}({\cal N}-\frac{1}
{2\sqrt{3}}{\cal N}^{\prime})-\frac{17}{126}{\cal P}+\frac{13}{252}{\cal Q} 
\nonumber \\
\mu_{\Xi^{*0}}^{(u)}&=&\frac{1}{3}{\cal M}-\frac{1}{63}{\cal P}
-\frac{17}{252}{\cal Q}  \nonumber \\
\mu_{\Xi^{*-}}^{(u)}&=&\frac{7}{24}{\cal M}-\frac{1}{6}({\cal N}-\frac{1}
{2\sqrt{3}}{\cal N}^{\prime})-\frac{11}{63}{\cal P}+\frac{1}{126}{\cal Q}  
\nonumber\\
\mu_{\Omega^{-}}^{(u)}&=&\frac{7}{24}{\cal M}-\frac{1}{6}({\cal N}-\frac{1}
{2\sqrt{3}}{\cal N}^{\prime})-\frac{3}{14}{\cal P}-\frac{1}{28}{\cal Q}.  
\label{mu10u}
\eea

Similarly the baryon decuplet magnetic moments in the s-flavor channels are 
given as follows
\bea
\mu_{\Delta}^{(s)}&=&-\frac{7}{48}{\cal M}+\frac{1}{12}({\cal N}-\frac{1}
{2\sqrt{3}}{\cal N}^{\prime})+\frac{2}{21}{\cal P}+\frac{5}{168}{\cal Q} 
\nonumber \\
\mu_{\Sigma^{*}}^{(s)}&=&-\frac{1}{6}{\cal M}+\frac{1}{126}{\cal P}
-\frac{1}{126}{\cal Q} 
\nonumber \\
\mu_{\Xi^{*}}^{(s)}&=&-\frac{3}{16}{\cal M}-\frac{1}{12}({\cal N}-\frac{1}
{2\sqrt{3}}{\cal N}^{\prime})-\frac{2}{21}{\cal P}-\frac{5}{84}{\cal Q}  
\nonumber\\
\mu_{\Omega}^{(s)}&=&-\frac{5}{24}{\cal M}-\frac{1}{6}({\cal N}-\frac{1}
{2\sqrt{3}}{\cal N}^{\prime})-\frac{3}{14}{\cal P}-\frac{1}{28}{\cal Q}.  
\label{mu10s}
\eea

In general all the baryon magnetic moments in the CBM also satisfy the 
model-independent relations in the u- and d-flavor components and the I-spin 
symmetry in the s-flavor channel of (\ref{udssymmetry}), as shown in 
(\ref{mu10u}) and (\ref{mu10s}).  Moreover one notes that the relations 
(\ref{udssymmetry}) are satisfied even in the multiquark decay constants 
$f_{\pi}\neq f_{K}$ do not affect the relations (\ref{udssymmetry}) in the 
u- and d-flavor channel without any strangeness and in the s-flavor channel 
with the same strangeness.


\subsection{Multiquark structure}


Until now we have considered the CBM in the adjoint representation where the
U-spin symmetry is broken only through the magnetic moment operators $\hat{%
\mu}^{i(a)}_{FSB}$ induced by the symmetry breaking derivative term. To take
into account the missing chiral symmetry breaking mass effect from ${\cal L}%
_{CSB}$ and ${\cal L}_{FSB}$, in this section we will treat nonperturbatively
the symmetry breaking mass terms via the higher dimensional IR channels
where the CBM can be handled in the Yabu-Ando scheme~\cite{yabu88} with the 
higher IR mixing in the baryon wave function to yield the minimal
multiquark structure with meson cloud {\it inside} the bag.

\begin{table}[t]
\caption{The baryon decuplet magnetic moments of the CBM in the U-spin 
symmetry broken case \cite{hong94} compared with the naive NRQM and the 
experimental data $*$}
\begin{center}
\begin{tabular}{crrrrrrrrrr}
\hline
R &$\mu_{\Delta^{++}}$ &$\mu_{\Delta^{+}}$ &$\mu_{\Delta^{0}}$ 
  &$\mu_{\Delta^{-}}$  &$\mu_{\Sigma^{*+}}$ &$\mu_{\Sigma^{*0}}$ 
  &$\mu_{\Sigma^{*-}}$ &$\mu_{\Xi^{*0}}$ &$\mu_{\Xi^{*-}}$
  &$\mu_{\Omega^{-}}$\\ 
\hline
0.00 &2.81 &1.22 &$-0.38$ &$-1.97$ &1.64 &$-0.17$ &$-1.88$ &0.14 &$-1.72$ 
     &$-1.45$\\
0.10 &2.87 &1.23 &$-0.40$ &$-2.03$ &1.70 &$-0.18$ &$-1.94$ &0.17 &$-1.77$ 
     &$-1.47$\\
0.20 &3.05 &1.30 &$-0.45$ &$-2.20$ &1.87 &$-0.20$ &$-2.10$ &0.22 &$-1.90$ 
     &$-1.54$\\
0.30 &3.24 &1.38 &$-0.49$ &$-2.36$ &2.04 &$-0.21$ &$-2.26$ &0.28 &$-2.04$ 
     &$-1.61$\\
0.40 &3.30 &1.40 &$-0.50$ &$-2.40$ &2.11 &$-0.21$ &$-2.31$ &0.30 &$-2.09$ 
     &$-1.62$\\
0.50 &3.43 &1.46 &$-0.52$ &$-2.49$ &2.23 &$-0.21$ &$-2.40$ &0.34 &$-2.17$ 
     &$-1.66$\\
0.60 &3.58 &1.52 &$-0.53$ &$-2.59$ &2.39 &$-0.21$ &$-2.50$ &0.40 &$-2.27$ 
     &$-1.70$\\
0.70 &3.71 &1.59 &$-0.54$ &$-2.67$ &2.55 &$-0.19$ &$-2.58$ &0.46 &$-2.34$ 
     &$-1.72$\\
0.80 &3.79 &1.63 &$-0.53$ &$-2.69$ &2.67 &$-0.17$ &$-2.60$ &0.51 &$-2.36$ 
     &$-1.70$\\
0.90 &3.81 &1.65 &$-0.52$ &$-2.68$ &2.74 &$-0.14$ &$-2.58$ &0.56 &$-2.33$ 
     &$-1.65$\\
1.00 &3.78 &1.65 &$-0.49$ &$-2.63$ &2.78 &$-0.11$ &$-2.52$ &0.60 &$-2.26$ 
     &$-1.57$\\
${\rm naive}$ &5.58 &2.79 &0.00 &$-2.79$ &3.11 &0.32 &$-2.47$ &0.64 &$-2.15$ 
     &$-1.83$\\
\hline
\end{tabular}
\end{center}
\par
{{${}^{*}$ For the experimental data $\mu_{\Delta^{++}}^{exp}=4.52\pm 0.50$ 
and $\mu_{\Omega^{-}}^{exp}=-1.94\pm 0.17 \pm 0.14$ we have referred to the 
Ref.~\cite{bosshard91} and Ref.~\cite{diehl91}, respectively.}}
\label{decuplet}
\end{table}
The possible SU(3) representations of the minimal multiquark Fock space are 
restricted by the Clebsch-Gordan series ${\bf 10}\oplus{\bf 27}\oplus{\bf 35}$ 
for the baryon decuplet with $Y_{R}=1$ and $J=\frac{3}{2}$ through the 
decomposition of the tensor product of the two IRs in the qqq$\bar{\rm q}$q 
so that the representation mixing coefficients in the eigenstate $|B\rangle=
\sum_{\lambda}C_{\lambda}^{B}|B\rangle^{\lambda}$ can be determined by 
diagonalizing the 3$\times$3 Hamiltonian matrix $h_{SB}$ given by (\ref{hsb}). 
 
Here one should note that in the Yabu-Ando approach the meson cloud, or 
$\bar{\rm q}$q content with all the possible flavor combinations to construct 
the pseudoscalar mesons inside the bag through the channel of qqq$\bar{\rm q}$q 
multiquark Fock space, contributes to the baryon decuplet magnetic moments 
since the baryon wave functions in the multiquark scheme of the CBM act 
nonperturbatively on the magnetic moment operators with both the quark and 
meson phase pieces in their inertia parameters.

The U-spin symmetry breaking effect shown in Figure~\ref{decupletfig} through 
the explicit 
operator $\hat{\mu}_{FSB}^{i(a)}$ and the multiquark structure yields meson 
cloud contributions to the baryon decuplet magnetic moments, comparable to 
those in the naive NRQM.  The vertical lines show that even though nature 
does not preserve the perfect Cheshire 
catness~\cite{nadkarni85,rho97ccp,rho99aip} at least in the SU(3) CBM, the 
model could be considered to be a good candidate which unifies the MIT bag and 
Skyrmion models with predictions almost independent of the bag radius.  One 
can also easily see in Figure~\ref{decupletfig} that the full symmetry breaking 
effects 
induce the magnetic moments of the baryon decuplet to pull the U-spin 
symmetric predictions back to the experimental data.  In Table~\ref{decuplet}, 
the SU(3) CBM predictions in the SU(3) symmetry breaking case in the multiquark 
structure are explicitly listed to be compared with the naive NRQM and the 
experimental data.  For the known experimental data we obtain $\nu_{\Delta
^{++}}^{cbm}=(1.01-1.37)\mu_{p}$ to be compared with the experimental value 
$\mu_{\Delta^{++}}^{exp}=(1.62\pm 0.18)\mu_{p}$ \cite{bosshard91} and the 
naive NRQM prediction $\mu_{\Delta^{++}}^{naive}=2\mu_{p}$.  Since the 
$\mu_{\Omega^{-}}$ could be dominantly achieved from the strange quark and 
kaon whose masses are kept in our massless chiral angle approximation, the 
prediction $\mu_{\Omega^{-}}=-(1.45-1.72)$ n.m. in the CBM seems to be fairly 
well consistent with the experimental data $\mu_{\Omega^{-}}^{exp}=-(1.94
\pm 0.17\pm 0.14)$ n.m.~\cite{diehl91} and the naive NRQM prediction 
$\mu_{\Omega^{-}}^{naive}=-1.83$ n.m..

\begin{figure}
\setlength{\unitlength}{1.0cm}
\begin{center}
\begin{picture}(12.0,8.0)(-1.0,-3.5)
\put(-1.0,-3){\vector(0,1){9.3}}
\put(-1.0,-3){\line(1,0){11.5}}
\multiput(-1.1,-2)(0,1){9}{\line(1,0){0.1}}
\multiput(-1.05,-2.5)(0,1){9}{\line(1,0){0.05}}
\put(-1.3,-0.1){0}
\put(-1.4,-1.1){-1}
\put(-1.4,-2.1){-2}
\put(-1.3,0.9){1}
\put(-1.3,1.9){2}
\put(-1.3,2.9){3}
\put(-1.3,3.9){4}
\put(-1.3,4.9){5}
\put(-1.3,5.9){6}
\put(-0.2,-3.5){$\Delta^{-}$}
\put(0.8,-3.5){$\Delta^{0}$}
\put(1.8,-3.5){$\Delta^{+}$}
\put(2.8,-3.5){$\Delta^{++}$}
\put(3.8,-3.5){$\Sigma^{*-}$}
\put(4.8,-3.5){$\Sigma^{*0}$}
\put(5.8,-3.5){$\Sigma^{*+}$}
\put(6.8,-3.5){$\Xi^{*-}$}
\put(7.8,-3.5){$\Xi^{*0}$}
\put(8.8,-3.5){$\Omega^{-}$}
\thicklines
\put(-0.3,-2.79){\line(1,0){0.6}}   
\put(0.7,0.00){\line(1,0){0.6}}     
\put(1.7,2.79){\line(1,0){0.6}}     
\put(2.7,5.58){\line(1,0){0.6}}     
\put(3.7,-2.47){\line(1,0){0.6}}    
\put(4.7,0.32){\line(1,0){0.6}}     
\put(5.7,3.11){\line(1,0){0.6}}     
\put(6.7,-2.15){\line(1,0){0.6}}    
\put(7.7,0.64){\line(1,0){0.6}}     
\put(8.7,-1.83){\line(1,0){0.6}}    
\thinlines
\put(3.0,4.02){\line(0,1){1.0}}      
\put(2.95,4.02){\line(1,0){0.1}}     
\put(2.95,5.02){\line(1,0){0.1}}     
\put(2.85,4.44){$\times$}            
\put(9.0,-2.16){\line(0,1){0.44}}    
\put(8.95,-2.16){\line(1,0){0.1}}    
\put(8.95,-1.72){\line(1,0){0.1}}    
\put(8.85,-2.02){$\times$}           
\thinlines
\put(0.25,-1.57){\line(0,1){0.34}}     
\put(0.20,-1.57){\line(1,0){0.1}}      
\put(1.20,0.00){\line(1,0){0.1}}       
\put(1.20,0.03){\line(1,0){0.1}}       
\put(2.25,1.23){\line(0,1){0.34}}      
\put(2.20,1.23){\line(1,0){0.1}}       
\put(2.20,1.57){\line(1,0){0.1}}       
\put(3.25,2.47){\line(0,1){0.67}}      
\put(3.20,2.47){\line(1,0){0.1}}       
\put(3.20,3.14){\line(1,0){0.1}}       
\put(4.25,-1.57){\line(0,1){0.34}}     
\put(4.20,-1.57){\line(1,0){0.1}}      
\put(4.20,-1.23){\line(1,0){0.1}}      
\put(5.25,0.00){\line(0,1){0.03}}      
\put(5.20,0.00){\line(1,0){0.1}}       
\put(5.20,0.03){\line(1,0){0.1}}       
\put(6.25,1.23){\line(0,1){0.34}}      
\put(6.20,1.23){\line(1,0){0.1}}       
\put(6.20,1.57){\line(1,0){0.1}}       
\put(7.25,-1.57){\line(0,1){0.34}}     
\put(7.20,-1.57){\line(1,0){0.1}}      
\put(7.20,-1.23){\line(1,0){0.1}}      
\put(8.25,0.00){\line(0,1){0.03}}      
\put(8.20,0.00){\line(1,0){0.1}}       
\put(8.20,0.03){\line(1,0){0.1}}       
\put(9.25,-1.57){\line(0,1){0.34}}     
\put(9.20,-1.57){\line(1,0){0.1}}      
\put(9.20,-1.23){\line(1,0){0.1}}      
\thicklines
\put(-0.25,-2.66){\line(0,1){0.69}}    
\put(0.75,-0.54){\line(0,1){0.16}}     
\put(1.75,1.22){\line(0,1){0.43}}      
\put(2.75,2.81){\line(0,1){1.00}}      
\put(3.75,-2.65){\line(0,1){0.70}}     
\put(4.75,-0.17){\line(0,1){0.16}}     
\put(5.75,1.59){\line(0,1){0.97}}      
\put(6.75,-2.36){\line(0,1){0.64}}     
\put(7.75,0.18){\line(0,1){0.46}}      
\put(8.75,-1.72){\line(0,1){0.27}}     
\thinlines
\put(-0.30,-2.66){\line(1,0){0.1}}     
\put(-0.30,-1.97){\line(1,0){0.1}}     
\put(0.70,-0.54){\line(1,0){0.1}}      
\put(0.70,-0.38){\line(1,0){0.1}}      
\put(1.70,1.22){\line(1,0){0.1}}       
\put(1.70,1.65){\line(1,0){0.1}}       
\put(2.70,2.81){\line(1,0){0.1}}       
\put(2.70,3.81){\line(1,0){0.1}}       
\put(3.70,-2.65){\line(1,0){0.1}}      
\put(3.70,-1.95){\line(1,0){0.1}}      
\put(4.70,-0.17){\line(1,0){0.1}}      
\put(4.70,-0.01){\line(1,0){0.1}}      
\put(5.70,1.59){\line(1,0){0.1}}       
\put(5.70,2.56){\line(1,0){0.1}}       
\put(6.70,-2.36){\line(1,0){0.1}}      
\put(6.70,-1.72){\line(1,0){0.1}}      
\put(7.70,0.18){\line(1,0){0.1}}       
\put(7.70,0.64){\line(1,0){0.1}}       
\put(8.70,-1.72){\line(1,0){0.1}}      
\put(8.70,-1.45){\line(1,0){0.1}}      
\end{picture}
\end{center}
\caption{Baryon decuplet magnetic moments.  The effective NRQM results with bag
 radius $0.0$ fm $\leq R \leq 1.0$ fm in the $U$-spin symmetric (thin vertical
lines) and symmetry broken (thick vertical lines) cases are compared with the
naive NRQM (thick lines) and the experimental data (thin vertical lines with a
cross).}
\label{decupletfig}
\end{figure}
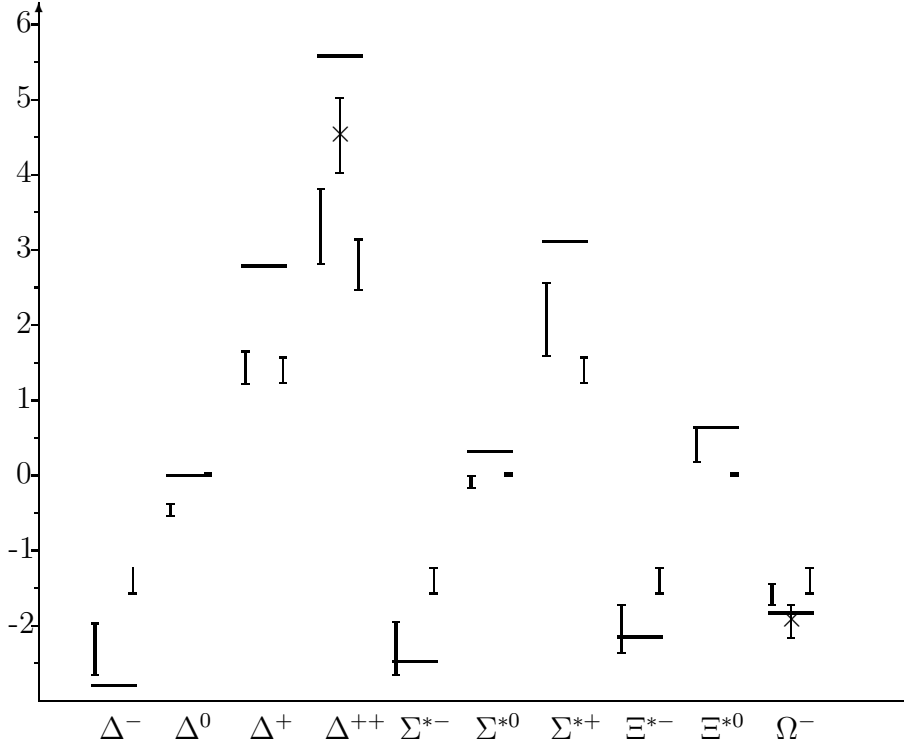

\begin{table}[h]
\caption{The strange flavor baryon decuplet magnetic moments in the naive and 
CBM  \cite{hong94}}
\begin{center}
\begin{tabular}{crrrr}
\hline
R &$\mu_{\Delta}^{(s)}$ &$\mu_{\Sigma^{*}}^{(s)}$ &$\mu_{\Xi^{*}}^{(s)}$ 
  &$\mu_{\Omega}^{(s)}$\\ 
\hline
0.00 &0.31 &$-0.21$ &$-0.64$  &$-1.08$\\
0.10 &0.31 &$-0.22$ &$-0.65$  &$-1.09$\\  
0.20 &0.33 &$-0.22$ &$-0.67$  &$-1.14$\\  
0.30 &0.34 &$-0.22$ &$-0.70$  &$-1.18$\\  
0.40 &0.35 &$-0.22$ &$-0.70$  &$-1.19$\\  
0.50 &0.37 &$-0.21$ &$-0.72$  &$-1.22$\\  
0.60 &0.39 &$-0.21$ &$-0.73$  &$-1.25$\\ 
0.70 &0.41 &$-0.20$ &$-0.74$  &$-1.26$\\ 
0.80 &0.44 &$-0.19$ &$-0.74$  &$-1.26$\\ 
0.90 &0.46 &$-0.18$ &$-0.74$  &$-1.25$\\ 
1.00 &0.48 &$-0.18$ &$-0.73$  &$-1.22$\\ 
${\rm naive}$ &0.00 &$-0.61$  &$-1.22$ &$-1.83$\\ 
\hline
\end{tabular}
\end{center}
\label{sdecuplet}
\end{table}

Since the Hamiltonian $h_{SB}$ has eigenstates degenerate with the 
isomultiplets in our approximation, where the I-spin symmetry breaking light 
quark masses are neglected so that the relations (\ref{udssymmetry}) are 
derived in the same strangeness sector, the multiquark structure in the 
q-flavor channels conserves the I-spin symmetry and model-independent relations 
(\ref{udssymmetry}).  The s-flavor magnetic moments $\mu_{B}^{(s)}$ in 
Table~\ref{sdecuplet}, reveal the stronger Cheshire catness than in $\mu_{B}$ 
and the pretty good consistency with the naive NRQM.

In Figure~\ref{decupletfig} and Table~\ref{decuplet}, the meson cloud 
contributions to the magnetic moments 
in the SU(3) effective NRQM are obtained with respect to the naive NRQM and 
experimental values.  With the help of the naive NRQM data, one could also 
easily see the meson cloud contributions, which are originated from the 
$\bar{\rm q}$q content and strange quarks inside the bag, as well as the 
massive kaons outside the bag.


\section{SAMPLE experiment and baryon strange form factors}
\setcounter{equation}{0}
\renewcommand{\theequation}{\arabic{section}.\arabic{equation}}



\subsection{SAMPLE experiment and proton strange form factor}



In this section, we consider the SAMPLE experiment and the corresponding 
theoretical paradigms in the chiral models to connect 
the chiral model predictions with the recent experimental data for 
the proton strange form factor.  As discussed in Introduction, there 
have been lots of theoretical predictions with varied values for the SAMPLE 
experimental results associated with the proton strange form factor 
through parity violating electron scattering.  Especially the positive value 
of the proton strange form factor predicted in the framework of the CBM is 
quite comparable to the recent SAMPLE experimental data.  

The SAMPLE experiment was performed at the MIT/Bates Linear Accelerator Center 
using a 200 MeV polarized electron beam incident on a liquid hydrogen target.  
The scattered electrons were detected in a large solid angle ($\sim 1.5$ sr) 
air {\v C}erenkov detector at backward angles $130^\circ < \theta < 170^\circ$.  
The parity-violating asymmetry $A$ was determined from the asymmetries in 
ratios of integrated detector signal to beam intensity for left- and 
right-handed beam pulses.  (For details of the SAMPLE experiment see 
Refs.~\cite{beise96,mueller97}.) 

On the other hand, there have been considerable discussions concerning the strangeness in hadron physics.  Beginning with Kaplan and Nelson's 
work~\cite{kaplan86} on the charged kaon condensation the theory of 
condensation in dense matter has become one of the central issues in nuclear 
physics and astrophysics together with the supernova collapse.  The $K^{-}$ 
condensation at a few times nuclear matter density was later 
interpreted~\cite{kaon2} in terms of cleaning of $\bar{\rm q}$q condensates 
from the quantum chromodynamics (QCD) vacuum by a dense nuclear matter and 
also was further theoretically investigated~\cite{lee95} in chiral phase 
transition.

Now, the internal structure of the nucleon is still a subject of
great interest to experimentalists as well as theorists.
In 1933, Frisch and Stern \cite{stern33} performed the first measurement
of the magnetic moment of the proton and obtained the earliest
experimental evidence for the internal structure of the nucleon.
However, it wasn't until 40 years later that the quark structure of
the nucleon was directly observed in deep inelastic electron scattering 
experiments. The development of QCD followed soon thereafter, and is now the 
accepted theory of the strong interactions governing the behavior of quarks 
and gluons associated with hadronic structure.  Nevertheless, we still lack a 
quantitative theoretical understanding of these properties (including 
the magnetic moments) and additional experimental information is 
crucial in our effort to understand the internal structure of the 
nucleons.  For example, a satisfactory quantitative understanding of 
the magnetic moment of the proton has still not been achieved, now 
more than 60 years after the first measurement was performed.

Quite recently, the SAMPLE experiment\cite{sample01,sample00} reported the
proton's neutral weak magnetic form factor, which has been suggested by the
neutral weak magnetic moment measurement through parity violating electron
scattering\cite{mck89,beise91}.  Moreover, McKeown\cite{mck} has shown that 
the strange form factor of proton should be positive by using the conjecture that 
the up-quark effects are generally dominant in the flavor dependence of 
the nucleon properties.  In fact, at a small momentum transfer 
$Q^2 = 0.1~{\rm (GeV/c)}^2$, the SAMPLE Collaboration obtained 
the positive experimental data for the proton strange magnetic form 
factor~\cite{sample01,sample00} 
\beq
G_M^s (Q^2 = 0.1 {\rm (GeV/c)}^2) = +0.14 \pm 0.29~{\rm (stat)} 
\pm 0.31~{\rm (sys)}.
\label{expdata}
\eeq
This positive experimental value is contrary to the negative values of the 
proton strange form factor which result from most of the model 
calculations~\cite{jaffe89,musolf94,koepf92,holstein90,park91,phatak94,
christov96,hammer96,ito95,weigel95,leinweber96,geiger96,musolf96a,musolf96b,
meissner97} except those of Hong, Park and Min~\cite{hong97,hong932} based on 
the SU(3) chiral bag model (CBM) \cite{gerry791, gerry792, jackson83, gerry84} 
and the recent predictions of the chiral quark soliton model~\cite{kim298} and 
the heavy baryon chiral perturbation theory~\cite{meissner00,vankolck00}.  
Recently the anapole moment effects associated with the parity violating 
electron scattering  have been intensively studied to yield more 
theoretical predictions~\cite{vankolck00,maekawa00,musolf00prd,musolf01nuc,bob01ph}.  
(For details of the anapole effects for instance see Ref.~\cite{bob01ph}.)  Through 
further investigations including gluon effects, one can also obtain somehow 
realistic predictions for the proton strange form factor.  

On the other hand, a number of parity-violating electron scattering experiments such as 
the SAMPLE experiment associated with a second deuterium measurement~\cite{sam2}, the 
HAPPEX experiment~\cite{happex}, the PVA4 experiment~\cite{pva4}, the G0 experiment~\cite{g0} 
and other recently approved parity violating measurements~\cite{jeff1,jeff2} at the Jefferson 
Laboratory, are planned for the near future.  (For details of the future experiments, 
see Ref~\cite{bob01ph}.) 

\begin{table}[t]
\caption{Electroweak quark couplings}
\begin{center}
\begin{tabular}{lrcc}
\hline
\noalign{\vskip3pt}
&&\multispan2{\hfil Z\hfil}\cr
\noalign{\vskip-12pt}
&&\multispan2{\hrulefill}\cr
Flavor &$\gamma$ &Vector & Axial Vector\\
\noalign{\hrule}
\noalign{\vskip6pt}
u&$\ {2\over3}$&
$\ {1\over4}-{2\over3}\sin^2\theta_W$&$-{1\over4}$\cr
d&$-{1\over3}$&$-{1\over4}+{1\over3}\sin^2\theta_W$&$\ {1\over4}$\cr
s&$-{1\over3}$&$-{1\over4}+{1\over3}\sin^2\theta_W$&$\ {1\over4}$\cr
\noalign{\vskip6pt}
\noalign{\hrule}
\end{tabular}
\end{center}
\label{couple}
\end{table}

Now we consider the form factors of the baryon octet with internal structure.  
If a particle is point-like, with no internal structure 
due to interactions other than EM, the photon couples to the 
EM current 
\beq
\hat{V}_{\gamma}^{\mu}=\frac{2}{3}\bar{u}\gamma^{\mu}u
 -\frac{1}{3}\bar{d}\gamma^{\mu}d-\frac{1}{3}\bar{s}\gamma^{\mu}s
\label{emcurrent}
\eeq
and according to the Feynman rules the matrix element of 
$\hat{V}_{\gamma}^{\mu}$ for the particle with transition from momentum 
state $p$ to momentum state $p+q$ is given by 
\beq
\langle  p+q|\hat{V}_{\gamma}^{\mu} |p\rangle = \bar{u}(p+q) \gamma^\mu u(p)
\label{feynman00}
\eeq
where $u(p)$ is the spinor for the particle states.  However if the particle has 
the internal structure caused by other interaction not given by QED, the Feynman 
rules cannot yield the explicit coupling of the particle to an external or internal photon line.  
The standard electroweak model couplings to the up, down and strange quarks 
are listed in Table~\ref{couple}.  The baryons are definitely extended objects 
with internal structure, for which the coupling constant can be described in terms of form factors which 
are real Lorentz scalar functions associated with the internal structure and fixed by the properties 
of the EM currents such as current conservation, covariance under Lorentz transformations 
and hermiticity.  The above matrix element is then generalized to have covariant decomposition
\beq
\langle p+q|\hat{V}_{\gamma}^{\mu} |p\rangle = \bar{u}(p+q)
\left[F_{1}^{\gamma}(q^2)\gamma^{\mu}
       +\frac{i}{2M_B}F_{2}^{\gamma}(q^2)\sigma^{\mu\nu} q_\nu\right]u(p)
\label{extptcleform}
\eeq
where $q$ is the momentum transfer and $\sigma^{\mu\nu}=\frac{i}{2}(
\gamma^{\mu}\gamma^{\nu}-\gamma^{\nu}\gamma^{\mu})$ and $M_{B}$ is the baryon
mass and $F_{1}^{\gamma}$ and $F_{2}^{\gamma}$ are the Dirac and Pauli EM 
form factors, which are Lorentz scalars and $p^{2}=(p+q)^{2}=M_{B}^{2}$ on 
shell so that they depend only on the Lorentz scalar variable $q^{2}$.  

With these form factors, the differential cross section in 
the laboratory system for electron scattering on the baryon is given as
\bea
\frac{d\sigma}{d\Omega}&=&\left(\frac{\alpha}{2E}\right)^{2}
\frac{\cos^{2}(\theta/2)}{\sin^{4}(\theta/2)}\frac{1}{1+2(E/m_{B})\sin^{2}
(\theta/2)}
\nonumber\\
& &\cdot\left(F_{1}^{2}(t)-\frac{t}{4m_{B}^{2}}F_{2}^{2}(t)
-\frac{t}{2m_{B}^{2}}(F_{1}(t)+F_{2}(t))^{2}\tan^{2}(\theta/2)\right)
\label{crosssection}
\eea
where $\alpha$ is the fine structure constant and $E$ and $\theta$ are the 
energy and scattering angle of the electron and $t=q^{2}$ is the Mandelstam 
variable.  In order to see the physical interpretation of these EM form 
factors, it is convenient to consider the matrix element (\ref{extptcleform}) 
in the reference frame with $\vec{p}+(\vec{p}+\vec{q})=0$ where one can have 
the rest frame in the vanishing $q^{2}$ limit.  In this rest frame of the 
baryon , we can associate the EM form factors at zero momentum transfer, 
$F_{1}(0)$ and $F_{2}(0)$, with the static properties of the baryon such as 
electric charge, magnetic moment and charge radius.

Next, we will also use the Sachs form factors, which are linear combinations of the 
Dirac and Pauli form factors 
\bea
G_{E}&=&F_{1}-\tau F_{2}\nonumber\\
G_{M}&=&F_{1}+F_{2}
\label{sachs}
\eea
where $\tau=-\frac{q^{2}}{4M_{B}^{2}}>0$.

The quark flavor structure of the form factors can be revealed by
writing the matrix elements of individual quark currents in terms of
form factors
\begin{equation}
\langle p+q |\bar q^f \gamma^\mu q^f | p\rangle \, \equiv  \, 
\bar u(p+q) \left[ F_1^f (q^2) \gamma^\mu
+  {i \over {2M_N}}  F_2^f (q^2)
\sigma^{\mu \nu} q_\nu \right] u(p)~ ; \ (f=u, d, s)
\end{equation}
which defines the form factors $F_1^f$ and $F_2^f$. Then using
definitions analogous to Eq. (\ref{sachs}), we can write
\begin{equation}
G_E^\gamma =  {2 \over 3} G_E^u
                        - {1 \over 3} G_E^d
                                 - {1 \over 3} G_E^s
\label{gegamma}
\end{equation}
with a similar expression for $G_M^\gamma$.

The neutral weak current operator is given by an expression analogous to
Eq. (\ref{emcurrent}) but with different coefficients:
\begin{equation}
\hat
V^\mu_Z = ({1 \over 4} - {2 \over 3} \sin^2 \theta_W) \bar u \gamma^\mu u
        +(- {1 \over 4} + {1 \over 3} \sin^2 \theta_W)  \bar d \gamma^\mu d
        +(- {1 \over 4} + {1 \over 3} \sin^2 \theta_W) \bar s \gamma^\mu s  \>.
\end{equation}
Here the coefficients depend on the weak mixing angle, which has recently
been determined \cite{datagroup9x} with high precision: 
$\sin^2 \theta_W = 0.2315\pm 0.0004\>.$  In direct analogy to Eq. 
(\ref{gegamma}), we have expressions for the neutral weak form factors $G_E^Z$ 
and $G_M^Z$ in terms of the different quark flavor components
\begin{equation}
G_{E,M}^{Z} = ({1 \over 4} - {2 \over 3} \sin^2 \theta_W) G_{E,M}^u
 +(- {1 \over 4} + {1 \over 3} \sin^2 \theta_W)  G_{E,M}^d
        +(- {1 \over 4} + {1 \over 3} \sin^2 \theta_W)  G_{E,M}^s  \>.
\label{gemz}
\end{equation}
An important point is that the form factors $G_{E,M}^{f}$ $(f=u$, $d$, $s)$ 
appearing in this expression are exactly the same as those in the EM form 
factors, as in Eq. (\ref{gegamma}).

Utilizing isospin symmetry, one then can eliminate the up and down quark 
contributions to the neutral weak form factors by using the proton and neutron
EM form factors and obtain the expressions
\begin{equation}
G_{E,M}^{Z,p} = ({1 \over 4} -  \sin^2 \theta_W) G_{E,M}^{\gamma,p}
                -{1 \over 4} G_{E,M}^{\gamma,n} -{1 \over 4} G_{E,M}^{s} \>.
\label{gemzp}
\end{equation}
This result shows how the neutral weak form factors are related to the EM form 
factors plus a contribution from the strange (electric or magnetic) form factor. 
Thus measurement of the neutral weak form factor will allow (after combination with the
EM form factors) determination of the strange form factor of interest. It should 
be mentioned that there are electroweak radiative corrections to the coefficients 
in Eq. (\ref{gemz}) due to processes such as those shown in Figure~\ref{eweak}.  
These are generally small corrections, of order 1-2\%, and can be reliably 
calculated~\cite{musolf90,musolf94a}.

\begin{figure}
\centerline{\epsfig{figure=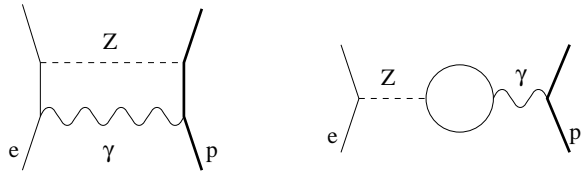,angle=270,width=3.0in}}
\vskip 0.2 in
\caption {Examples of amplitudes contributing to electroweak radiative
corrections to the coefficients in Eq. (\ref{gemz}).}
\label{eweak}
\end{figure}

The EM form factors present in Eq. (\ref{gemzp}) are very 
accurately known (1-2 \%) for the proton in the momentum transfer
region $Q^2 < 1$ (GeV/c)${}^2$. The neutron form factors are not known
as accurately as the proton form factors (the electric form factor $G_E^n$
is at present rather poorly constrained by experiment), although
considerable work to improve our knowledge of these
quantities is in progress.  Thus, the present lack of knowledge of the
neutron form factors will not significantly hinder the interpretation of
the neutral weak form factors.

The properties of the Sachs form factors $G_E$ and $G_M$ near $Q^2=0$ are 
of particular interest in that they represent static physical properties 
of the baryon.  Namely, at zero momentum transfer, one can have the relations 
between the EM form factors and the static physical quantities of the baryon octet, 
namely $G_{E}(0)=Q$ and $G_{M}(0)=\mu$ where $Q$ and $\mu$ are nothing but 
the electric charge and magnetic moment operators of the baryon.  The $F_{2}(0)$ 
is thus interpreted as the anomalous magnetic moments of the baryon octet $\mu^{an}=\mu-Q$. 

In the strange flavor sector, the fractional EM charge $Q^{s}$ of the baryon
can be obtained from the strange flavor fractional EM charges in the baryon
to yield $Q^{s}_{N}=0$, $Q_{\Lambda}^{s}=Q_{\Sigma}^{s}=-\frac{1}{3}$ and
$Q_{\Xi}^{s}=-\frac{2}{3}$.  The strange flavor anomalous magnetic moments
degenerate in isomultiplets can then be easily given by $\mu^{an(s)}=
\mu^{(s)}-Q^{s}$ so that the strange form factors at zero
momentum transfer defined as $F_{2}^{(s)}=-3\mu^{an(s)}$ can be calculated
to yield
\begin{equation}
F_{2B}^{s}(0) = G_{M}^{s}(0)-G_{E}^{s}(0)
\label{f2bs0gmge}
\end{equation}
Since the nucleon has no net strangeness, we find 
$G_E^s(0)=0$. However, one can express the slope of $G_E^s$ at $Q^2=0$ in the 
usual fashion in terms of a ``strangeness radius'' 
$r_s$
\begin{equation}
r^2_s\equiv -6\left[dG_E^s/dQ^2\right]_{Q^2=0}.
\end{equation}

Now we consider the parity-violating asymmetry for elastic scattering of 
right- vs.~left-handed electrons from nucleons at backward scattering angles, 
which is quite sensitive to $G_M^Z$ as discussed in 
Ref.~\cite{mck89,beck89,averett99}.  The SAMPLE experiment measured the 
parity-violating asymmetry in the elastic scattering of 200 MeV polarized 
electrons at backward angles with an average 
$Q^2 \simeq 0.1$(GeV/c)${}^2$. For $G_M^s=0$, the expected asymmetry in 
the SAMPLE experiment is about $-7\times 10^{-6}$ or $-7$ ppm, and the 
asymmetry depends linearly on $G_M^s$.  The neutral weak axial form factor 
$G_A^Z$ contributes about 20\% to the asymmetry in the SAMPLE experiment.  In 
parity-violating electron scattering $G_A^Z$ is modified by a substantial 
electroweak radiative correction.  The corrections were estimated 
in~\cite{musolf90,musolf94a}, but there is considerable uncertainty in the 
calculations.  The uncertainty in these radiative corrections substantially 
limits the ability to determine $G_M^s$, as will be discussed below.

The elastic scattering asymmetry for the proton is measured to yield
\begin{equation}
A = -4.92 \pm 0.61 \pm 0.73 \> {\rm ppm} 
\label{eqaaa}
\end{equation}
where the first uncertainty is statistical and the second is the 
estimated systematic error.  This value is in good agreement with the 
previous reported measurement~\cite{mueller97}.

On the other hand, the quantities $G_{E,M}^Z$ for the proton can be
determined via elastic parity-violating electron 
scattering~\cite{mck89,beise91}.  The difference in cross sections for 
right and left handed incident
electrons arises from interference of the EM and
neutral weak amplitudes, and so contains products of EM
and neutral weak form factors. At the mean kinematics of the experiment 
($Q^{2}=0.1$ (GeV/c)$^{2}$ and $\theta=146.1^{\circ}$), the theoretical 
asymmetry for elastic scattering from the proton is given by
\beq
A=(-5.72+3.49~G_{M}^{s}+1.55~G_{A}^{e}(T=1))~{\rm ppm},
\label{asymmp}  
\eeq
where 
\beq
G_{A}^{e}=G_{A}^{Z}+\eta F_{A}+R^{e},
\eeq
where $G_{A}^{Z}$ is the contribution from a single Z-exchange, as would 
be measured in neutrino-proton elastic scattering, given as  
\beq
G_{A}^{Z}=-(1+R_{A}^{1})G_{A}+R_{A}^{0}+G_{A}^{s},
\label{gaz00}
\eeq
and $\eta=\frac{8\pi\sqrt{2}\alpha}{1-4\sin^{2}\theta_{W}}=3.45$ with 
the fine-structure constant $\alpha$, and 
$F_{A}$ is the nucleon anapole moment~\cite{zel57} and $R^{e}$ is 
a radiative correction.  Here $G_A$ is the charged current nucleon form 
factor: we use $G_A=G_A(0)/(1+{Q^2\over M_A^2})^2$, with
$G_A(0)=-(g_A/g_V) = 1.267\pm 0.035$~\cite{datagroup9x} and
$M_A=1.061\pm 0.026$~(GeV/c)~\cite{garvey93}. $G_A^s(Q^2=0)=\Delta s
=-0.12\pm 0.03$~\cite{lipkin99},
and $R_A^{0,1}$ are the isoscalar and isovector
axial radiative corrections. The radiative
corrections were estimated by Ref.~\cite{musolf90} to be
$R_A^1=-0.34$ and $R_A^0=-0.12$, but with nearly
100\% uncertainty.\footnote{The notation used here is
$R_A^0=(1/2)(3F-D)R_A^{T=0}$, where $\sqrt{3}R_A^{T=0}=-0.62$
in Ref.~\cite{musolf94a}} 

For the case of a deuterium target, a separate measurement was performed 
with the same apparatus, where both elastic and quasi-elastic scattering 
from the deuteron were measured due to the large energy acceptance of the 
detector.  Based on the appropriate fractions of the yield, the elastic 
scattering and threshold electrodisintegration contributions were 
estimated to change the measured asymmetry by only about 1\%.  The 
asymmetry for the deuterium is measured to yield 
\beq
A=-6.79\pm 0.64 \pm 0.51~{\rm ppm}.
\eeq
On the other hand, the theoretical asymmetry for the deuterium is given by
\beq
A=(-7.27+0.75~G_{M}^{s}+1.78~G_{A}^{e}(T=1))~{\rm ppm}.
\label{asymmd}  
\eeq
Note that in this case the expected asymmetry is $-8.8 {\rm ppm}$ 
again assuming zero strange quark contribution and the axial corrections of 
Ref.~\cite{zhu00}.  

Combining this measurement with the previously reported hydrogen 
asymmetry~\cite{sample00} and with the expressions in Eqs. (\ref{asymmp}) and 
(\ref{asymmd}) leads to the two sets of diagonal bans in Figure~\ref{samdata}.  The inner 
portion of each band corresponds to the statistical error, and the outer 
portion corresponds to statistical and systematic errors combined in 
quadrature.  The best experimental value for the strange magnetic form 
factor is given by (\ref{expdata}).
 
As noted in recent papers~\cite{mckeown99bk,dong98} most model calculations 
tend to produce negative values of $G_M^s(0)$, typically about $-0.3$.  A 
recent calculation using lattice QCD techniques (in the quenched 
approximation) reports a result $G_{M}^{s}(0)= -0.36 \pm 0.20$~\cite{dong98}. 
A recent study using a constrained Skyrme-model Hamiltonian that fits the 
baryon magnetic moments yields a positive value of $G_M^s(0)= +0.37$~\cite{hong97}.

\begin{figure}
\centerline{\psfig{figure=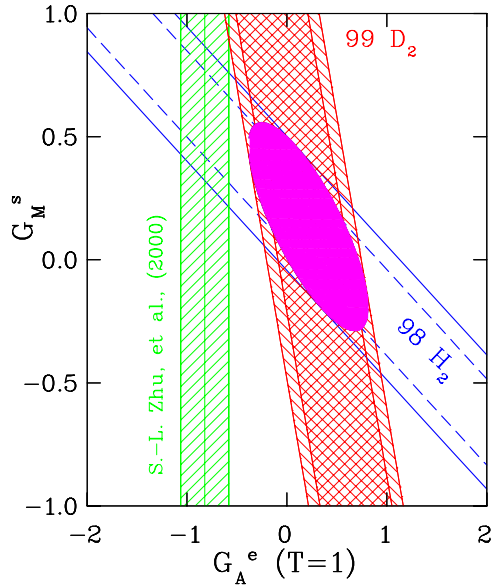,width=2.5in}}
\vskip 0.2 in
\caption{A result of a combined analysis of the data from the two 
SAMPLE measurements.  The two error bands from the hydrogen 
experiment~\cite{sample00} and the deuterium experiment are indicated.  
The inner hatched region includes the statistical error and the outer 
represents the systematic uncertainty added in quadrature.  The ellipse 
represents the allowed region for both form factors at the $1\sigma$ level.  
Also plotted is the estimate of the isovector axial $e-N$ form factor 
$G_{A}^{e}~(T=1)$, obtained by using the anapole form factor and radiative 
corrections of Zhu et al.~\cite{zhu00}.}
\label{samdata}
\end{figure}


\subsection{Strange form factors of baryons in chiral models}


In this section we will revisit the symmetry breaking mass effects to 
investigate the V-spin symmetric Coleman-Glashow sum rules \cite{hong932} in 
the framework of the perturbative scheme, where the representation mixing 
coefficients can be obtained in the quantum mechanical perturbation theory, 
differently from the Yabu-Ando approach discussed in the previous section 
with the direct diagonalization.

In the perturbative method, the Hamiltonian is split up into $H=H_{0}+H_{SB}$ 
where $H_{0}$ is the SU(3) flavor symmetric part given by 
(\ref{chiralham}) and the symmetry breaking part is described by
\beq
H_{SB}=m(1-D^{8}_{88})
\label{hsbper}
\eeq
with $m$ the inertia parameter corresponding to $\omega$ of 
(\ref{omegapara}) in the Yabu-Ando method where the Hamiltonian has been 
divided into the representation independent and dependent parts.

Provided one includes the representation mixing as in the previous section, 
the baryon wave function is described in terms of the higher representation
\beq
|B\rangle=|B\rangle^{8}-C_{\bar{10}}^{B}|B\rangle^{\bar{10}}
               -C_{27}^{B}|B\rangle^{27}
\label{bragle}
\eeq
where the representation mixing coefficients are explicitly calculated as
\beq
C_{\lambda}^{B}=\frac{_{\lambda}\langle B|H_{SB}|B\rangle_{8}}
                  {E_{\lambda}-E_{8}}
\label{clambdab}
\eeq
with the eigenvalues $E_{\lambda}$ and eigenfunctions $|B\rangle^{\lambda}=
\Phi_{B}^{\lambda}\otimes|{\rm intrinsic}\rangle$ of the equation 
$H_{0}|B\rangle^{\lambda}=E_{\lambda}|B\rangle^{\lambda}$.  Here 
$\Phi_{B}^{\lambda}$ is the collective wavefunction discussed above and the 
intrinsic state degenerate to all the baryons is described by a Fock state 
of the quark operator and the classical meson configuration.

Using the octet wavefunctions with the higher representation mixing 
coefficients (\ref{clambdab}), the additional hyperfine structure of the 
magnetic moment spectrum in the quantum mechanical perturbative scheme is 
given by
\beq
\delta\mu_{B}^{i}=-2\sum_{\lambda = \bar{10},27}\frac{
_{8}\langle B|\hat{\mu}^{i}|B\rangle_{\lambda}{}_{\lambda}\langle B|H_{SB}
|B\rangle_{8}}{E_{\lambda}-E_{8}}
\label{deltamubi}
\eeq
up to the first order of $m$, the strength of the symmetry breaking in 
(\ref{hsbper}).  It is interesting here to note that one has the off-diagonal 
matrix elements of the magnetic moment operators $\hat{\mu}^{i}$ with 
higher representations $\bar{\bf 10}$ and ${\bf 27}$, differently from the 
diagonal matrix elements of the chiral symmetric magnetic moments in the 
section 3.1.  This fact is presumably related to the existence of 
{\it exotic} states \cite{jezabek87} belonging to $\bar{\bf 10}$ and 
${\bf 27}$, which decay to the initial states in ${\bf 8}$ through the 
channel of the operator $D_{88}^{8}$ related to the symmetry breaking mass 
effects.

One can then obtain the V-spin symmetry relations in the perturbative 
corrections of the octet magnetic moments
\bea
\delta\mu_{p}&=&\delta\mu_{\Xi^{-}}=m{\cal I}_{2}
(\frac{2}{125}{\cal M}+\frac{8}{1125}({\cal N}-2{\cal N}^{\prime}))\nonumber\\
\delta\mu_{n}&=&\delta\mu_{\Sigma^{-}}=m{\cal I}_{2}
(\frac{31}{750}{\cal M}-\frac{46}{1125}({\cal N}-\frac{21}{23}
{\cal N}^{\prime}))\nonumber\\
\delta\mu_{\Sigma^{+}}&=&\delta\mu_{\Xi^0}=m{\cal I}_2
(\frac{1}{125}{\cal M}+\frac{4}{1125}({\cal N}-2{\cal N}^\prime ))\nonumber\\
\delta\mu_{\Lambda}&=&m{\cal I}_2
(\frac{9}{500}{\cal M}+\frac{1}{125}({\cal N}-2{\cal N}^\prime ))\nonumber\\
\delta\mu_{\Sigma^0}&=&m{\cal I}_2 (\frac{37}{1500}{\cal M}-\frac{7}{375}
({\cal N}-\frac{17}{21}{\cal N}^\prime ))
\label{deltamus}
\eea
where the operator $\hat{\mu}^{i}_{FSB}$ is neglected due to its small 
contributions.

Here one notes that the above V-spin symmetric relations come from the SU(3) 
group theoretical fact that the matrix elements of the operators in 
(\ref{deltamubi}), such as 
$\langle 8|D_{38}^{8}+\frac{1}{\sqrt{3}}D_{88}^{8}|\lambda\rangle\langle
\lambda |D_{88}^{8}|8\rangle$, have degeneracy for the V-spin multiplets 
$(p$, $\Xi^{-})$, $(n$, $\Sigma^{-}$) and ($\Xi^{0}$, $\Sigma^{+}$) as in the 
U-spin symmetry of the section 3.1.  Also as in the Yabu-Ando approach since 
the baryon wavefunctions in the multiquark structure of the CBM act on the 
magnetic moment operators with the quark and meson phase contributions in 
their inertia parameters, one could have the meson cloud content in the 
qqq$\bar{\rm q}$q multiquark Fock subspace of the chiral bag.




Now we consider the form factors of the baryon octet with internal structure 
in the framework of the CBM.  The baryons in the CBM are definitely extended 
objects with internal structure characterized by the bag radius and dressed 
by the meson cloud.  As discussed before, the $F_{2}(0)$ is interpreted as 
the anomalous magnetic moments of the baryon octet $\mu^{an}=\mu-Q$ whose 
numerical values can be easily obtained from Table~\ref{form} by subtracting 
the corresponding electric charges.  Here one should note that the EM currents 
$J_{EM}^{\mu}$ obtainable from (\ref{jvmua}) are conserved as mentioned before 
and the charge density operator is a constant of motion so that the EM charge 
operator can be quantized in a conventional way even though the EM charge density 
is modified due to the derivative dependent symmetry breaking terms.

In the strange flavor sector, the {\it strange} form factors \cite{jaffe89} at 
zero momentum transfer can be calculated from Eqs. (\ref{muss}) and (\ref{f2bs0gmge}) 
to yield
\begin{equation}
\begin{array}{ll}
F^{(s)}_{2N}(0) = -3\mu^{(s)}_N, &
F^{(s)}_{2\Lambda}(0) = -3\mu^{(s)}_\Lambda-1, \\
F^{(s)}_{2\Sigma}(0) = -3\mu^{(s)}_\Sigma-1, &
F^{(s)}_{2\Xi}(0) = -3\mu^{(s)}_\Xi -2.
\end{array}
\label{f8s}
\end{equation}
Note that the I-spin symmetric relation in Eq. (\ref{udsym}) can be
expressed in a simpler form as
\begin{equation}
F^{(u)}_{2B}(0) = F^{(d)}_{2,\bar{B}}(0).
\label{udsym2}
\end{equation}
Now the baryon octet strange form factors in Eq. (\ref{f8s}) can be explicitly 
splitted into three pieces as follows
\begin{equation}
F_{2B}^{(s)}=F_{2B}^{(s),0}({\cal M},{\cal N},{\cal N}^{\prime})
+F_{2B}^{(s),1}({\cal P},{\cal Q})+\delta F_{2B}^{(s),2}(m{\cal I}_{2}).
\label{f2b3pieces}
\eeq
\begin{table}[t]
\caption{The strange form factors of baryon octet}
\begin{center}
\begin{tabular}{crrrrrrrr}
\hline
    & \multicolumn{1}{c}{$F^{(s),0}_{2N}$}
    & \multicolumn{1}{c}{$\delta F^{(s),1}_{2N}$ }
    & \multicolumn{1}{c}{$\delta F^{(s),2}_{2N}$}
    & \multicolumn{1}{c}{$F^{(s)}_{2N}$}
    & \multicolumn{1}{c}{$F^{(s)}_{2\Lambda}$}
    & \multicolumn{1}{c}{$F^{(s)}_{2\Xi}$}
    & \multicolumn{1}{c}{$F^{(s)}_{2\Sigma}$}
    & \multicolumn{1}{c}{$F^{0}_{2}$}\\
\hline
Fit   & 0.16  & 0.28 & $-0.07$
      & 0.37 & 1.37 & 1.22    & $-0.99$ &0.26 \\
 CBM  & $-0.19$ &$-0.12$ & 0.61
      & 0.30 & 0.49 & 0.25    & $-1.54$ &$-0.67$ \\
  SM  & $-0.13$ &$-0.09$ & 0.20
      &$-0.02$ &0.51 & 0.09   & $-1.74$ &$-0.67$ \\
\hline
\end{tabular}
\end{center}
\label{form}
\end{table}

In the adjoint representation, one can obtain the CS and explicit current 
FSB contributions to the strange form factors  
\bea
F_{2N}^{(s),0}&=&\frac{7}{20}{\cal M}-\frac{1}{15}({\cal N}+\frac{1}{2}
{\cal N}^{\prime})\nonumber\\
F_{2\Lambda}^{(s),0}&=&\frac{9}{20}{\cal M}+\frac{1}{5}({\cal N}+\frac{1}{2}
{\cal N}^{\prime})-1\nonumber\\
F_{2\Xi}^{(s),0}&=&\frac{3}{5}{\cal M}+\frac{4}{15}({\cal N}+\frac{1}{2}
{\cal N}^{\prime})-2\nonumber\\
F_{2\Sigma}^{(s),0}&=&\frac{11}{20}{\cal M}-\frac{1}{5}({\cal N}+\frac{1}{2}
{\cal N}^{\prime})-1
\label{f2baryon0}
\eea
and 
\beq
\begin{array}{ll}
F_{2N}^{(s),1}=-\frac{1}{15}{\cal P}-\frac{1}{30}{\cal Q}, &
F_{2\Lambda}^{(s),1}=\frac{1}{5}{\cal P}+\frac{1}{10}{\cal Q}\\
F_{2\Xi}^{(s),1}=\frac{1}{3}{\cal P}+\frac{1}{15}{\cal Q}, &
F_{2\Sigma}^{(s),1}=-\frac{11}{45}{\cal P}-\frac{1}{18}{\cal Q}.\\
\end{array}
\label{f2baryon1}
\eeq
Treated in the quantum mechanical perturbative scheme of the previous section, 
the representation mixing coefficients from the multiquark structure can be 
explicitly given as  
\bea
\delta F_{2N}^{(s),2}&=&m{\cal I}_{2}(-\frac{43}{750}{\cal M}+\frac{38}{1125}
{\cal N}-\frac{26}{1125}{\cal N}^{\prime})\nonumber\\
\delta F_{2\Lambda}^{(s),2}&=&m{\cal I}_{2}(-\frac{9}{250}{\cal M}-\frac{2}{125}
{\cal N}+\frac{4}{125}{\cal N}^{\prime})\nonumber\\
\delta F_{2\Xi}^{(s),2}&=&m{\cal I}_{2}(-\frac{3}{125}{\cal M}-\frac{4}{375}
{\cal N}+\frac{8}{375}{\cal N}^{\prime})\nonumber\\
\delta F_{2\Sigma}^{(s),2}&=&m{\cal I}_{2}(-\frac{37}{750}{\cal M}
+\frac{14}{375}{\cal N}-\frac{34}{1125}{\cal N}^{\prime}).
\label{deltaf2octet}
\eea

Next using the flavor singlet vector currents $J_{V}^{\mu 0}$, which can be 
constructed by replacing $\hat{Q}_{a}$ by $1$ in (\ref{jvmua}), instead of the 
EM currents in the matrix element (\ref{extptcleform}) we can also obtain the 
flavor singlet form factors \cite{jaffe89, park90, beise91}
\beq
F_{2}^{0}=\frac12 {\cal M}-1
\label{f2o}
\eeq
which are degenerate with all the baryon octet even in the multiquark structure 
regardless of whether one uses the Yabu-Ando or perturbative methods.

In Table~\ref{form} one can acquire the numerical values for the strange form 
factors and flavor singlet form factors.


\section{Unification of chiral bag model with other models}
\setcounter{equation}{0}
\renewcommand{\theequation}{\arabic{section}.\arabic{equation}}


\subsection{Connection to naive nonrelativistic quark model}


Until now we have considered the static properties such as the magnetic 
moments and form factors of the baryon octet in the CBM which unifies the 
MIT bag and Skyrmion models with the bag radius parameter.  In this section 
we will relate the CBM with the naive NRQM by investigating the 
model-independent sum rules in the magnetic moments, which have been already 
derived in the CBM in the previous sections for the baryon octet and decuplet 
to have a clue for the unification of the naive NRQM into the CBM.

In the naive NRQM, the wave function of a baryon consists of several degrees 
of freedom \cite{huang82}
\beq
\psi ({\rm baryon})=\psi ({\rm space})\psi({\rm spin})\psi ({\rm flavor})
\psi({\rm color})
\label{psibaryonnaive}
\eeq
where the spatial wave function is symmetric in the ground state, and the 
spin state can either be completely symmetric ($J=\frac{3}{2}$) or of mixed 
symmetry ($J=\frac12$), and there are $3^{3}$ flavor combinations which can be 
reshuffled into irreducible representations of SU(3)\footnote{In terms of 
group theory, the combination of three quark flavors yield a decuplet, a 
singlet and two octets since the direct product of three fundamental 
representations of SU(3) decomposes according to the Clebsch-Gordan series 
${\bf 3}\otimes {\bf 3}\otimes {\bf 3}={\bf 1}\oplus {\bf 8}\oplus {\bf 8}
\oplus {\bf 10}$.}, and the color wave function is antisymmetric and 
degenerate to all the baryons since every naturally occurring baryon is a 
color singlet, and the full baryon wave function is antisymmetric under the 
interchange of any two quarks.

The baryon octet wave function is then constructed by the nontrivial 
spin/flavor wave function of the form
\beq
\psi({\rm baryon~octet})=\frac{\sqrt{2}}{3}(\psi_{12}({\rm spin})\psi_{12}
({\rm flavor})+{\rm two~ other~ terms})
\label{psibaryon8}
\eeq
where $\psi_{ij} ({\rm spin})$ and $\psi_{ij}({\rm flavor})$ are the states 
with mixed symmetry such that their product is completely symmetric in quarks 
$i$ and $j$.  Since each quark has the intrinsic spin in itself the baryon 
magnetic moments in the naive NRQM are obtained by linearly adding the 
magnetic angular momentum quantum number of the wave function.  The baryon 
octet magnetic moments and the $\Lambda\Sigma^{0}$ transition matrix element 
can then be constructed in terms of linear vector sum of the three constituent 
quark magnetic moments $\mu_{q}$ (q=u,d,s) \cite{perkins87} 
\bea
\mu_{p}&=&\frac{1}{3}(4\mu_{u}-\mu_{d}),
~~~~\mu_{n}=\frac{1}{3}(-\mu_{u}+4\mu_{d})\nonumber\\   
\mu_{\Sigma^{-}}&=&\frac{1}{3}(4\mu_{d}-\mu_{s}),
~~~~\mu_{\Sigma^{0}}=\frac{1}{3}(2\mu_{u}+2\mu_{d}-\mu_{s}),
\nonumber\\
\mu_{\Sigma^{+}}&=&\frac{1}{3}(4\mu_{u}-\mu_{s}),
~~~~\mu_{\Xi^{-}}=\frac{1}{3}(-\mu_{d}+4\mu_{s}),\nonumber\\
\mu_{\Xi^{0}}&=&\frac{1}{3}(-\mu_{u}+4\mu_{s}),
~~~~\mu_{\Lambda}=\mu_{s}\nonumber\\
\frac{1}{\sqrt{3}}\mu_{\Lambda\Sigma^{0}}&=&\frac{1}{3}(-4\mu_{u}
-\mu_{d}),
\label{munaive8}
\eea
where $\mu_{q}=Q_{q}(m_{N}/m_{q})$ in unit of nuclear magnetons (=$e\hbar/2
m_{N}c$) with $m_{q}$ the q-flavor quark mass and $m_{N}$ the nucleon mass.  
Here one notes that due to $\mu_{u}/\mu_{d}=-2$ one has the ratio $\mu_{n}/
\mu_{p}=-\frac{2}{3}$ comparable to the experimental value $-0.69$ and the 
CBM prediction $-\frac{3}{4}$ in the leading order of $N_{c}$.

Since the d- and s-flavor charges are degenerate in the SU(3) EM charge 
operator $\hat{Q}_{EM}$, the baryon magnetic moments in the SU(3) flavor 
symmetric limit with the chiral symmetry breaking masses $m_{u}=m_{d}=m_{s}$ 
satisfy the U-spin symmetric Coleman-Glashow sum rules in the naive NRQM, the 
analogy of the U-spin symmetry relations (\ref{magmomcs}) in the CBM
\bea
\mu_{\Sigma^{+}}&=&\mu_{p}=\frac{m_{N}}{m_{u}}\nonumber\\
\mu_{\Xi^{0}}&=&\mu_{n}=-\frac{2}{3}\frac{m_{N}}{m_{u}}\nonumber\\
\mu_{\Xi^{-}}&=&\mu_{\Sigma^{-}}=-\frac{1}{3}\frac{m_{N}}{m_{u}}\nonumber\\
\mu_{\Sigma^{0}}&=&-\mu_{\Lambda}=\frac{1}{3}\frac{m_{N}}{m_{u}}
\label{musumrulesnaive}
\eea
and the other Coleman-Glashow sum rule (\ref{sumboctet}) for the summation of 
the magnetic moments over all the octet baryons.  The naive NRQM also 
predicts the other sum rules (\ref{muls02}) and the relations of the hyperon 
and transition magnetic moments in terms of the nucleon magnetic moments 
(\ref{musums}) in the CBM.

Using the projection operators (\ref{projectionop}) one can easily see that 
the nucleon magnetic moments in the u-flavor channel of the naive NRQM are 
given by $\mu_{p}^{(u)}=\frac{4}{3}Q_{u}(m_{N}/m_{u})$ and $\mu_{n}^{(u)}=
-\frac{1}{3}Q_{u}(m_{N}/m_{u})$, and the d-flavor components of nucleon 
magnetic moments are given by (\ref{udssymmetry}) as in the CBM, but 
$\mu_{N}^{(s)}=0$ due to the absence of the strange quarks in the nucleon of 
the naive NRQM.  In general, one can easily see that the SU(3) flavor 
components of hyperon magnetic moments also satisfy the identities 
(\ref{udssymmetry}) in the naive NRQM.

In the more general SU(3) flavor symmetry breaking case with $m_{u}=m_{d}
\neq \mu_{s}$ one can easily see that the baryon octet magnetic moments fulfill 
the Coleman-Glashow sum rule (\ref{musigma0}), since $\mu_{\Sigma^{+}}+
\mu_{\Sigma^{-}}$ is independent of the third component of the isospin, and 
the last model-independent relation in (\ref{musums}) and the identities in 
(\ref{udssymmetry}) hold since they are the relations derived in the same 
strangeness sector.

Together with the above model-independent Coleman-Glashow sum rules shared 
by two models, the CBM predictions propose the unification of the naive NRQM 
and the CBM which has the meson cloud, around the quarks of the naive NRQM, 
located both in the quark and meson phases.  In other words, the CBM can be 
phenomenologically proposed as an effective NRQM in the adjoint 
representation and model-independent relations and Cheshire cat properties 
are shown to support the effective NRQM conjecture with meson cloud.

In Table 2 the SU(2) CBM predictions \cite{gerry80} are explicitly listed 
to be compared with the naive NRQM and SU(3) CBM so that the pure kaon 
contributions to the baryon magnetic moments can be explicitly 
calculated with respect to the naive NRQM.

Next starting with the symmetric spin configuration in the ground state with 
symmetric $\psi ({\rm space})$ one can have the spin-$\frac{3}{2}$ baryon 
decuplet wave function in the naive NRQM with the symmetric flavor state 
to yield
\beq
\psi({\rm baryon~decuplet})=\psi_{s}({\rm spin})\psi_{s}({\rm flavor}).
\label{psibaryon10}
\eeq

In the naive NRQM the baryon magnetic moments are then obtained as the linear 
sum of the three constituent quark magnetic moments, similarly to 
(\ref{munaive8})
\bea
\mu_{\Delta^{-}}&=&3\mu_{d},
~~~~\mu_{\Delta^{0}}=\mu_{u}+2\mu_{d}\nonumber\\   
\mu_{\Delta^{+}}&=&2\mu_{d}+\mu_{d},
~~~~\mu_{\Delta^{++}}=3\mu_{u},
\nonumber\\
\mu_{\Sigma^{*-}}&=&2\mu_{u}+\mu_{s},
~~~~\mu_{\Sigma^{*0}}=\mu_{u}+\mu_{d}+\mu_{s},\nonumber\\
\mu_{\Sigma^{*+}}&=&2\mu_{u}+\mu_{s},
~~~~\mu_{\Xi^{*-}}=\mu_{d}+2\mu_{s}\nonumber\\
\mu_{\Xi^{*0}}&=&\mu_{u}+2\mu_{s},
~~~~\mu_{\Omega^{-}}=3\mu_{s}.
\label{munaive10}
\eea

In the SU(3) flavor symmetric limit with the chiral symmetry breaking 
masses $m_{u}=m_{d}=m_{s}$, the decuplet baryons with the EM charge $Q_{EM}$ 
are described by \cite{beg64}
\beq
\mu_{B}=Q_{EM}\frac{m_{N}}{m_{u}}
\label{mubqemnaive}
\eeq
and satisfy the U-spin symmetry relations (\ref{uspin10}) and (\ref{cg10}).

On the other hand the $\Delta$ magnetic moments in the u- and s-flavor 
channels are given by
\beq
\mu_{\Delta}^{(u)}=(Q+1)\frac{2}{3}\frac{m_{N}}{m_{u}},
~~~~\mu_{\Delta}^{(s)}=0,
\label{mudeltaus}
\eeq
and in general all the baryon decuplet magnetic moments fulfill the 
model-independent relations (\ref{udssymmetry}) in the u- and d-flavor 
components and the I-spin symmetry in the s-flavor channel where the 
isomultiplets have the degenerate strangeness number.

Finally one should note that the other sum rules (\ref{sumrule1}) and 
(\ref{sumrule2}) and the identities in (\ref{udssymmetry}) hold even in the 
SU(3) flavor symmetry breaking case ($m_{u}=m_{d}\neq m_{s}$) since they are 
the relations derived in the same strangeness sector.

The above model-independent sum rules in the baryon decuplet satisfied by the 
naive NRQM and the CBM support the effective NRQM conjecture as in the baryon 
octet.  The effective NRQM conjecture discussed in the baryon octet and 
decuplet support the possibility of the unification of the CBM with the naive 
NRQM while the Cheshire catness suggests another clue to the unification of 
the CBM with the Skyrmion model.  In the next section we will proceed to 
consider the other plausible unification of the CBM with the NJL model, chiral 
perturbation theory, CK model and chiral quark soliton model.


\subsection{Connection to other models}


So far the chiral soliton model such as the Skyrmion model have been 
constructed mainly on the basis of the low-energy meson phenomenology 
since the effective meson Lagrangian underlying QCD is not known.  There has 
been some progress in deriving effective meson Lagrangian 
either directly from QCD \cite{cahill85} or from the quark flavor 
dynamics~\cite{dhar85} of the NJL model \cite{nambu61}.  Especially it has been claimed 
that the Skyrmion model can be derived \cite{ebert86, reinhardt89} from the 
NJL model in the limit of large vector and axial-vector meson masses.  
Consequently one may claim that there can be plausibility in the unification 
of the CBM with the NJL model.

Next in the strong chiral symmetry breaking limit the Yabu-Ando approach to 
the Skyrmion model has suggested \cite{yabu88} the mass formula similar to 
the one derived in the bound state scheme in CK model so that one may conclude 
that the perturbation and bound state schemes are two extreme limits of the 
Yabu-Ando approach.  Similarly, in the large limit of the symmetry 
breaking strength $\omega$ of (\ref{omegapara}), the CBM results are 
comparable to those of Refs.~\cite{kunz89, oh91} estimated in the bound state 
scheme of CK model.

Finally, in the chiral quark soliton model \cite{kim97}, the baryon decuplet 
magnetic moments satisfy the model independent sum rules (\ref{uspin10}) and 
(\ref{cg10}) as in the naive NRQM.  Moreover one can easily see that the CBM 
shares with the naive NRQM and chiral quark soliton model the following sum 
rules
\begin{eqnarray}
-4\mu_{\Delta^{++}}+6\mu_{\Delta^{+}}+3\mu_{\Sigma^{*+}}-6\mu_{\Sigma^{*0}}
+\mu_{\Omega^{-}}&=&0  \label{sum1} \\
-2\mu_{\Delta^{++}}+3\mu_{\Delta^{+}}+2\mu_{\Sigma^{*+}}-4\mu_{\Sigma^{*0}}
+\mu_{\Xi^{*-}}&=&0  \label{sum2} \\
-\mu_{\Delta^{++}}+2\mu_{\Delta^{+}}-2\mu_{\Sigma^{*0}} +\mu_{\Xi^{*0}}&=&0
\label{sum3} \\
\mu_{\Delta^{++}}-2\mu_{\Delta^{+}}+\mu_{\Delta^{0}}&=&0  
\label{sumrules10cq}
\end{eqnarray}
and
\begin{equation}
\mu_{\Delta^{0}}-\mu_{\Sigma^{*-}}=\mu_{\Sigma^{*+}}-\mu_{\Xi^{*0}} 
=\frac{1 }{2}(\mu_{\Delta^{+}}-\mu_{\Xi^{*-}}) =\frac{1}{3}(\mu_{\Delta^{++}}
-\mu_{\Omega^{-}}).  
\label{vspins10cq}
\eeq
These sum rules also suggest the possibility of unification of the CBM with 
the naive NRQM and chiral quark soliton model.


\section{Improved Dirac quantization of Skyrmion model}
\setcounter{equation}{0}
\renewcommand{\theequation}{\arabic{section}.\arabic{equation}}


\subsection{Modified mass spectrum in SU(2) Skyrmion}


In this section, we will apply the Batalin-Fradkin-Tyutin (BFT) method to the 
Skyrmion to obtain the modified mass spectrum of the baryons by including the Weyl ordering 
correction.  We will next canonically quantize the SU(2) Skyrme model by using
the Dirac quantization method, which will be shown to be consistent with the 
BFT one after the adjustable parameters are introduced to define 
the generalized momenta without any loss of generality \cite{hong991}.

Now we start with the SU(2) Skyrmion Lagrangian of the form
\begin{equation}
L_{SM}=\int{\rm d}r^{3}\left[-\frac{1}{4}f_{\pi}^{2}{\rm tr}(l_{\mu}l^{\mu})
+\frac{1} {32e^{2}}{\rm tr}[l_{\mu},l_{\nu}]^{2}\right]
\label{sklagrangian}
\end{equation}
where $l_{\mu}=U^{\dagger}\partial_{\mu}U$ and $U$ is an SU(2) matrix 
satisfying the boundary condition $\lim_{r \rightarrow \infty} U=I$ so that 
the pion field vanishes as $r$ goes to infinity.  

On the other hand, in the Skyrmion model, since the hedgehog ansatz has 
maximal or spherical symmetry, it is easily seen that spin plus isospin equals 
zero, so that isospin transformations and spatial rotations are related to 
each other.  Furthermore spin and isospin states can be treated by collective 
coordinates $a^{\mu}=(a^{0},\vec{a})$ $(\mu=0,1,2,3)$
corresponding to the spin and isospin rotations
\begin{equation}
A(t) = a^{0}+i\vec{a}\cdot\vec{\tau},
\end{equation}
which is the time dependent collective variable defined on the SU(2)$_{F}$ 
group manifold and is related with the zero modes associated with the 
collective rotation (\ref{collective}) in the SU(3) CBM.  With the hedgehog 
ansatz described in section 1.4 and the collective rotation $A(t)\in$ SU(2), 
the chiral field can be given by $U(\vec{x},t)=A(t)U_{0}(\vec{x})
A^{\dagger}(t)=e^{i\tau_{a}R_{ab}\hat{x}_{b}f(r)}$ where $R_{ab}=\frac{1}{2}
{\rm tr} (\tau_{a}A\tau_{b}A^{\dagger})$ and the Skyrmion Lagrangian can be 
written as
\begin{equation}
L_{SM}=-M_{0}+2{\cal I}_{10}\dot{a}^{\mu}\dot{a}^{\mu}   
\label{sklagrangianoo}
\end{equation}
where $M_{0}$ and ${\cal I}_{10}$ are the static mass and the moment of 
inertia given as
\bea
M_{0}&=&\frac{2\pi f_{\pi}}{e}\int_{0}^{\infty}{\rm d}z~z^{2}\left(\left(
\frac{d\theta}{dz}\right)^{2}+\left(2+2\left(\frac{d\theta}{dz}\right)^{2}
+\frac{\sin^{2}\theta}{z^{2}}\right)\frac{\sin^{2}\theta}{z^{2}}\right)
\label{skstaticmass}\\
{\cal I}_{10}&=&\frac{8\pi}{3e^{3}f_{\pi}}\int_{0}^{\infty}{\rm d}z~z^{2}
\sin^{2}\theta \left(1+\left(\frac{d\theta}{dz}\right)^{2}
+\frac{\sin^{2}\theta}{z^{2}}\right)
\label{skinertiamom}
\eea
with the dimensionless quantity $z=ef_{\pi}r$.  

Introducing the canonical momenta $\pi^{\mu}=4{\cal I}_{10}\dot{a}^{\mu}$ 
conjugate to the collective coordinates $a^{\mu}$ one can then obtain the 
canonical Hamiltonian
\begin{equation}
H=M_{0}+\frac{1}{8{\cal I}_{10}}\pi^{\mu}\pi^{\mu}.  
\label{hamilsu2}
\end{equation}
and the spin and isospin operators 
\begin{eqnarray}
J^{i}&=&\frac{1}{2}(a^{0}\pi^{i}-a^{i}\pi^{0}-\epsilon_{ijk}a^{j}\pi^{k}), 
\nonumber \\
I^{i}&=&\frac{1}{2}(a^{i}\pi^{0}-a^{0}\pi^{i}-\epsilon_{ijk}a^{j}\pi^{k}).
\label{spinsu2}
\end{eqnarray}

On the other hand our system has the second class constraints\footnote{%
Here one notes that, due to the commutator $\{\pi^{\mu},\Omega_{1}\}
=-2a^{\mu}$, one can obtain the algebraic relation $\{\Omega_1,H\}={\frac{1}{%
2{\cal {I}}}}\Omega_2$.} 
\begin{equation}
\Omega_{1} = a^{\mu}a^{\mu}-1\approx 0,~~~\Omega_{2} = a^{\mu}\pi^{\mu}
\approx 0,  
\label{omega2su2}
\end{equation}
to yield the Poisson algebra with $\epsilon^{12}=-\epsilon^{21}=1$
\begin{equation}
\Delta_{k k^{\prime}}=\{\Omega_{k},\Omega_{k^{\prime}}\} = 2\epsilon^{k
k^{\prime}}a^{\mu}a^{\mu}  
\label{deltasu2}.
\end{equation}

We now recapitulate the construction of the first class SU(2) 
Hamiltonian.  Following the BFT formalism \cite{batalin86, hong991, oliveira97} 
we introduce two auxiliary fields $(\theta,\pi_{\theta})$ with the Poisson 
brackets
\begin{equation}
\{\theta, \pi_{\theta}\}=1.
\label{phisu2}
\end{equation}
One can then obtain the first class constraints
\begin{equation}
\tilde{\Omega}_{1}=\Omega_{1}+2\theta,~~~
\tilde{\Omega}_{2}=\Omega_{2}-a^{\mu}a^{\mu}\pi_{\theta},
\label{1stconst}
\end{equation}
satisfying the first class constraint Lie algebra $\{\tilde{\Omega}_{i},\tilde{\Omega}_{j}\}=0$.
Demanding that they are strongly involutive in the extended phase space, i.e.,
$\{\tilde{\Omega}_{i}, \tilde{{\cal F}}\}=0$, one can construct the first 
class BFT physical fields $\tilde{{\cal F}}=(\tilde{a}^{\mu}, 
\tilde{\pi}^{\mu})$ corresponding to the original fields ${\cal F}=(a^{\mu},
\pi^{\mu})$, as a power series of the auxiliary fields 
$(\theta, \pi_{\theta})$ 
\begin{eqnarray}
\tilde{a}^{\mu}&=&a^{\mu}\left(\frac{a^{\mu}a^{\mu}+2\theta}
{a^{\mu}a^{\mu}}\right)^{1/2}
\nonumber \\
\tilde{\pi}^{\mu}&=&(\pi^{\mu}-a^{\mu}\pi_{\theta})\left(\frac{a^{\mu}a^{\mu}}{
a^{\mu}a^{\mu}+2\theta}\right)^{1/2}.
\label{pitildesu2}
\end{eqnarray}

As discussed in Ref. \cite{kim98}, any functional ${\cal K}(\tilde{{\cal F}})$
of the first class fields $\tilde{{\cal F}}$ is also first class, namely, 
$\tilde{{\cal K}}({\cal F};\Phi )={\cal K}(\tilde{{\cal F}})$.  Using the property, 
we construct a first-class Hamiltonian in terms of the above BFT physical variables.  
The result is
\begin{equation}
\tilde{H}=M_{0}+\frac{1}{8{\cal I}_{10}}\tilde{\pi}^{\mu}\tilde{\pi}^{\mu}.
\label{htildesu2}
\end{equation}
We then directly rewrite this Hamiltonian in terms of the original as well as
auxiliary fields \cite{hong001}
\begin{equation}
\tilde{H}=M_{0}+\frac{1}{8{\cal I}_{10}}(\pi^{\mu}-a^{\mu}\pi_{\theta})
(\pi^{\mu}-a^{\mu}\pi_{\theta})\frac{a^{\nu}a^{\nu}}{a^{\nu}a^{\nu}+2\theta},
\label{hctsu2}
\end{equation}
which is also strongly involutive with the first class constraints 
$\{\tilde{\Omega}_{i},\tilde{H}\}=0$.  However, with the first class 
Hamiltonian (\ref{hctsu2}), one cannot naturally
generate the first class Gauss' law constraint from the time evolution of
the primary constraint $\tilde{\Omega}_{1}$. Now, by introducing an additional 
term proportional to the first class constraints $\tilde{\Omega}_{2}$ into $%
\tilde{H}$, we obtain an equivalent first class Hamiltonian
\begin{equation}
\tilde{H}^{\prime}=\tilde{H}+\frac{1}{4{\cal I}_{10}}\pi_{\theta} 
\tilde{\Omega}_{2},
\label{hctpsu2}
\end{equation}
which naturally generates the Gauss' law constraint
\begin{equation}
\{\tilde{\Omega}_{1},\tilde{H}^{\prime}\}=\frac{1}{2{\cal I}_{10}}
\tilde{\Omega}_{2},~~~
\{\tilde{\Omega}_{2},\tilde{H}^{\prime}\}=0. 
\label{gasu2}
\end{equation}
Here one notes that $\tilde{H}$ and $\tilde{H}^{\prime}$ act on physical
states in the same way since such states are annihilated by the first class
constraints.

\begin{table}[t]
\caption{The static properties of baryons in the standard and Weyl ordering
corrected (WOC) Skyrmions compared with experimental data.  The 
quantities used as input parameters are indicated by $*$.}
\begin{center}
\begin{tabular}{crrr}
\hline
Quantity  &Standard  &WOC &Experiment\\
\hline
$M_{N}$ &939 {\rm MeV}$^{*}$ &939 {\rm MeV}$^{*}$ &939 {\rm MeV}\\
$M_{\Delta}$ &1232 {\rm MeV}$^{*}$ &1232 {\rm MeV}$^{*}$ &1232 {\rm MeV}\\
$f_{\pi}$ &64.5 {\rm MeV} &63.2 {\rm MeV} &93.0 {\rm MeV}\\
e &5.44 &5.48 &$-$\\
$\langle r^{2}\rangle^{1/2}_{M,I=0}$  &0.92 {\rm fm} &0.94 {\rm fm}
                                      &0.81 {\rm fm}\\
$\langle r^{2}\rangle^{1/2}_{M,I=1}$  &$\infty$ &$\infty$ &0.80 {\rm fm}\\
$\langle r^{2}\rangle^{1/2}_{I=0}$  &0.59 {\rm fm} &0.60 {\rm fm}
                                    &0.72 {\rm fm}\\
$\langle r^{2}\rangle^{1/2}_{I=1}$  &$\infty$ &$\infty$ &0.88 {\rm fm}\\
$\mu_{p}$ &1.87 &1.89 &2.79\\
$\mu_{n}$ &$-1.31$ &$-1.32$ &$-1.91$\\
$\mu_{\Delta^{++}}$ &3.72 &3.75 &4.7$-$6.7\\
$\mu_{N\Delta}$ &2.27 &2.27 &3.29\\
$\mu_{p}-\mu_{n}$ &3.18 &3.21 &4.70\\
\hline
\end{tabular}
\end{center}
\label{bft}
\end{table}

Using the first class constraints in this Hamiltonian (\ref{hctpsu2}), one
can obtain the Hamiltonian of the form
\begin{equation}
\tilde{H}^{\prime}=M_{0}+\frac{1}{8{\cal I}_{10}}(a^{\mu}a^{\mu}\pi^{\nu}
\pi^{\nu}-a^{\mu}\pi^{\mu}a^{\nu}\pi^{\nu}).  
\label{htilde2su2}
\end{equation}
Following the symmetrization procedure, the first class Hamiltonian yields
the slightly modified energy spectrum with the Weyl ordering correction 
\cite{lee81,hong991,oliveira97,hong001}
\begin{equation}
\langle\tilde{H}^{\prime}\rangle=M_{0}+\frac{1}{2{\cal I}_{10}}\left[I(I+1)
+\frac{1}{4}\right]
\label{nhtsu2}
\end{equation}
where $I$ is the isospin quantum number of baryons.

Next, using the Weyl ordering corrected energy spectrum (\ref{nhtsu2}), we 
easily obtain the hyperfine structure of the nucleon and delta hyperon masses 
to yield the static mass and the moment of inertia 
\begin{equation}
M_{0}=\frac{1}{3}(4M_{N}-M_{\Delta}),~~~
{\cal I}_{10}=\frac{3}{2}(M_{\Delta}-M_{N})^{-1}.
\label{massessu2}
\end{equation}
Substituting the experimental values $M_{N}=939$ MeV and $N_{\Delta}=1232$ MeV 
into Eq. (\ref{massessu2}) and using the expressions (\ref{skinertiamom}), one 
can predict the pion decay constant $f_{\pi}$ and the Skyrmion parameter $e$ 
as follows
$$
f_{\pi}=63.2~{\rm MeV},~~~e=5.48.
$$  
With these fixed values of $f_{\pi}$ and $e$, one can then proceed to yield the 
predictions for the other static properties of the baryons.  The isoscalar and 
isovector mean square (magnetic) charge radii and the baryon and transition 
magnetic moments are contained in Table~\ref{bft}, together with the 
experimental data and the standard Skyrmion 
predictions~\cite{adkins83, zahed86, liu83}.\footnote{For the 
delta magnetic moments, we use the experimental data of Nefkens et 
al.~\cite{nefkens78}.} It is remarkable that the effects of Weyl ordering 
correction in the baryon energy spectrum are propagated through the model 
parameters $f_{\pi}$ and $e$ to modify the predictions of the baryon static 
properties. 




Moreover, one can show that, by fixing a free adjustable parameter $c$ introduced 
to define generalized momenta, the baryon energy eigenvalues obtained by the 
standard Dirac method are consistent with the above BFT result.  To be more specific, 
we can obtain the modified quantum energy spectrum of the baryons~\cite{hong991} 
\begin{equation}
\langle H_{N}\rangle=M_{0}+\frac{1}{8{\cal I}_{10}}[l(l+2)+\frac{9}{4}-c^{2}]
\label{hwcapp}
\end{equation}
which is consistent with the BFT result (\ref{nhtsu2}) if the adjustable parameter $c$ is fixed 
with the values $c=\pm\frac{\sqrt{5}}{2}$.  Here one notes that these values for 
the parameter $c$ relate the Dirac bracket scheme with the BFT one to yield the 
desired quantization in the SU(2) Skyrmion model so that one can achieve the 
unification of these two different formalisms.  (For details see Ref.~\cite{hong991}.) 




On the other hand, we can obtain the BRST invariant Lagrangian in the 
framework of the BFV formalism \cite{fradkin75, fujiwara90, bizdadea95} which 
is 
applicable to theories with the first class constraints by introducing two 
canonical sets of ghosts and anti-ghosts together with auxiliary fields.  Following 
the procedure in Appendix C.1, one can arrive at the BRST invariant 
Lagrangian~\cite{hong001}
\begin{eqnarray}
L_{eff}&=&-M_{0}+\frac{2{\cal I}_{10}}{1-2\theta}\dot{a}^{\mu}\dot{a}^{\mu} 
-\frac{2{\cal I}_{10}}{(1-2\theta)^{2}}\dot{\theta}^{2} 
-2{\cal I}_{10}(1-2\theta)^{2}(B+2\bar{{\cal C}}{\cal C})^{2}  \nonumber \\
& &-\frac{\dot{\theta}\dot{B}}{1-2\theta} +\dot{\bar{{\cal C}}}\dot{{\cal C}},
\label{leff2app}
\end{eqnarray}
which is invariant under the BRST transformation
\begin{eqnarray}
\delta_{B}a^{\mu}&=&\lambda a^{\mu}{\cal C},~~~ \delta_{B}\theta=-\lambda
(1-2\theta){\cal C},  \nonumber \\
\delta_{B}\bar{{\cal C}}&=&-\lambda B,~~~ \delta_{B}{\cal C}=\delta_{B}B=0.
\label{brstskapp}
\end{eqnarray}
Here ${\cal C}$ ($\bar{{\cal C}}$) and $B$ are the (anti-)ghosts and the corresponding 
auxiliary fields.  (For details see Appendix C.1.)


\subsection{Phenomenology in SU(3) Skyrmion}


Now let us consider the hyperfine splittings for the SU(3) 
Skyrmion~\cite{witten83,mazur84,jezabek87} which has been studied in two main 
schemes as discussed in the previous chapters.  Firstly, the
SU(3) cranking method exploits rigid rotation of the Skyrmion in the collective
space of SU(3) Euler angles with full diagonalization of the flavor symmetry
breaking (FSB) terms~\cite{hong932,hong931,hong94}. Especially, Yabu and 
Ando~\cite{yabu88}
proposed the exact diagonalization of the symmetry breaking terms by
introducing higher irreducible representation mixing in the baryon wave
function, which was later interpreted in terms of the multiquark
structure~\cite{kim89,lee89} in the baryon wave function. Secondly, Callan and
Klebanov~\cite{callan85} suggested an interpretation of baryons containing a
heavy quark as bound states of solitons of the pion chiral Lagrangian with
mesons. In their formalism, the fluctuations in the strangeness direction are
treated differently from those in the isospin 
directions~\cite{callan85,scoccola88}.

In order to generalize the standard flavor symmetric (FS) SU(3) Skyrmion 
rigid rotator approach~\cite{kleb94,kleb96} to the SU(3) Skyrmion case with 
the pion mass and FSB terms, we will now investigate the chiral breaking pion mass and 
FSB effects on $c$ the ratio of the strange-light to light-light interaction 
strengths and $\bar{c}$ that of the strange-strange to light-light.

Now we start with the SU(3) Skyrmion Lagrangian of the form
\begin{eqnarray}
 {\cal L}&=&-\frac{1}{4}f_{\pi}^{2}{\rm tr}(l_{\mu}l^{\mu}) +\frac{1}{32e^{2}}%
 {\rm tr}[l_{\mu},l_{\nu}]^{2}+{\cal L}_{WZW}  \nonumber \\
 & &+\frac{1}{4}f_{\pi}^{2}{\rm tr}M(U+U^{\dag}-2)+{\cal L}_{FSB},  \nonumber
 \\
 {\cal L}_{FSB}&=&\frac{1}{6}(f_{K}^{2}m_{K}^{2}-f_{\pi}^{2}m_{\pi}^{2}) {\rm %
tr}((1-\sqrt{3}\lambda_{8})(U+U^{\dag}-2))  \nonumber \\
& &-\frac{1}{12}(f_{K}^{2}-f_{\pi}^{2}){\rm tr} ((1-\sqrt{3}%
\lambda_{8})(Ul_{\mu}l^{\mu} +l_{\mu}l^{\mu}U^{\dag})),
\label{lagfsbapp}
\end{eqnarray}
where $f_{\pi}$ ($f_{K}$) and $e$ are the pion (kaon) decay constants and the 
dimensionless Skyrme parameter as before.  Here 
$l_{\mu}=U^{\dag}\partial_{\mu}U$ with an SU(3) matrix $U$ and $M$ is 
proportional to the quark mass matrix given by
\[
M={\rm diag}~ (m_{\pi}^{2},~ m_{\pi}^{2},~ 2m_{K}^{2}-m_{\pi}^{2}),
\]
where $m_{\pi}=138$ MeV and $m_{K}=495$ MeV. Note that ${\cal L}_{FSB}$ is
the FSB correction term due to the relations $m_{\pi}\neq m_{K}$ and
$f_{\pi}\neq f_{K}$~\cite{pari91,hong932} and the Wess-Zumino-Witten (WZW) term~\cite{witten83} is
described by the action
\beq
\Gamma_{WZW}=-\frac{iN_{c}}{240\pi^{2}}\int_{{\sf M}}{\rm d}^{5}r\epsilon^{\mu%
\nu \alpha\beta\gamma}{\rm tr}(l_{\mu}l_{\nu}l_{\alpha}l_{\beta}l_{\gamma}),
\label{wzwsu3}
\eeq
where $N_{c}$ is the number of colors and the integral is done on the
five-dimensional manifold ${\sf M}=V\times S^{1}\times I$ with the
three-space volume $V$, the compactified time $S^{1}$ and the unit interval $%
I$ needed for a local form of WZW term.  Here note that we have used the 
three-space volume $V$ instead of $\bar{V}$ of the CBM case.

Using Eq. (\ref{nhtsu3}) in Appendix C.2 and following the Klebanov and 
Westerberg's quantization scheme \cite{kleb94} for the strangeness flavor 
direction in the BFT formalism, one can obtain the Hamiltonian of the form
\begin{eqnarray}
H&=&M_{0}+\frac{1}{2}\Gamma _{0}m_{\pi }^{2}+\frac{1}{2{\cal I}_{10}}(%
\vec{I}^{2}+\frac{1}{4})+\frac{N_{c}}{8{\cal I}_{20}^{\prime }}(\mu_{K} -1)
a^{\dag }a
\nonumber \\
&&+\left[ \frac{1}{2{\cal I}_{10}}-\frac{1}{4{\cal I}_{20}^{\prime }\mu_{K} }%
\left(1+\frac{\Gamma _{2}}{{\cal I}_{10}}\right) (\mu_{K}-1)\right] 
a^{\dag }\vec{I}\cdot \vec{\tau}a  \nonumber \\
&&+\left[\frac{1}{8{\cal I}_{10}}-\frac{1}{8{\cal I}_{20}^{\prime}\mu_{K}^{2}}
\left(1+\frac{\Gamma _{2}}{{\cal I}_{10}}\mu_{K}\right.\right.  \nonumber \\
&&\left.\left.-\frac{\Gamma _{2}^{2}+2{\cal I}_{10} (\Gamma _{1}-\Gamma _{2})%
}{4{\cal I}_{10}{\cal I}_{20}^{\prime}}(\mu_{K}-1)\right) 
(\mu_{K}-1)\right](a^{\dag}a)^{2},   
\label{htildesu3app}
\end{eqnarray}
where
$$
\mu_{K}=\left( 1+\frac{\chi ^{2}m_{K}^{2}-m_{\pi }^{2}
+\Gamma _{3}/\Gamma _{0}}{m_{0}^{2}}\right) ^{1/2},~~~
m_{0}=\frac{N_{c}}{4(\Gamma _{0}{\cal I}_{20}^{\prime })^{1/2}}
$$
and $a^{\dag }$ is creation operator for constituent strange quarks and we
have ignored the irrelevant creation operator $b^{\dag}$ for strange
antiquarks~\cite{kleb94}. Then, introducing the angular momentum of the strange
quarks $\vec{J}_{s}=\frac{1}{2}a^{\dag }\vec{\tau}a$, one can rewrite the 
Hamiltonian (\ref{htildesu3app}) as
\begin{equation}
H=M_{0}+\frac{1}{2}\Gamma _{0}m_{\pi }^{2}+\omega a^{\dag }a+\frac{1}{2%
{\cal I}_{10}}\left( \vec{I}^{2}+2c\vec{I}\cdot \vec{J}_{s}+\bar{c}\vec{J}%
_{s}^{2}+\frac{1}{4}\right)  
\label{hjssu3app}
\end{equation}
where
\begin{eqnarray}
\omega &=&\frac{N_{c}}{8{\cal I}_{20}^{\prime }}(\mu_{K}-1),  \nonumber \\
c&=&1-\frac{{\cal I}_{10}}{2{\cal I}_{20}^{\prime}\mu_{K}}
\left(1+\frac{\Gamma_{2}}{{\cal I}_{10}}\right) (\mu_{K} -1),  \nonumber \\
\bar{c}&=&1-\frac{{\cal I}_{10}}{{\cal I}_{20}^{\prime }\mu_{K}^{2}}\left( 1+%
\frac{\Gamma _{2}}{{\cal I}_{10}}\mu_{K} -\frac{\Gamma _{2}^{2}
+2{\cal I}_{10}(\Gamma _{1}-\Gamma _{2})}{4{\cal I}_{10}
{\cal I}_{20}^{\prime }}(\mu_{K}-1)\right) (\mu_{K}-1).  \nonumber
\end{eqnarray}
Here note that the FSB effects are included in $c$ and $\bar{c}$, through
$\Gamma_{1}$, $\Gamma_{2}$, ${\cal I}_{20}^{\prime}$ and $\chi$ and
$\Gamma_{3}$ in $\mu_{K}$.

\begin{table}[h]
\caption{The values of $c$ and $\bar{c}$ in the massless pion and massive pion
rigid rotator approaches to the SU(3) Skyrmions compared with experimental
data. For the rigid rotator approaches, both the predictions in the flavor
symmetric (FS) case and flavor symmetry breaking (FSB) one are listed.}
\begin{center}
\begin{tabular}{lcc}
\hline
Source & $c$ & $\bar{c}$ \\ \hline
Rigid rotator, massless and FS  & 0.92 & 0.86 \\
Rigid rotator, massless and FSB & 0.82 & 0.69 \\
Rigid rotator, massive and FS   & 0.79 & 0.66 \\
Rigid rotator, massive and FSB  & 0.67 & 0.56 \\
Experiment & 0.67 & 0.27 \\ \hline
\end{tabular}
\end{center}
\label{ccbar}
\end{table}

The Hamiltonian (\ref{hjssu3app}) then yields the structure of the hyperfine
splittings as follows
\begin{eqnarray}
\delta M &=&\frac{1}{2{\cal I}_{10}}\left[cJ(J+1)+(1-c) \left( I(I+1)-\frac{%
Y^{2}-1}{4}\right)\right.  \nonumber \\
&&\left.+(1+\bar{c}-2c)\frac{Y^{2}-1}{4}+\frac{1}{4}(1+\bar{c}-c)\right],
\end{eqnarray}
where $\vec{J}=\vec{I}+\vec{J}_{s}$ is the total angular momentum of the
quarks, and $c$ and $\bar{c}$ are the modified quantities due to the
existence of the FSB effect as shown above.

Now using the experimental values of the pion and kaon decay constants
$f_{\pi}=93$ MeV and $f_{K}=114$ MeV, we fix the value of the Skyrmion
parameter $e$ to fit the experimental data of $c_{exp}=0.67$ to
yield the predictions for the values of $c$ and $\bar{c}$
\begin{equation}
c=0.67,~~~\bar{c}=0.56
\end{equation}
which are contained in Table~\ref{ccbar}, together with the experimental data 
and the SU(3) rigid rotator predictions without pion mass.  For the massless and
massive rigid rotator approaches we have used the above values for the decay 
constants $f_{\pi}$ and $f_{K}$ to obtain both the predictions in the FS and 
FSB cases.  As a result, we have explicitly shown that the more realistic 
physics considerations via the pion mass and the FSB terms improve both the 
$c$ and $\bar{c}$ values, as shown in Table~\ref{ccbar}~\cite{hong01prd}.


\subsection{Berry phase and Casimir energy in SU(3) Skyrmion} 


Now we investigate the relations between the Hamiltonian
(\ref{hjssu3app}) and the Berry phases~\cite{berry84}.  In the Berry phase 
approach to the SU(3) Skyrmion, the Hamiltonian takes the simple 
form~\cite{rho96cnd}
\begin{equation}
H^{*}=\epsilon_{K}+\frac{1}{8{\cal
I}_{1}}(\vec{R}^{2}-2g_{K}\vec{R}\cdot
\vec{T}_{K}+g_{K}^{2}\vec{T}_{K}^{2})  \label{h*}
\end{equation}
where $\epsilon_{K}$ is the eigenenergy in the $K$ state, $g_{K}$
is the Berry charge, $\vec{R}$ ($\vec{L}$) is the right (left)
generators of the group $SO(4)\approx SU(2)\times SU(2)$ and
$\vec{T}_{K}$ is the angular
momentum of the "slow" rotation. We recall that 
$\vec{I}=\frac{\vec{L}}{2} =-\frac{\vec{R}}{2}$ and 
$\vec{L}^{2}=\vec{R}^{2}$ on $S^{3}$.
Applying the BFT scheme to the Hamiltonian (\ref{h*}) we can
obtain the Hamiltonian of the form
\begin{equation}
\tilde{H}^{*}=\epsilon_{K}+\frac{1}{2{\cal
I}_{1}}(\vec{I}^{2}+g_{K}\vec{I}
\cdot\vec{T}_{K}+(\frac{g_{K}}{2})^{2}\vec{T}_{K}^{2}+\frac{1}{4}).
\label{ht*}
\end{equation}
In the case with the relation $\bar{c}=c^{2}$, the Hamiltonian
(\ref{hjssu3app}) is equivalent to $\tilde{H}^{*}$ in the Berry phase
approach where the corresponding physical quantities can be read
off as follows
\begin{equation}
\epsilon_{K}=M_{0}+\frac{1}{2}\Gamma_{0}m_{\pi}^{2}+\omega a^{\dag}a,~~
\vec{T}_{K}=\vec{J}_{s},~~g_{K}=2c.
\label{rels}
\end{equation}
The same case with the Hamiltonian (\ref{ht*}) follows from the
quark model and the bound state approach with the quartic terms in
the kaon field neglected. In fact, the strange-strange
interactions in the Hamiltonian (\ref{hjssu3app}) break these relations
to yield the numerical values of $\bar{c}$ in Table~\ref{ccbar}.

Next, the baryon mass spectrum in the chiral models can be described in 
powers of $N_{c}$ as follows,
\beq
H=E_{1}N_{c}+E_{0}N_{c}^{0}+E_{-1}N_{c}^{-1}+\cdots
\eeq
where the ellipsis stands for the contributions from the higher 
order terms of $N_{c}^{-1}$.  Note that, for instance in Eq. 
(\ref{hjssu3app}), $E_{1}$ and $E_{-1}$ correspond to 
$M_{0}+\frac{1}{2}\Gamma _{0}m_{\pi }^{2}+\omega a^{\dag }a$ and the terms 
from the rotational degrees of freedom associated with the moment of inertia 
$1/{\cal I}_{10}$, respectively.  Moreover, in fitting the values of the pion 
and kaon decay constants $f_{\pi}$ and $f_{K}$ and the value of the Skyrmion
parameter $e$ as in the numerical evaluations of Table~\ref{bft} and 
Table~\ref{ccbar} for instance, we have missed the Casimir effect 
contributions, with which one can improve the predictions to obtain more 
realistic phenomenology.

Now, in order to take into account the missing order $N_{c}^{0}$ effects, we 
consider the Casimir energy contributions to the 
Hamiltonian (\ref{hjssu3app}).  The Casimir energy originated from the meson 
fluctuation can be given by the phase shift formula~\cite{mou93,park98k}
\begin{eqnarray}
E_{0}&=&\frac{1}{2\pi}\sum_{i=\pi, K}\left(\int_{0}^{\infty}{\rm d}p
\left[-\frac{p}{\sqrt{p^{2}+m_{i}^{2}}}(\delta (p)-\bar{a}_{0} p^{3}
-\bar{a}_{1}p) +\frac{\bar{a}_{2}}{\sqrt{p^{2}+\mu^{2}}}\right]\right.
\nonumber \\
& &\left.-\frac{3}{8}\bar{a}_{0}m_{i}^{4} \left(\frac{3}{4}+\frac{1}{2}\ln \frac{\mu^{2}%
}{m_{i}^{2}}\right) +\frac{1}{4}\bar{a}_{1} m_{i}^{2}\left(1+\ln\frac{\mu^{2}}{m_{i}^{2}%
}\right) -m_{i}\delta (0)\right)+\cdots
\nonumber
\end{eqnarray}
where the ellipsis denotes the contributions from the counter
terms and the bound states (if any). Here $\mu$ is the energy
scale and $\delta (p)$ is the phase shift with the momentum $p$
and the coefficients $\bar{a}_{i}$ $(i=0,1,2)$ are defined by the
asymptotic expansion of $\delta ^{\prime} (p)$, namely,
$\delta^{\prime}(p)=3\bar{a}_{0}p^{2}+\bar{a}_{1}-\frac{\bar{a}_{2}}
{p^{2}}+\cdots$.  Even though the Casimir energy correction does
not contribute to the ratios $c$ and $\bar{c}$ since these ratios
are associated with the order $1/N_{c}$ piece of the Hamiltonian
(\ref{hjssu3app}), these effects are significant in the baryon mass 
itself~\cite{mou93,park98k} given in Eqs. (\ref{yabuham}) and 
(\ref{hjssu3app}), and also seems to be significant in other physical 
quantities such as the H dibaryon mass~\cite{kleb96}.    

Now, we would like to briefly comment on numerical estimation of the 
Casimir energy.  Even though it is difficult to determine the magnitude of the 
Casimir energy due to the ambiguity in using the derivative expansion in the chiral soliton 
models, the magnitude is known to depend on the dynamical details of the 
Lagrangian and loop corrections and its sign is estimated to be negative.  
The preliminary calculations produce the Casimir energy with range 
$-(200-1000)$ MeV~\cite{zahed86prd} and later the more reliable estimations 
yield $-(500-600)$ MeV~\cite{holzwarth92}.    


\section{Superqualiton model}
\setcounter{equation}{0}
\renewcommand{\theequation}{\arabic{section}.\arabic{equation}}


\subsection{Color-flavor-locking phase and Q-matter}


So far we have studied the phenomenology of hadron physics without 
introducing matter density degrees of freedom.  In this section, 
we consider the possibilities of the applications of the chiral models such 
as superqualiton model to the dense matter physics.  Here note that one can have 
somewhat intriguing similarity between the hadron-quark 
continuity~\cite{schafer99ef} and the CCP.  In other words, Sch\"afer and Wilczek 
proposed that the three-flavor color-flavor locking (CFL) operative at asymptotic 
density continues upto the chiral transition density, in which case there will 
be hadron-quark continuity since there will be a one-to-one mapping between 
hadrons and quark/gluons.

Now, we consider quark matter with a finite baryon number described by QCD with a chemical 
potential, which is to restrict the system to have a fixed baryon number,
\begin{equation}
{\cal L}={\cal L}_{\rm QCD}-\mu \bar\psi_i\gamma^0\psi_i,
\end{equation}
where $\bar\psi_i\gamma^0\psi_i$ is
the quark number density and equal chemical potentials are assumed for
different flavors, for simplicity.
The ground state in the CFL phase is nothing but
the Fermi sea where all quarks are gaped by Cooper-pairing;
the octet has a gap $\Delta$ while the singlet has $2\Delta$.
Equivalently, this system can be described in terms of bosonic
degrees of freedom, which are small fluctuations of Cooper pairs.
Following Ref.~\cite{hong99dk},
we introduce bosonic variables, defined as
\begin{equation}
{U_L}_{ai}(x)\equiv\lim_{y\to x}{\left|x-y\right|^{\gamma_m}
\over\Delta(p_F)}\,\epsilon_{abc}\epsilon_{ijk}
\psi^{bj}_{L}(-\vec v_F,x)\psi^{ck}_{L}(\vec v_F,y),
\end{equation}
where $\gamma_m$ ($\sim\alpha_s$) is the anomalous dimension of the
diquark field and $\psi(\vec v_F,x)$ denotes a quark field with
momentum close to a Fermi momentum $\mu\vec v_F$~\cite{hong00tn}.
Similarly, we define $U_R$ in terms of right-handed quarks to describe
the small fluctuations of the condensate of right-handed quarks.
Since the bosonic fields, $U_{L,R}$, are colored, they will interact
with gluons. In fact, the colored massless excitations will
constitute the longitudinal components of gluons through Higgs
mechanism. Thus, the low-energy effective Lagrangian density
for the bosonic fields in the CFL phase can be written as
\bea
{\cal L}_{\rm eff}&=&\left[\frac{1}{4}{F}^{2}{\rm tr}(\partial_{\mu}
  U_{L}^{\dag}\partial^{\mu}U_{L})
  +n_L{\cal L}_{WZW}
 +(L\leftrightarrow R)\right]+{\cal L}_m
 \nonumber\\
& &-\frac{1}{4}F_{\mu\nu}^{A}F^{\mu\nu A}
 +g_{s}G_{\mu}^{A}J^{\mu A}+\cdots,
\label{efflagqual}
\end{eqnarray}
where ${\cal L}_m$ is the meson mass term and the ellipsis
denotes the higher order terms in the derivative
expansion, including mixing terms between $U_L$ and $U_R$.
The gluons couple to the bosonic fields through a minimal coupling
with a conserved current, given as
\bea
J^{A\mu}&=&{i\over 2}F^2{\rm tr}~U_L^{-1}T^A\partial^{\mu}U_L+
{1\over 24\pi^2}\epsilon^{\mu\nu\rho\sigma}
{\rm tr}~T^AU_L^{-1}\partial_{\nu}U_LU_L^{-1}\partial_{\rho}U_L
U_L^{-1}\partial_{\sigma}U_L
\nonumber\\
& &+(L\leftrightarrow R)+\cdots,
\label{jamuqual}
\eea
where the ellipsis denotes the currents from the higher order
derivative terms in Eq.~(\ref{efflagqual}).
$F$ is a quantity analogous to the pion decay constant,
calculated to be $F\sim\mu$ in the CFL color
superconductor~\cite{son00cm}. The Wess-Zumino-Witten (WZW)
term~\cite{witten83} is described by the action (\ref{wzwsu3}) 
in the previous chapter.  The coefficients of the WZW term in the 
effective Lagrangian, (\ref{efflagqual}), have been shown to be 
$n_{L,R}=1$ by matching the flavor anomalies~\cite{hong99dk}, 
which is later confirmed by an explicit calculation~\cite{nowak00wa}.

Among the small fluctuations of condensates, the colorless
excitations correspond to genuine Nambu-Goldstone (NG) bosons,
which can be described by a color singlet combination of
$U_{L,R}$~\cite{hong00ei,casalbuoni99wu}, given as
\begin{equation}
\Sigma_i^j\equiv U_{Lai}U_R^{*aj}.
\end{equation}
The NG bosons transform under the $SU(3)_L\times SU(3)_R$
chiral symmetry as
\begin{equation}
\Sigma\mapsto g_L\Sigma g_R^{\dagger},\quad {\rm with}\quad
g_{L,R}\in SU(3)_{L,R}.
\end{equation}

Since the chiral symmetry is explicitly broken by current quark
mass, the instanton effects, and the electromagnetic interaction,
the NG bosons will get mass, which has been calculated by various
groups~\cite{son00cm,hong00ei,rho00xf}.  Here we focus on the meson mass 
due to the current strange quark mass ($m_s)$, since it will be dominant for 
the intermediate density. Then, the meson mass term is simplified as
\begin{equation}
{\cal L}_m=C\, {\rm tr}(M^T\Sigma)\cdot {\rm tr} (M^*\Sigma^{\dagger})+
O(M^4),
\label{m}
\end{equation}
where $M={\rm diag}(0,0,m_s)$ and
$C\sim \Delta^4/\mu^2\,\cdot\ln(\mu^2/\Delta^2)$. (Note that in general there
will be two more mass terms quadratic in $M$. But, they all
vanish if we neglect the current mass of up and down quarks
and also the small color-sextet component of the Cooper
pair~\cite{hong00ei}.)

Now, let us try to describe the CFL color superconductor in
terms of the bosonic variables. We start with the effective
Lagrangian described above, which is good at low energy, without
putting in the quark fields. As in the Skyrmion model of baryons,
we anticipate the gaped quarks
come out as solitons, made of the bosonic degrees of freedom.
That the Skyrme picture can be realized in the CFL color superconductor
is already shown in~\cite{hong99dk}, but there the mass of the
soliton is not properly calculated. Here, by identifying the
correct ground state of the CFL superconductor in the bosonic
description, we find the superqualitons have same quantum numbers
as quarks with mass of the order of gap, showing that they are
really the gaped quarks in the CFL color superconductor.
Furthermore, upon quantizing the zero modes of the soliton, we find
that high spin excitations of the soliton have energy of order
of $\mu$, way beyond the scale where the effective bosonic
description is applicable, which we interpret as the absence of
high-spin quarks, in agreement with the fermionic description.
It is interesting to note that, as we will see below, by calculating
the soliton mass in the bosonic description, one finds the coupling
and the chemical potential dependence of the Cooper-pair gap,
at least numerically, which gives us a complementary way,
if not better, of estimating the gap.

As the baryon number (or the quark number) is conserved,
though spontaneously broken,~\footnote{The spontaneously
broken baryon number
just means that the states in the Fock space do not have
a well-defined baryon number. But, still the baryon number
current is conserved in the operator sense~\cite{coleman85}.}
the ground state in the bosonic description should
have the same baryon (or quark) number as the ground state
in the fermionic description.
Under the $U(1)_Q$ quark number symmetry,
the bosonic fields transform as
\begin{equation}
U_{L,R}\mapsto e^{i\theta Q}U_{L,R}e^{-i\theta Q}=e^{2i\theta}U_{L,R},
\end{equation}
where $Q$ is the quark number operator,
given in the bosonic description as
\begin{equation}
Q=i\int {\rm d}^3x~{F^2\over4}
{\rm tr}\left[U_L^{\dagger}\partial_tU_L-\partial_tU_L^{\dagger}U_L
+\left(L\leftrightarrow R\right)\right],
\end{equation}
neglecting the quark number coming from the WZW term,
since the ground state has no nontrivial topology.
The energy in the bosonic description is
\begin{equation}
E=\int{\rm d}^3x{F^2\over 4}
{\rm tr}\left[\left|\partial_tU_L\right|^2
+\left|\vec\nabla U_L\right|^2
+\left(L\leftrightarrow R\right)\right]+E_m+\delta E,
\end{equation}
where $E_m$ is the energy due to meson mass and $\delta E$ is the
energy coming from the higher derivative terms. Assuming
the meson mass energy is positive and $E_{m}+\delta E\ge0$, which
is reasonable because $\Delta/F\ll1$,
we can take, dropping the positive terms due to the spatial derivative,
\begin{equation}
E\ge \int{\rm d}^3x{F^2\over 4}
{\rm tr}\left[\left|\partial_tU_L\right|^2
+\left(L\leftrightarrow R\right)\right](\equiv E_Q).
\end{equation}
Since for any number $\alpha$
\begin{equation}
\int{\rm d}^3x~{\rm tr}\left[\left|U_L+\alpha i\partial_tU_L
\right|^2+\left(L\leftrightarrow R\right)\right]\ge0,
\end{equation}
we get a following Schwartz inequality,
\begin{equation}
Q^2\le  I\,E_Q,
\label{boundqual}
\end{equation}
where we defined
\begin{equation}
I={F^2\over 4}\int{\rm d}^3x\,{\rm tr}\left[U_LU_L^{\dagger}
+\left(L\leftrightarrow R\right)\right].
\end{equation}
Note that the lower bound in Eq.~(\ref{boundqual})
is saturated for $E_Q=\omega Q$ or
\begin{equation}
U_{L,R}=e^{i\omega t} \quad{\rm with}\quad \omega={Q\over I}.
\end{equation}
The ground state of the color superconductor, which has the lowest
energy for a given quark number $Q$, is nothing but so-called
$Q$-matter, or the interior of very large
$Q$-ball~\cite{coleman85ki,hong98ur}. Since in the fermionic
description the system has the quark number $Q=\mu^3/\pi^2\int{\rm
d}^3x=\mu^3/\pi^2\cdot I/F^2$, we find, using
$F\simeq0.209\mu$~\cite{son00cm},
\begin{equation}
\omega={1\over\pi^2}\left({\mu\over F}\right)^3 F
\simeq2.32\mu.
\label{fpi}
\end{equation}
By passing, we note that numerically $\omega$ is very close to
$4\pi F$.  The ground state of the system in the bosonic description is a
$Q$-matter whose energy per unit quark number is $\omega$.
Now, let us suppose we consider creating a $Q=1$ state out of
the ground state. In the fermionic description,
this corresponds that we excite a gaped quark in the Fermi sea into
a free state, which costs energy at least $2\Delta$.
In the bosonic description, this amounts to creating a superqualiton
out of the $Q$-matter, while reducing the quark number of the
$Q$-matter by one. Therefore, since, reducing the quark number of
the $Q$-matter by one, we gain energy $\omega$, the energy cost to
create a gaped quark from the ground state in the bosonic
description is
\begin{equation}
\delta {\cal E}=M_Q-\omega,
\label{deltae}
\end{equation}
where $M_Q$ is the energy of the superqualiton configuration.  
From the relation that $2\Delta=M_Q-\omega$, we later estimate
numerically the coupling and the chemical potential dependence
of the Cooper gap.


\subsection{Bosonization of QCD at high density}


It is sometimes convenient to describe a system of interacting fermions in 
terms of bosonic variables, since often in that description the interaction 
of elementary excitations becomes weak and perturbative approaches are 
applicable~\cite{stone94ys}.  Now, we attempt to bosonize cold quark matter 
of three light flavors, where the low-lying energy states are bosonic.

Following the Skyrme picture of baryons in QCD at low density,
we now investigate how gaped quarks in high density QCD are realized
in its bosonic description with the Lagrangian given in
Eq.~(\ref {efflagqual})~\cite{hong99dk,hong01plb}.
Assuming the maximal symmetry in the superqualiton,
we seek a static configuration for the field $U_{L}$ which is the
$SU(2)$ hedgehog in color-flavor in $SU(3)$ as in (\ref{hedgehog})
\begin{equation}
U_{Lc}(\vec{x})=\left(
\begin{array}{cc}
e^{i\vec{\tau}\cdot\hat{x}\theta (r)} & 0 \\
0 & 1
\end{array}
\right)
\label{uqual}
\end{equation}
where $\theta (r)$ is the chiral angle determined by minimizing the static
mass $M_{0}^{Q}$ given below and for unit winding number we take
$\lim_{r\rightarrow \infty}\theta (r)=0$ and $\theta (0)=\pi$.
The static configuration for the other fields are described as
\begin{equation}
U_{R}=0,~~G_{0}^{A}=\frac{x^{A}}{r}\omega (r),~~G_{i}^{A}=0.
\end{equation}

Now we consider the zero modes of the $SU(3)$ superqualiton as follows
\begin{equation}
U(\vec{x},t)={\cal A}(t)U_{Lc}(\vec{x}){\cal A}(t)^{\dag}.
\label{uxt}
\end{equation}
The Lagrangian for the zero modes is then given by
\begin{equation}
L=-M_{0}^{Q}+\frac{1}{2}I_{ab}{\rm tr}({\cal A}^{\dag}\dot{\cal A}
\frac{\lambda_{a}}{2}){\rm tr}({\cal A}^{\dag}\dot{\cal A}
\frac{\lambda_{b}}{2})-\frac{i}{2}{\rm tr}(Y{\cal A}^{\dag}\dot{\cal A}),
\end{equation}
where $I_{ab}$ is an invariant tensor on ${\cal M}=SU(3)/U(1)$
and $Y$ is the hypercharge
\[
Y=\frac{\lambda_{8}}{\sqrt{3}}=\frac{1}{3}\left(
\begin{array}{ccc}
1 & 0 & 0\\
0 & 1 & 0\\
0 & 0 & -2\\
\end{array}
\right).
\]
Using the above static configuration, we obtain the static mass $M_{0}$ and
the tensor $I_{ab}$ as follows
\bea
M_{0}^{Q}&=&\frac{4\pi}{3}F^{2}\int_{0}^{\infty}{\rm d}r\left[\frac12 r^{2}\left(\frac{d\theta}{dr}\right)^{2}
+\sin^{2}\theta
\right.
\nonumber\\
& &\left.+\frac{\alpha_{s}}{2\pi^{3}F^{2}}
\left(\frac{\theta-\sin\theta\cos\theta-\pi}{2r}\right)^{2}e^{-2m_{E}r}\right],
\nonumber\\
I_{ab}&=&-\frac{32\pi}{9}F^{2}\int_{0}^{\infty}{\rm d}r
r^{2}\sin^{2}\theta = -4I_{1},
~~~~~~~~~~~~~(a=b=1,2,3)\nonumber\\
&=&-\frac{8\pi}{3} F^{2}\int_{0}^{\infty}{\rm d}r r^{2}(1-\cos \theta)
= -4I_{2},
~~~~~~~~(a=b=4,5,6,7)\nonumber\\
&=&0,
~~~~~~~~~~~~~~~~~~~~~~~~~~~~~~~~~~~~~~~~~~~~~~~~~~~~(a=b=8)
\end{eqnarray}
where $\alpha_{s}$ is the strong coupling constant and 
$m_{E}=\mu(6\alpha_{s}/\pi)^{1/2}$ is the electric screening mass for the 
gluons.

As in Appendix C.2, since ${\cal A}$ belongs to $SU(3)$, 
${\cal A}^{\dag}\dot{\cal A}$ is anti-Hermitian and traceless to be expressed 
as a linear combination of $i\lambda_{a}$ as follows
\[
{\cal A}^{\dag}\dot{\cal A}=iFv^{a}\lambda_{a}=iF\left(
\begin{array}{cc}
\vec{v}\cdot\tau +\nu 1 & V \\
V^{\dag} & -2\nu\\
\end{array}
\right)
\]
where $\vec{v}$, $V$ and $\nu$ are given by Eq. (\ref{vs}).  The Lagrangian is 
then expressed as
\begin{equation}
L=-M_{0}^{Q}+2F^{2}I_{1}\vec{v}^{2}+2F^{2}I_{2}V^{\dag}V+\frac{1}{3}N_{c}F\nu.
\end{equation}

In order to separate the SU(2) rotations from the deviations into strange
directions, we write the time-dependent rotations as in Eq. (\ref{curlya}).  
Furthermore, we exploit the time-dependent collective coordinates 
$a^{\mu}=(a^{0},\vec{a})$ $(\mu=0,1,2,3)$ as in the SU(2) 
Skyrmion~\cite{adkins83}, via $A(t) = a^{0}+i\vec{a}\cdot\vec{\tau}$, and the 
small rigid oscillations $S$ described by Eqs. (\ref{stexp}) and (\ref{curlyd}).

After some algebra, one can obtain the relations among the variables in
(\ref{vs}) and the SU(2) collective coordinates $a^{\mu}$ and the strange
deviations $D$ such as
\begin{eqnarray}
F\nu &=&\frac{i}{2}(D^{\dag}\dot{D}-\dot{D}^{\dag}D)-D^{\dag}(a^{0}\vec
{\dot{a}}-\dot{a}^{0}\vec{a}+\vec{a}\times\vec{\dot{a}})\cdot\vec{\tau}D
\nonumber\\
& &-\frac{i}{3}(D^{\dag}\dot{D}-\dot{D}^{\dag}D)D^{\dag}D+\cdots,
\end{eqnarray}
to yield the superqualiton Lagrangian to order $1/N_{c}$
\begin{eqnarray}
L&=&-M_{0}^{Q}+2I_{1}\dot{a}^{\mu}\dot{a}^{\mu}
+4I_{2}\dot{D}^{\dag}\dot{D}+\frac{i}{6}N_{c}(D^{\dag}\dot{D}
-\dot{D}^{\dag}D)-4I_{2}m_{K}^{2}D^{\dag}D\nonumber\\
& &+2i(I_{1}-2I_{2})\{D^{\dag} (a^{0}\vec{\dot{a}}
-\dot{a}^{0}\vec{a}+\vec{a}\times\vec{\dot{a}})\cdot\vec{\tau}\dot{D}
\nonumber\\
& &-\dot{D}^{\dag}(a^{0}\vec{\dot{a}}-\dot{a}^{0}\vec{a} +\vec{a}\times\vec{%
\dot{a}})\cdot\vec{\tau}D\}
-\frac{1}{3}N_{c}D^{\dag}(a^{0}\vec{\dot{a}}
-\dot{a}^{0}\vec{a}+\vec{a}\times\vec{\dot{a}})\cdot\vec{\tau}D  \nonumber \\
& &+2\left(I_{1}-\frac{4}{3}I_{2}\right)(D^{\dag}D)(\dot{D}^{\dag}\dot{D})
-\frac{1}{2}\left(I_{1}-\frac{4}{3}I_{2}\right)(D^{\dag}\dot{D}+\dot{D}^{\dag}D)^{2}
\nonumber \\
& &+2I_{2}(D^{\dag}\dot{D}-\dot{D}^{\dag}D)^{2}-\frac{i}{9}N_{c}(D^{\dag}%
\dot{D}-\dot{D}^{\dag}D)D^{\dag}D\nonumber\\
& &+\frac{8}{3}I_{2}m_{K}^{2}(D^{\dag}D)^{2}
\label{lagqual}
\end{eqnarray}
where we have included the kaon mass terms proportional to the strange quark
mass which is not negligible.

The momenta $\pi^{\mu}$ and $\pi_{s}^{\alpha}$, conjugate to the
collective coordinates $a^{\mu}$ and the strange deviation
$D_{\alpha}^{\dag}$ are given by
\begin{eqnarray}
\pi^{0}&=&4I_{1}\dot{a}^{0}-2i(I_{1}-2I_{2})
(D^{\dag}\vec{a}\cdot\vec{\tau}\dot{D}-\dot{D}^{\dag}\vec{a}\cdot\vec{\tau}%
D) +\frac{1}{3}N_{c}D^{\dag}\vec{a}\cdot\vec{\tau}D  \nonumber \\
\vec{\pi}&=&4I_{1}\vec{\dot{a}}+2i(I_{1}-2I_{2})
\{D^{\dag}(a^{0}\vec{\tau}-\vec{a}\times\vec{\tau})\dot{D} -\dot{D}%
^{\dag}(a^{0}\vec{\tau}-\vec{a}\times\vec{\tau})D\}  \nonumber \\
& &-\frac{1}{3}N_{c}D^{\dag}(a^{0}\vec{\tau}-\vec{a}\times\vec{\tau})D  
\nonumber \\
\pi_{s}&=&4I_{2}\dot{D}-\frac{i}{6}N_{c}D-2i(I_{1}-2I_{2})
(a^{0}\vec{\dot{a}}-\dot{a}^{0}\vec{a}+\vec{a}\times\vec{\dot{a}}) 
\cdot\vec{\tau}D  \nonumber \\
& &+2\left(I_{1}-\frac{4}{3}I_{2}\right)(D^{\dag}D)\dot{D} -\left(I_{1}-%
\frac{4}{3}I_{2}\right)(D^{\dag}\dot{D}+\dot{D}^{\dag}D)D  \nonumber \\
& &-4I_{2}(D^{\dag}\dot{D}-\dot{D}^{\dag}D)D+\frac{i}{9}N_{c}(D^{\dag}D)D
\nonumber
\end{eqnarray}
satisfying the Poisson brackets $\{a^{\mu},\pi^{\nu}\}=\delta^{\mu\nu}$, 
$\{D_{\alpha}^{\dag},\pi_{s}^{\beta}\}=\{D^{\beta},\pi_{s,\alpha}^{\dag}\}
=\delta_{\alpha}^{\beta}$.

Performing Legendre transformation, we obtain the Hamiltonian to order $1/N_{c}$
as follows
\begin{eqnarray}
H&=&M_{0}^{Q}+\frac{1}{8I_{1}}\pi^{\mu}\pi^{\mu}
+\frac{1}{4I_{2}}\pi_{s}^{\dag}\pi_{s}-i\frac{{N}_{c}}{24I_{2}}
(D^{\dag}\pi_{s}-\pi_{s}^{\dag}D) +\left(\frac{N_{c}^{2}}{144I_{2}}\right.\nonumber\\
& &\left.+4I_{2}m_{K}^{2}\right)D^{\dag}D+i\left(\frac{1}{4I_{1}}
-\frac{1}{8I_{2}}\right)
\{D^{\dag} (a^{0}\vec{\pi}-\vec{a}\pi^{0}+\vec{a}\times\vec{\pi})
\cdot\vec{\tau}\pi_{s}  \nonumber
\\
& &-\pi_{s}^{\dag}(a^{0}\vec{\pi}-\vec{a}\pi^{0} +\vec{a}\times\vec{\pi}%
)\cdot\vec{\tau}D\} +\frac{N_{c}}{24I_{2}}D^{\dag}(a^{0}\vec{\pi}-\vec{a}%
\pi^{0} +\vec{a}\times\vec{\pi})\cdot\vec{\tau}D  \nonumber \\
& &+\left(\frac{1}{2I_{1}}-\frac{1}{3I_{2}}\right)(D^{\dag}D)
(\pi_{s}^{\dag}\pi_{s})+\left(\frac{1}{12I_{2}}-\frac{1}{8I_{1}}\right)
(D^{\dag}\pi_{s}+\pi_{s}^{\dag}D)^{2}  \nonumber \\
& &-\frac{1}{8I_{2}}\left(D^{\dag}\pi_{s}-\pi_{s}^{\dag}D\right)^{2} 
-i\frac{N_{c}}{24I_{2}}(D^{\dag}\pi_{s}-\pi_{s}^{\dag}D)(D^{\dag}D)  
\nonumber \\
& &+\left(\frac{N_{c}^{2}}{108I_{2}}-\frac{8}{3}I_{2}m_{K}^{2}
\right)(D^{\dag}D)^{2}.
\label{hamilqual}
\end{eqnarray}

Applying the BFT scheme~\cite{batalin86,wtkim94,hong991} to the above 
result with the auxiliary fields $(\theta, \pi_{\theta})$, one can obtain 
the first class Hamiltonian
\begin{eqnarray}
\tilde{H}&=&M_{0}^{Q}+\frac{1}{8I_{1}}
(\pi^{\mu}-a^{\mu}\pi_{\theta})(\pi^{\mu}-a^{\mu}\pi_{\theta})
\frac{a^{\nu}a^{\nu}}
{a^{\nu}a^{\nu}+2\theta}\nonumber \\
& &+\frac{1}{4I_{2}} \pi_{s}^{\dag}\pi_{s}-i\frac{N_{c}}{24I_{2}}%
(D^{\dag}\pi_{s}-\pi_{s}^{\dag}D) +\left(\frac{N_{c}^{2}}
{144I_{2}}+4I_{2}m_{K}^{2}\right)
D^{\dag}D  \nonumber \\
& &+i\left(\frac{1}{4I_{1}}-\frac{1}{8I_{2}}\right)\{D^{\dag} (a^{0}\vec{%
\pi}-\vec{a}\pi^{0}+\vec{a}\times\vec{\pi}) \cdot\vec{\tau}\pi_{s}
\nonumber \\
& &-\pi_{s}^{\dag}(a^{0}\vec{\pi}-\vec{a}\pi^{0} +\vec{a}\times\vec{\pi}%
)\cdot\vec{\tau}D\} +\frac{N_{c}}{24I_{2}}D^{\dag}(a^{0}\vec{\pi}-\vec{a}%
\pi^{0} +\vec{a}\times\vec{\pi})\cdot\vec{\tau}D  \nonumber \\
& &+\cdots
\end{eqnarray}
where the ellipsis stands for the strange-strange interaction terms of order
$1/N_{c}$ which can be readily read off from Eq. (\ref{hamilqual}).

Following the Klebanov and Westerberg's quantization scheme \cite{kleb94} for 
the strangeness flavor direction one can obtain the Hamiltonian of the form
\begin{equation}
\tilde{H}=M_{0}^{Q}+\nu_{K}  a^{\dag}a +\frac{1}
{2I_{1}} \left(\vec{I}^{2}+2c\vec{I}\cdot\vec{J}_{s}+\bar{c}\vec{J}_{s}^{2}
+\frac{1}{4}\right)  \label{hjsqual}
\end{equation}
where $\vec{I}$ and $\vec{J}_{s}$ are the isospin and angular momentum for
the strange quarks and
$$
\nu_{K}=\frac{N_{c}}{24I_{2}}(\mu_{Q} -1),~~
c=1-\frac{I_{1}}{2I_{2}\mu_{Q}}(\mu_{Q} -1),~~
\bar{c}=1-\frac{I_{1}}{I_{2}\mu_{Q}^{2}}(\mu_{Q} -1)  
$$
with
$$
\mu_{Q}=\left(1+\frac{m_{K}^{2}}{m_{Q}^{2}}\right)^{1/2},~~~~~
m_{Q}=\frac{N_{c}}{24I_{2}}.
$$
Here note that $a^{\dag}$ is creation operator for constituent strange quarks 
and the factor $\frac{1}{4}$ originates from BFT corrections
\cite{hong991}, which are applicable to only u- and d-superqualitons.  The
Hamiltonian (\ref{hjsqual}) then yields the mass spectrum of superqualiton
as follows~\cite{hong01plb}
\begin{eqnarray}
M_{Q}&=&M_{0}^{Q}-(Y-\frac{1}{3})\nu_{K}+\frac{1}{2I_{1}}\left[ cJ(J+1)
+(1-c)I(I+1)\right.\nonumber\\
& &\left.+(\bar{c}-c)\frac{(Y-1/3)(Y-7/3)}{4}+\frac{1}{4}\delta_{I,1/2}\right]
\end{eqnarray}
with the total angular momentum of the quark $\vec{J}=\vec{I}+\vec{J}_{s}$.

Unlike creating Skyrmions out of Dirac vacuum, in dense matter
the energy cost to create a superqualiton should be compared
with the Fermi Sea. By creating a superqualiton, we have to remove
one quark in the Fermi sea  since the total baryon number
has to remain unchanged.  Similar to Cooper pair mechanism \cite{cooper56},
from Eq. (\ref{deltae}),
the twice of u- and s-superqualiton masses are then given by
\begin{eqnarray}
2M_{u}&=&M_{0}^{Q}+\frac{1}{2I_{1}}-\omega\nonumber\\
2M_{s}&=&M_{0}^{Q}+\nu_{K}+\frac{3}{8I_{1}}\bar{c}-\omega
\end{eqnarray}
to yield the predictions for the values of $M_{u}(=M_{d})$ and $M_{s}$
\begin{equation}
\begin{array}{lll}
M_{u}=0.079\times 4\pi F, &~~~M_{s}=0.081\times 4\pi F, &~~~{\rm for}~m_{K}/F=0.1\\
M_{u}=0.079\times 4\pi F, &~~~M_{s}=0.089\times 4\pi F, &~~~{\rm for}~m_{K}/F=0.3\\
M_{u}=0.079\times 4\pi F, &~~~M_{s}=0.109\times 4\pi F, &~~~{\rm for}~m_{K}/F=0.8,
\end{array}
\end{equation}
which are comparable to the Cooper gap~\cite{hong01plb,rajagopal00hep}.  

\begin{table}[t]
\caption{The dependence of superqualiton masses on the coupling $\alpha_{s}$ 
with $m_{K}/F=0.3$}
\begin{center}
\begin{tabular}{crrrr}
\hline
$\alpha_{s}$  &$M_{Q}(u)/4\pi F$  &$M_{Q}(s)/4\pi F$
              &$M_{u}/4\pi F$     &$M_{s}/4\pi F$\\
\hline
  0.050    &1.040    &1.061    &0.078    &0.089\\
  0.100    &1.040    &1.061    &0.078    &0.089\\
  0.150    &1.041    &1.061    &0.079    &0.089\\
  0.200    &1.041    &1.061    &0.079    &0.089\\
  0.250    &1.041    &1.061    &0.079    &0.089\\
  0.300    &1.041    &1.062    &0.079    &0.089\\
  0.350    &1.041    &1.062    &0.079    &0.089\\
  0.400    &1.042    &1.062    &0.079    &0.089\\
  0.450    &1.042    &1.062    &0.079    &0.089\\
  0.500    &1.042    &1.062    &0.079    &0.089\\
  0.550    &1.042    &1.062    &0.079    &0.089\\
  0.600    &1.042    &1.062    &0.079    &0.089\\
  0.650    &1.042    &1.062    &0.079    &0.090\\
  0.700    &1.042    &1.063    &0.079    &0.090\\
  0.750    &1.042    &1.062    &0.079    &0.090\\
  0.800    &1.042    &1.063    &0.079    &0.090\\
  0.850    &1.042    &1.063    &0.079    &0.090\\
  0.900    &1.042    &1.063    &0.079    &0.090\\
  0.950    &1.042    &1.063    &0.080    &0.090\\
  1.000    &1.043    &1.063    &0.080    &0.090\\
\hline
\end{tabular}
\end{center}
\label{qualiton}
\end{table}
To see if the estimated superqualiton mass is indeed the Cooper gap,
one needs to compare our numerical results with the analytic expression
for the coupling dependence of the gap.
In Table~\ref{qualiton} we show the dependence of superqualiton
masses on the strong coupling constant $\alpha_{s}$.
By fitting the numerical results with the gap as, in the unit of
$4\pi F$,
\begin{equation}
\log(M_u)=a \log (\alpha_s)+b\alpha_s^{-1/2}+c.
\end{equation}
We get $a=0.00085$, $b=-0.00233$, and $c=0.1193$.
This is very different from the analytic expression obtained in
the literature,
\begin{equation}
\Delta\sim{\mu\over g_s^5}
\exp\left(-{3\pi^2\over\sqrt{2}g_s}\right).
\label{weak}
\end{equation}
As suggested in Ref.~\cite{rajagopal00rs}, the weak coupling
result, Eq.~(\ref{weak}), is applicable only when the coupling is
extreme small or the chemical potential is very large.
In our numerical analysis, we are unable to probe this region.

\vskip 0.5cm 
\noindent 
{\Large {\bf Acknowledgments}}
\vskip 0.5cm 
\noindent 
STH would like to deeply thank R.D. McKeown for the warm hospitality 
at the Kellogg Radiation Laboratory, Caltech where a part of this 
work has been done.  The authors are grateful to G.E. Brown, D.K. 
Hong, W.T. Kim, Y.W. Kim, B.H. Lee, H.K. Lee, S.H. Lee, R.D. McKeown, C.M. 
Maekawa, D.P. Min, B.Y. Park and M. Rho for helpful discussions.  
The work of STH was supported by Grant No. 2000-2-11100-002-5 from 
the Basic Research Program of the Korea Science Engineering Foundation 
and that of Y.J. Park was supported by the Korea Research Foundation 
Grant No. KRF-2000-015-DP0070.      

\newpage
\begin{appendix}


\section{Spin symmetries in the SU(3) group}
\setcounter{equation}{0}
\renewcommand{\theequation}{A.\arabic{equation}}


In order to discuss the I-, U- and V-spin symmetries associated with the 
SU(3) group, we will briefly review the root diagram approach to the 
construction of the Lie algebra of the SU(3) group which has eight 
generators.

Since the rank of the SU(3) group is two, one can have the Cartan 
subalgebra \cite{georgi82, wybourne74}, the set of two commuting generators 
$H_{i}$ ($i=1,2$) corresponding to $\lambda_{3}$ and $\lambda_{8}$
\beq
[ H_{1},~H_{2}]=0,
\label{h1h2}
\eeq
and the other generators $E_{\alpha}$ ($\alpha =\pm 1,\pm 2,\pm 3$) satisfying 
the commutator relations
\begin{eqnarray}
\left[ H_{i},~E_{\alpha}\right]&=&e^{\alpha}_{i}E_{\alpha}\nonumber\\
\left[ E_{\alpha},~E_{\beta}\right]&=&N_{\alpha\beta}E_{\alpha+\beta}\nonumber\\
\left[ E_{\alpha},~E_{-\alpha}\right]&=&e^{\alpha}_{i}H_{i}
\label{commutatorshe}
\end{eqnarray}
where $e^{\alpha}_{i}$ is the $i$-th component of the root vector 
$\hat{e}^{\alpha}$ in a two dimensional root space and $N_{\alpha\beta}$ is 
a normalization constant to be fixed.

Normalizing the root vectors such that $\sum_{\alpha}e_{i}^{\alpha}e_{j}
^{\alpha}=\delta_{ij}$, one can choose the root vectors
\bea
\hat{e}^{1}&=&-\hat{e}^{-1}=\left(\frac{1}{\sqrt{3}},0\right)\nonumber\\
\hat{e}^{2}&=&-\hat{e}^{-2}=\left(\frac{1}{2\sqrt{3}},\frac12\right)\nonumber\\
\hat{e}^{3}&=&-\hat{e}^{-3}=\left(-\frac{1}{2\sqrt{3}},\frac12\right)
\label{rootvectors}
\eea
as illustrated in Figure~\ref{root} where one has two simple roots $\hat{e}^{2}$ and 
$\hat{e}^{-3}$ of the equal length separated by an angle $\frac{2\pi}{3}$ so 
that one can obtain the Dynkin diagram \cite{georgi82, wybourne74} for the 
su(3) Lie algebra given by $\circ\hbox{\hskip -0.5em}-\hbox{\hskip -1.0em}-
\hbox{\hskip -0.5em}\circ$.

\begin{figure}
\setlength{\unitlength}{1.0cm}
\begin{center}
\begin{picture}(8,5)(-4,-2.5)
\put(4.3,0){$H_{1}$}
\put(0,3.3){$H_{2}$}
\put(3.3,0.35){$\hat{e}^{1}$}
\put(-3.3,0.35){$\hat{e}^{-1}$}
\put(1.8,2.5){$\hat{e}^{2}$}
\put(-1.85,2.5){$\hat{e}^{3}$}
\put(1.8,-2.7){$\hat{e}^{-3}$}
\put(-1.85,-2.7){$\hat{e}^{-2}$}
\thinlines
\put(-4.0,0.0){\vector(1,0){8.0}}
\put(0.0,-3.0){\vector(0,1){6.0}}
\thicklines
\put(0.0,0.0){\vector(1,0){3.0}}
\put(0.0,0.0){\vector(-1,0){3.0}}
\put(0.0,0.0){\vector(2,3){1.5}}
\put(0.0,0.0){\vector(-2,3){1.5}}
\put(0.0,0.0){\vector(2,-3){1.5}}
\put(0.0,0.0){\vector(-2,-3){1.5}}
\end{picture}\end{center}
\caption{Root diagram for SU(3) group.  The simple root vectors $\hat{e}^{2}$ 
and $\hat{e}^{-3}$ can produce all the other root vectors through the 
operations of addition and $\hat{e}^{\alpha}=-\hat{e}^{-\alpha}$.}
\label{root}
\end{figure}
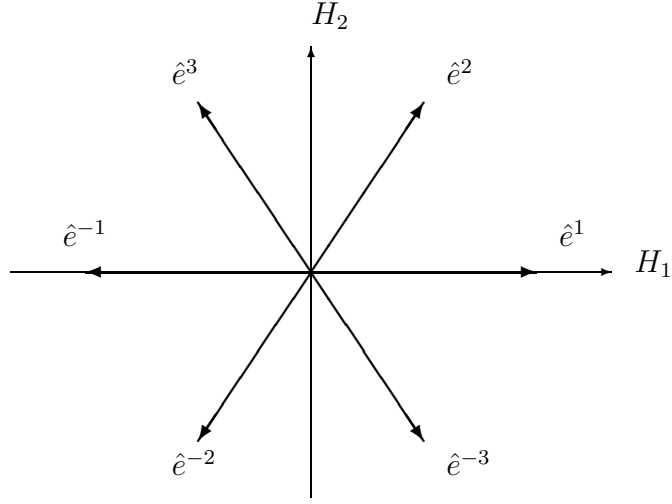

Substituting the root vectors in Figure~\ref{root} normalized as in 
(\ref{rootvectors}) into the relations (\ref{h1h2}) and (\ref{commutatorshe}) 
one can easily derive the commutator relations
\beq
\begin{array}{ll}
\left[H_{1},~H_{1}\right]=0,
&\left[H_{1},~E_{1}\right]=\frac{1}{\sqrt{3}}E_{1},\\
\left[H_{1},~E_{2}\right]=\frac{1}{2\sqrt{3}}E_{2},
&\left[H_{1},~E_{3}\right]=-\frac{1}{2\sqrt{3}}E_{3},\\
\left[H_{2},~E_{1}\right]=0,
&\left[H_{2},~E_{2}\right]=\frac{1}{2}E_{2},\\
\left[H_{2},~E_{3}\right]=\frac12 E_{3},
&\left[E_{1},~E_{-1}\right]=\frac{1}{\sqrt{3}}H_{1},\\
\left[E_{2},~E_{-2}\right]=\frac{1}{2\sqrt{3}}H_{1}+\frac12 H_{2},
&\left[E_{3},~E_{-3}\right]=-\frac{1}{2\sqrt{3}}H_{1}+\frac12 H_{2},\\
\left[E_{1},~E_{3}\right]=\frac{1}{\sqrt{6}}E_{2},
&\left[E_{2},~E_{-3}\right]=-\frac{1}{2\sqrt{3}}H_{1}+\frac12 H_{2},\\
\left[E_{1},~E_{3}\right]=\frac{1}{\sqrt{6}}E_{2},
&\left[E_{2},~E_{-3}\right]=\frac{1}{\sqrt{6}}E_{1},\\
\left[E_{-1},~E_{2}\right]=\frac{1}{\sqrt{6}}E_{3}.
&{} 
\end{array}
\label{hhhes}
\eeq

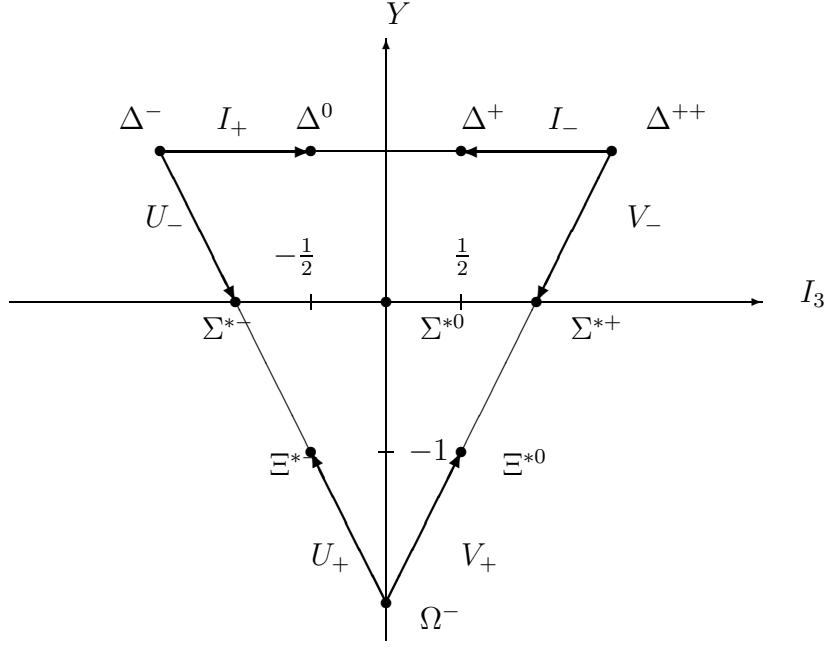
\begin{figure}
\setlength{\unitlength}{1.0cm}
\begin{center}
\begin{picture}(10,7)(-5,-4)
\thinlines
\put(5.5,0){$I_{3}$}
\put(0,3.7){$Y$}
\put(3,2){\circle*{0.15}}
\put(3.45,2.3){$\Delta^{++}$}
\put(-3,2){\circle*{0.15}}
\put(-3.55,2.3){$\Delta^{-}$}
\put(-1,2){\circle*{0.15}}
\put(-1.20,2.3){$\Delta^{0}$}
\put(1,2){\circle*{0.15}}
\put(1,2.3){$\Delta^{+}$}
\put(1,-2){\circle*{0.15}}
\put(1.55,-2.3){$\Xi^{*0}$}
\put(-1,-2){\circle*{0.15}}
\put(-1.55,-2.3){$\Xi^{*-}$}
\put(2,0){\circle*{0.15}}
\put(2.45,-0.5){$\Sigma^{*+}$}
\put(0,0){\circle*{0.15}}
\put(0.45,-0.5){$\Sigma^{*0}$}
\put(-2,0){\circle*{0.15}}
\put(-2.45,-0.5){$\Sigma^{*-}$}
\put(0,-4){\circle*{0.15}}
\put(0.45,-4.35){$\Omega^{-}$}
\put(-5,0){\vector(1,0){10.0}}
\put(0.0,-4.5){\vector(0,1){8.0}}
\put(-3,2){\line(1,0){6}}
\put(-3,2){\line(1,-2){3}}
\put(3,2){\line(-1,-2){3}}
\put(0.9,0.5){$\frac{1}{2}$}
\put(1,-0.1){\line(0,1){0.2}}
\put(-1.5,0.5){$-\frac{1}{2}$}
\put(-1,-0.1){\line(0,1){0.2}}
\put(0.3,-2.1){$-1$}
\put(-0.1,-2){\line(1,0){0.2}}
\put(-2.25,2.3){$I_{+}$}
\put(2.15,2.3){$I_{-}$}
\put(3.2,1){$V_{-}$}
\put(-3.2,1){$U_{-}$}
\put(1,-3.5){$V_{+}$}
\put(-1,-3.5){$U_{+}$}
\thicklines
\put(-3,2){\vector(1,0){2}}
\put(-3,2){\vector(1,-2){1}}
\put(3,2){\vector(-1,0){2}}
\put(3,2){\vector(-1,-2){1}}
\put(0,-4){\vector(1,2){1}}
\put(0,-4){\vector(-1,2){1}}
\end{picture}\end{center}
\caption{Spin symmetry operations in the baryon decuplet.}
\label{spin}
\end{figure}

Associating the root vectors $H_{i}$ ($i=1,2$) and $E_{\alpha}$ ($\alpha=
\pm1,\pm 2,\pm 3$) with the physical operators $Y$, $I_{3}$, $I_{\pm}$, 
$U_{\pm}$ and $V_{\pm}$ through the definitions
\bea
H_{1}&=&\frac{1}{\sqrt{3}}I_{3},~~~~~H_{2}=\frac12 Y\nonumber\\
E_{\pm}&=&\frac{1}{\sqrt{6}}I_{\pm},~~~~E_{\pm 3}=\frac{1}{\sqrt{6}}U_{\pm}
\nonumber\\
E_{\pm 2}&=&\frac{1}{\sqrt{6}}V_{\pm}
\label{h1i3}
\eea
one can use the commutator relations (\ref{hhhes}) to yield the explicit 
expressions for the eigenvalue equations of the spin operators in the SU(3) 
group \cite{deswart63}
\bea
I_{+}|Y,I,I_{3}\rangle&=&((I-I_{3})(I+I_{3}+1))^{\frac12}|Y,I,I_{3}+1\rangle
\nonumber\\
U_{+}|Y,I,I_{3}\rangle&=&(a_{+}(I-I_{3}+1))^{\frac12}|Y+1,I+\frac12,
I_{3}-\frac12\rangle \nonumber\\
& &-(a_{-}(I+I_{3}))^{\frac12}|Y+1,I-\frac12, I_{3}-\frac12\rangle \nonumber\\
V_{+}|Y,I,I_{3}\rangle&=&(a_{+}(I+I_{3}+1))^{\frac12}|Y+1,I+\frac12,
I_{3}+\frac12\rangle \nonumber\\
& &+(a_{-}(I-I_{3}))^{\frac12}|Y+1,I-\frac12, I_{3}+\frac12\rangle 
\label{iuvplus}
\eea
where the de Swart phase convention~\cite{deswart63} is used and 
\bea
a_{+}&=&\frac{(Y_{+}+\frac{1}{3}(p-q)+1)
(Y_{+}+\frac{1}{3}(p+2q)+2)(-Y_{+}+\frac{1}{3}(2p+q))}{(2I+1)(2I+2)}
\nonumber\\
a_{-}&=&\frac{(Y_{-}+\frac{1}{3}(p-q))
(Y_{-}+\frac{1}{3}(p+2q)+1)(Y_{-}-\frac{1}{3}(2p+q)-1)}{2I(2I+1)}
\label{applusaminus}
\eea
with $Y_{\pm}=\frac12 Y\pm I$.  Here $p$ and $q$ are nonnegative coefficients 
needed to construct bases for the IR $D(p,q)$ of SU(3) group.  The dimension 
${\bf n}$ of $D(p,q)$, namely the number of the basis vectors is then given by 
$(1+p)(1+q)(1+\frac12 (p+q))$ \cite{deswart63} so that one can denote the IRs 
of interest as follows,
\beq
\begin{array}{lll}
{\bf 1}=D(0,0),&{\bf 3}=D(1,0),&\bar{\bf 3}=D(0,1),\\ 
{\bf 81}=D(1,1),&{\bf 10}=D(3,0),&\bar{\bf 10}=D(0,3),\\ 
\bar{\bf 27}=D(2,2),&{\bf 35}=D(4,1),&\bar{\bf 35}=D(1,4),\\ 
{\bf 28}=D(6,0),&{\bf 64}=D(3,3),&\bar{\bf 81}=D(5,2),\\ 
\bar{\bf 81}=D(2,5). &{}  &{}
\end{array}
\label{dimensions}
\eeq

Using the relations for the raising spin operators (\ref{iuvplus}) and the 
similarly constructed relations for the lowing spin operators $I_{-}$, 
$U_{-}$ and $V_{-}$ one can derive the isoscalar factors \cite{deswart63} of 
the SU(3) group for the Clebsch-Gordan series which have been used in the 
previous sections.  In Figure~\ref{spin} is depicted the spin symmetry operation 
diagram for the decuplet baryons.


\section{Inertia parameters in the chiral bag model}
\setcounter{equation}{0}
\renewcommand{\theequation}{B.\arabic{equation}}


\subsection{Angular part of the matrix element}


In this section, we will derive the explicit expression of the quark phase 
inertia parameters in the CBM, which are to some extent abstractly described 
in the above sections, by considering one of the most complicated quantity 
${\cal N}^{\prime}$ whose meson phase contribution is already explicitly 
given in the previous section.  (For the other inertia parameters, see the 
Refs. \cite{hong932, park88, park90, hong92}.)  To obtain the explicit 
description of the quark phase inertia parameter ${\cal N}^{\prime}_{q}$, we 
will first calculate the angular part of the matrix element ${}_{h}\langle 
m|\lambda_{4}V_{3}|n\rangle_{s}$ in this section.

Now one notes that the vector operator $V_{i}=\epsilon_{ijk}x_{j}\gamma^{0}
\gamma^{k}$ can be given in terms of $v_{i}=\epsilon_{ijk}\hat{r}_{j}
\sigma_{k}$ as follows
\beq
V_{i}=\left(
\begin{array}{cc}
0 & rv_{i}\\
rv_{i} & 0\\
\end{array}
\right)
\label{vimatrix}
\eeq
where the unit vectors $\hat{r}_{i}$ ($i=1,2,3$) can be expressed in terms of 
the spherical harmonics $Y_{l,m}(\theta,\phi )$
\bea
\hat{r}_{1}&=&\sin\theta\cos\phi =\left(\frac{2\pi}{3}\right)^{\frac12}
              (Y_{1,-1}-Y_{1,1})\nonumber\\
\hat{r}_{2}&=&\sin\theta\sin\phi =i\left(\frac{2\pi}{3}\right)^{\frac12}
              (Y_{1,-1}+Y_{1,1})\nonumber\\
\hat{r}_{1}&=&\cos\theta =\left(\frac{4\pi}{3}\right)^{\frac12}
              Y_{1,0}.
\label{harmonics}
\eea

Acting the unit vector operators on the eigenstate of the angular momentum 
$|j,m_{j}\rangle$, one can obtain the identities
\bea
\hat{r}_{1,2}|j,m_{j}\rangle&=&\left(\frac{(j-m_{j}+1)(j-m_{j}+2)}
          {4(2j+1)(2j+3)}\right)^{\frac12}|j+1,m_{j}-1\rangle
          \nonumber\\
& &\mp\left(\frac{(j+m_{j}+1)(j+m_{j}+2)}
          {4(2j+1)(2j+3)}\right)^{\frac12}|j+1,m_{j}+1\rangle
          \nonumber\\
& &-\left(\frac{(j+m_{j}-1)(j+m_{j})}
          {4(2j-1)(2j+1)}\right)^{\frac12}|j-1,m_{j}-1\rangle
          \nonumber\\
& &\pm\left(\frac{(j-m_{j}-1)(j-m_{j})}
          {4(2j-1)(2j+1)}\right)^{\frac12}|j-1,m_{j}+1\rangle
          \nonumber\\
\hat{r}_{3}|j,m_{j}\rangle&=&\left(\frac{(j-m_{j}+1)(j+m_{j}+1)}
          {(2j+1)(2j+3)}\right)^{\frac12}|j+1,m_{j}\rangle
          \nonumber\\
& &+\left(\frac{(j-m_{j})(j+m_{j})}
          {(2j-1)(2j+1)}\right)^{\frac12}|j-1,m_{j}\rangle.
\label{r12jmj}
\eea

Now the angular parts of the s-quark eigenstates with $\kappa^{\prime}=
\pm 1$ corresponding to $j=l\pm \frac12$ are given in terms of the quantum 
number $j$ and $m_{j}$ and spin states $|\uparrow\rangle$ and $|\downarrow
\rangle$
\bea
|j,m_{j}\rangle_{+1}&=&\left(\frac{j+m_{j}}{2j}\right)^{\frac12}
    |j-\frac12,m_{j}-\frac12\rangle|\uparrow\rangle \nonumber\\
& &+\left(\frac{j-m_{j}}{2j}\right)^{\frac12}
    |j-\frac12,m_{j}+\frac12\rangle|\downarrow\rangle \nonumber\\
|j,m_{j}\rangle_{-1}&=&-\left(\frac{j-m_{j}+1}{2j+2}\right)^{\frac12}
     |j+\frac12,m_{j}-\frac12\rangle|\uparrow\rangle\nonumber\\
& &+\left(\frac{j+m_{j}+1}{2j+2}\right)^{\frac12}
     |j+\frac12,m_{j}+\frac12\rangle|\downarrow\rangle
\label{jmjplus1}
\eea
which satisfy the relations
\bea
{}_{i^{\prime}}\langle j^{\prime},m_{j^{\prime}}|j,m_{j}\rangle_{i}&=&
\delta_{i}^{i^{\prime}}\delta_{j}^{j^{\prime}}\delta_{m_{j}}^{m_{j^{\prime}}}
\nonumber\\
\vec{\sigma}\cdot\hat{r}_{i}|j,m_{j}\rangle_{+1}&=&-|j,m_{j}\rangle_{-1}.
\label{iprime}
\eea

Applying the identities (\ref{harmonics}) and (\ref{r12jmj}) to the s-quark 
eigenstate angular parts (\ref{jmjplus1}) one can evaluate the following 
relations
\bea
v_{3}|j,m_{j}\rangle_{+1}&=&-i\left(\frac{j-m_{j}+1}{2j+2}\right)^{\frac12}
  \frac{j-m_{j}}{2j}|j+\frac12,m_{j}-\frac12\rangle|\uparrow\rangle \nonumber\\
& &-i\left(\frac{j+m_{j}+1}{2j+2}\right)^{\frac12}
  \frac{j+m_{j}}{2j}|j+\frac12,m_{j}+\frac12\rangle|\downarrow\rangle\nonumber\\
v_{3}|j,m_{j}\rangle_{-1}&=&i\left(\frac{j+m_{j}}{2j}\right)^{\frac12}
  \frac{j+m_{j}+1}{2j+2}|j-+\frac12,m_{j}-\frac12\rangle|\uparrow\rangle 
\nonumber\\
& &-i\left(\frac{j-m_{j}}{2j}\right)^{\frac12}
  \frac{j-m_{j}+1}{2j+2}|j-\frac12,m_{j}+\frac12\rangle|\downarrow\rangle
\label{v3v3}
\eea
which are crucial in the following calculation of the angular part of the 
matrix element involved in the inertia parameter in the quark phase
\bea
{}_{1}\langle K,m_{K}|\lambda_{4}v_{3}|j,m_{j}\rangle_{+1}|s\rangle&=&
i\left(\frac{K+m_{K}}{2K+2}\right)^{\frac12}\frac{K-m_{K}+1}{2K+1}
\delta_{j+\frac12}^{K}\delta_{m_{j}+\frac12}^{m_{K}}
\nonumber\\
{}_{2}\langle K,m_{K}|\lambda_{4}v_{3}|j,m_{j}\rangle_{+1}|s\rangle&=&
{}_{4}\langle K,m_{K}|\lambda_{4}v_{3}|j,m_{j}\rangle_{-1}|s\rangle
\nonumber\\
& &
\hskip -3.0cm
=i\left(\frac{K+m_{K}}{2K}\right)^{\frac12}\frac{2K(2m_{K}-1)}{(2K-1)(2K+1)}
\delta_{j+\frac12}^{K}\delta_{m_{j}+\frac12}^{m_{K}}
\nonumber\\
{}_{3}\langle K,m_{K}|\lambda_{4}v_{3}|j,m_{j}\rangle_{+1}|s\rangle&=&
-{}_{1}\langle K,m_{K}|\lambda_{4}v_{3}|j,m_{j}\rangle_{-1}|s\rangle
\nonumber\\
& &
\hskip -3.0cm
=i\left(\frac{K-m_{K}+1}{2K+2}\right)^{\frac12}\frac{(2K+2)(2m_{K}-1)}
{(2K+1)(2K+3)}\delta_{j-\frac12}^{K}\delta_{m_{j}+\frac12}^{m_{K}}
\nonumber\\
{}_{4}\langle K,m_{K}|\lambda_{4}v_{3}|j,m_{j}\rangle_{+1}|s\rangle&=&
{}_{3}\langle K,m_{K}|\lambda_{4}v_{3}|j,m_{j}\rangle_{-1}|s\rangle =0
\nonumber\\
{}_{2}\langle K,m_{K}|\lambda_{4}v_{3}|j,m_{j}\rangle_{-1}|s\rangle&=&
i\left(\frac{K-m_{K}+1}{2K}\right)^{\frac12}\frac{K+m_{K}}
{2K+1}\delta_{j-\frac12}^{K}\delta_{m_{j}+\frac12}^{m_{K}}
\nonumber\\
\label{kmklambda}
\eea
with $|s\rangle=(0,0,1)^{T}$ the s-quark eigenstate in the SU(3) flavor space.  
Here one can easily see that the angular parts of the hedgehog quark 
eigenstates, constructed with $|j,m_{j}\rangle_{\pm 1}$ and the isospin 
eigenstates $|\Uparrow\rangle=(1,0,0)^{T}$ and $|\Downarrow\rangle=
(0,1,0)^{T}$, are given by
\bea
|K,m_{K}\rangle_{1}&=&-\left(\frac{K-m_{K}+1}{2K+2}\right)^{\frac12}
  |K+\frac12, m_{K}-\frac12\rangle_{+1}|\Uparrow\rangle \nonumber\\
& &-\left(\frac{K+m_{K}+1}{2K+2}\right)^{\frac12}
  |K+\frac12, m_{K}+\frac12\rangle_{+1}|\Downarrow\rangle \nonumber\\
|K,m_{K}\rangle_{2}&=&-\left(\frac{K+m_{K}}{2K}\right)^{\frac12}
  |K-\frac12, m_{K}-\frac12\rangle_{-1}|\Uparrow\rangle \nonumber\\
& &-\left(\frac{K-m_{K}}{2K}\right)^{\frac12}
  |K-\frac12, m_{K}+\frac12\rangle_{-1}|\Downarrow\rangle \nonumber\\
|K,m_{K}\rangle_{3}&=&-\left(\frac{K-m_{K}+1}{2K+2}\right)^{\frac12}
  |K+\frac12, m_{K}+\frac12\rangle_{-1}|\Uparrow\rangle \nonumber\\
& &+\left(\frac{K+m_{K}+1}{2K+2}\right)^{\frac12}
  |K+\frac12, m_{K}+\frac12\rangle_{-1}|\Downarrow\rangle \nonumber\\
|K,m_{K}\rangle_{4}&=&\left(\frac{K+m_{K}}{2K}\right)^{\frac12}
  |K-\frac12, m_{K}-\frac12\rangle_{+1}|\Uparrow\rangle \nonumber\\
& &+\left(\frac{K-m_{K}}{2K}\right)^{\frac12}
  |K-\frac12, m_{K}+\frac12\rangle_{+1}|\Downarrow\rangle
\label{kmk1234}
\eea
which fulfill the relations
\bea
{}_{i^{\prime}}\langle K^{\prime},m_{K^{\prime}}|K,m_{K}\rangle_{i}&=&
\delta_{i}^{i^{\prime}}\delta_{j}^{j^{\prime}}\delta_{m_{K}}^{m_{K^{\prime}}}
\nonumber\\
\vec{\sigma}\cdot\hat{r}_{i}|K,m_{K}\rangle_{i}&=&(-1)^{i}|K,m_{K}\rangle_{i+2}.
\label{iprimek}
\eea
Here one notes that $|K,m_{K}\rangle_{1}$ and $|K,m_{K}\rangle_{2}$ ($|K,m_{K}
\rangle_{3}$ and $|K,m_{K}\rangle_{4}$) have the quantum number $\kappa=+1$ 
($\kappa=-1$) where $\kappa=P(-1)^{K}$. 


\subsection{Quark phase inertia parameter}


In this section we will combine the angular part of the matrix element 
derived in the previous section with the radial wavefunctions of the quark 
eigenstates so that one can calculate the quark phase inertia parameter 
${\cal N}^{\prime}_{q}$.

Now the unperturbed hedgehog and strange quark eigenstates in terms of the 
quantum numbers $\kappa$ and $\kappa^{\prime}$ are described as follows
\bea
\psi_{m}^{0h}&=&c_{1}n_{1}\left(
\begin{array}{c}
j_{K}(\varepsilon_{m}r)\\
i\vec{\sigma}\cdot\hat{r}j_{K+1}(\varepsilon_{m}r)\\
\end{array}
\right)
|K,m_{K}\rangle_{1}
\nonumber\\
& &-c_{2}n_{2}\left(
\begin{array}{c}
-j_{K}(\varepsilon_{m}r)\\
i\vec{\sigma}\cdot\hat{r}j_{K-1}(\varepsilon_{m}r)\\
\end{array}
\right)
|K,m_{K}\rangle_{2}
~~~~~~~~~~~~~~~~~{\rm for~\kappa=+1}
\nonumber\\
\psi_{m}^{0h}&=&-c_{1}n_{1}\left(
\begin{array}{c}
-j_{K+1}(\varepsilon_{m}r)\\
i\vec{\sigma}\cdot\hat{r}j_{K}(\varepsilon_{m}r)\\
\end{array}
\right)
|K,m_{K}\rangle_{3}
\nonumber\\
& &+c_{2}n_{2}\left(
\begin{array}{c}
j_{K-1}(\varepsilon_{m}r)\\
i\vec{\sigma}\cdot\hat{r}j_{K}(\varepsilon_{m}r)\\
\end{array}
\right)
|K,m_{K}\rangle_{4}
~~~~~~~~~~~~~~~~~~~~{\rm for~\kappa=-1}
\nonumber\\
\psi_{n}^{0s}&=&n_{1}^{\prime}\left(
\begin{array}{c}
j_{l}(\omega_{n}r)\\
i\vec{\sigma}\cdot\hat{r}j_{l+1}(\omega_{n}r)\\
\end{array}
\right)
|j,m_{j}\rangle_{+1}|s\rangle
\nonumber
~~~~~~~~~~~~~~~~~~~~~{\rm for~\kappa^{\prime}=+1}
\nonumber\\
\psi_{n}^{0s}&=&-n_{2}^{\prime}\left(
\begin{array}{c}
-j_{l}(\omega_{n}r)\\
i\vec{\sigma}\cdot\hat{r}j_{l-1}(\omega_{n}r)\\
\end{array}
\right)
|j,m_{j}\rangle_{-1}|s\rangle
~~~~~~~~~~~~~~~~~~~{\rm for~\kappa^{\prime}=-1}
\nonumber\\
\label{psim0hn}
\eea
where $j_{K}(\varepsilon_{m}r)$ and $j_{l}(\omega_{n}r)$ are the spherical 
Bessel functions with the energy eigenvalues $\varepsilon_{m}$ and 
$\omega_{n}$, respectively, and the constants $c_{1}$ and $c_{2}$ are the 
normalization constants satisfying $c_{1}^{2}+c_{2}^{2}=1$ and the constants 
$n_{1}$ and $n_{2}$ are normalized as 
\bea
n_{1}^{-2}R^{-3}E_{m}&=&E_{m}(j_{K}^{2}(E_{m})+j_{K+1}^{2}(E_{m}))
-2(K+1) j_{K}(E_{m}) j_{K+1}(E_{m})\nonumber\\
n_{2}^{-2}R^{-3}E_{m}&=&E_{m}(j_{K}^{2}(E_{m})+j_{K-1}^{2}(E_{m}))
-2Kj_{K}(E_{m})j_{K-1}(E_{m})
\label{n1minus2}
\eea
and the other constants $n_{1}^{\prime}$ and $n_{2}^{\prime}$ also satisfy the 
above conditions with $(\Omega_{n}=\omega_{n}R,l)$ instead of $(E_{m}=
\varepsilon_{m}R,K)$.

Using the angular parts of the matrix elements (\ref{kmklambda}) and the full 
quark eigenstate wavefunctions (\ref{psim0hn}), one can now calculate the 
matrix element ${}_{h}\langle m|\lambda_{4}V_{3}|n\rangle_{s}$ as below
\bea
{}_{h}\langle m|\lambda_{4}V_{3}|n\rangle_{s}&=&
\eta\left(\frac{K-m_{K}+1}{2K+2}\right)^{\frac12}\{-c_{1}N_{1}N_{1}^{\prime}
\frac{(2K+2)(2m_{K}-1)}{(2K+1)(2K+3)}(\iota_{1}+\iota_{2})
\nonumber\\
& &+c_{2}N_{2}N_{1}^{\prime}\frac{K+m_{K}}{2K+1}\left(\frac{K+1}{K}
\right)^{\frac12}\iota_{1}\}
\delta_{j-\frac12}^{K}\delta_{m_{j}+\frac12}^{m_{K}}
\nonumber\\ 
& &+\eta\left(\frac{K+m_{K}+1}{2K}\right)^{\frac12}\{c_{1}N_{1}N_{2}^{\prime}
\frac{K-m_{K}+1}{2K+1}\left(\frac{K}{K+1}\right)^{\frac12}\iota_{3}
\nonumber\\
& &+c_{2}N_{2}N_{2}^{\prime}\frac{2K(2m_{K}-1)}{(2K-1)(2K+1)}
(\iota_{3}-\iota_{4})\}
\delta_{j+\frac12}^{K}\delta_{m_{j}+\frac12}^{m_{K}}
\nonumber\\ 
& &+\eta\left(\frac{K+m_{K}}{2K}\right)^{\frac12}\{c_{1}N_{1}N_{1}^{\prime}
\frac{K-m_{K}+1}{2K+1}\left(\frac{K}{K+1}\right)^{\frac12}\iota_{3}
\nonumber\\
& &+c_{2}N_{2}N_{1}^{\prime}\frac{2K(2m_{K}-1)}{(2K-1)(2K+1)}
(\iota_{3}+\iota_{4})\}
\delta_{j+\frac12}^{K}\delta_{m_{j}+\frac12}^{m_{K}}
\nonumber\\ 
& &+\eta\left(\frac{K-m_{K}+1}{2K+2}\right)^{\frac12}\{-c_{1}N_{1}N_{2}^{\prime}
\frac{(2K+2)(2m_{K}-1)}{(2K+1)(2K+3)}(\iota_{1}-\iota_{2})
\nonumber\\
& &+c_{2}N_{2}N_{2}^{\prime}\frac{K+m_{K}}{2K+1}\left(\frac{K+1}{K}
\right)^{\frac12}\iota_{1}\}
\delta_{j-\frac12}^{K}\delta_{m_{j}+\frac12}^{m_{K}}
\label{matrixelement4}
\eea
where $\eta=j_{K}(E_{m})j_{K}^{\prime}(E_{n})/|j_{K}(E_{m})j_{K}^{\prime}
(E_{n})|$, $N_{1}=R^{\frac{3}{2}}j_{K}(E_{n})n_{1}$ and 
$N_{2}=R^{\frac{3}{2}}j_{K}(E_{n})n_{2}$ and $N_{1}^{\prime}$ and 
$N_{2}^{\prime}$ are similarly defined for the strange quark eigenstates.  The 
radial integrations here are given as
\bea
\iota_{2}&=&\frac{\int_{0}^{R}{\rm d}r r^{3}j_{K+1}(\varepsilon_{m}r)
j_{K}(\omega_{n}r)}{R^{3}j_{K}(E_{m})j_{K}(\Omega_{n})}
\nonumber\\
\iota_{3}&=&\frac{\int_{0}^{R}{\rm d}r r^{3}j_{K}(\varepsilon_{m}r)
j_{K-1}(\omega_{n}r)}{R^{3}j_{K}(E_{m})j_{K-1}(\Omega_{n})}
\nonumber\\
\iota_{4}&=&\frac{\int_{0}^{R}{\rm d}r r^{3}j_{K-1}(\varepsilon_{m}r)
j_{K}(\omega_{n}r)}{R^{3}j_{K}(E_{m})j_{K}(\Omega_{n})}.
\label{iotas}
\eea
 
Similarly one can calculate the other matrix element 
${}_{h}\langle m|\lambda_{4}|n\rangle_{s}$ which is also needed for the 
inertia parameter ${\cal N}^{\prime}_{q}$
\bea
{}_{h}\langle m|\lambda_{4}| n\rangle_{s}&=&
\eta\left(\frac{K-m_{K}+1}{2K+2}\right)^{\frac12}c_{1}N_{1}N_{1}^{\prime}
\frac{1-v_{1}}{E_{m}-\Omega_{n}}
\delta_{j-\frac12}^{K}\delta_{m_{j}+\frac12}^{m_{K}}
\nonumber\\
& &+\eta\left(\frac{K+m_{K}}{2K}\right)^{\frac12}c_{2}N_{2}N_{2}^{\prime}
\frac{1+v_{2}}{E_{m}-\Omega_{n}}
\delta_{j+\frac12}^{K}\delta_{m_{j}+\frac12}^{m_{K}}
\nonumber\\
& &+\eta\left(\frac{K+m_{K}}{2K}\right)^{\frac12}c_{2}N_{2}N_{1}^{\prime}
\frac{1-v_{2}}{E_{m}-\Omega_{n}}
\delta_{j+\frac12}^{K}\delta_{m_{j}+\frac12}^{m_{K}}
\nonumber\\
& &+\eta\left(\frac{K-m_{K}+1}{2K+2}\right)^{\frac12}c_{1}N_{1}N_{2}^{\prime}
\frac{1+v_{1}}{E_{m}-\Omega_{n}}
\delta_{j-\frac12}^{K}\delta_{m_{j}+\frac12}^{m_{K}}
\label{matrixelement44}
\eea
where $v_{1}=j_{K+1}(E_{m})/j_{K}(E_{m})$ and $v_{2}=j_{K+1}(E_{m})/
j_{K}(E_{m})$.

Combining the above two matrix elements (\ref{matrixelement4}) and 
(\ref{matrixelement44}) one can obtain the explicit expression for the quark 
phase inertia parameter ${\cal N}^{\prime}_{q}$
\bea
& &\frac{1}{R}\sum_{m,n}\frac{
{}_{h}\langle m|\lambda_{4}|n\rangle_{s}
{}_{s}\langle n|\lambda_{4}V_{3}|m\rangle_{h}}
{\varepsilon_{m}-\omega_{n}} 
\nonumber\\
&=&\sum_{m,n,K}\frac{1-v_{1}}{(E_{m}-\Omega_{n})^{2}}c_{1}N_{1}N_{1}^{\prime}
\{K_{+}c_{2}N_{2}N_{1}^{\prime}\iota_{1}
+\frac{K+1}{3}c_{1}N_{1}N_{1}^{\prime}(\iota_{1}+\iota_{2})\}
\nonumber\\
& &+\sum_{m,n,K}\frac{1+v_{2}}{(E_{m}-\Omega_{n})^{2}}c_{2}N_{2}N_{2}^{\prime}
\{K_{-}c_{1}N_{1}N_{2}^{\prime}\iota_{3}
+\frac{K}{3}c_{2}N_{2}N_{2}^{\prime}(\iota_{3}-\iota_{4})\}
\nonumber\\
& &+\sum_{m,n,K}\frac{1-v_{2}}{(E_{m}-\Omega_{n})^{2}}c_{2}N_{2}N_{1}^{\prime}
\{K_{-}c_{1}N_{1}N_{2}^{\prime}\iota_{3}
+\frac{K}{3}c_{2}N_{2}N_{1}^{\prime}(\iota_{3}+\iota_{4})\}
\nonumber\\
& &+\sum_{m,n,K}\frac{1+v_{1}}{(E_{m}-\Omega_{n})^{2}}c_{1}N_{1}N_{2}^{\prime}
\{K_{+}c_{2}N_{2}N_{2}^{\prime}\iota_{1}
+\frac{K+1}{3}c_{1}N_{1}N_{2}^{\prime}(\iota_{1}-\iota_{2})\}
\nonumber\\
\label{1overr}
\eea
where $K_{+}$ and $K_{-}$ are defined as 
$$
K_{+}=\left(\frac{K+1}{K}\right)^{\frac12}\frac{K}{3},~~
K_{-}=\left(\frac{K}{K+1}\right)^{\frac12}\frac{K+1}{3}
$$
and the summation over the index $m_{K}$ has been carried out.  Here one 
notes that the summation indices $m$ and $n$ of the left hand side are 
understood as the shorthand of the sets of the quantum numbers $(K,m_{K},
\kappa,m)$ and $(j,m_{j},\kappa^{\prime},n)$ associated with the hedgehog 
and strange quark eigenstates, respectively.


\section{Batalin-Fradkin-Tyutin quantization scheme}
\setcounter{equation}{0}
\renewcommand{\theequation}{C.\arabic{equation}}


\subsection{BRST symmetries in Skyrmion model}


In this section we will obtain the BRST invariant Lagrangian in the framework 
of the BFV formalism \cite{fradkin75, fujiwara90, bizdadea95} which is 
applicable to theories with the first class constraints by introducing two 
canonical sets of ghosts and anti-ghosts together with auxiliary fields 
$({\cal C}^{i},\bar{{\cal P}}_{i})$, $({\cal P}^{i},\bar{{\cal C}}_{i})$, 
$(N^{i},B_{i})$, $(i=1,2)$ which satisfy the super-Poisson algebra
\footnote{Here one notes that the BRST symmetry can be also constucted by 
using the residual gauge symmetry interpretation of the BRST invariance~\cite{yee93,yee94}.}
\beq
\{{\cal C}^{i},\bar{{\cal P}}_{j}\}=\{{\cal P}^{i},\bar{{\cal C}}_{j}\}
=\{N^{i},B_{j}\}=\delta_{j}^{i}.
\eeq
Here the super-Poisson bracket is defined as
\beq
\{A,B\}=\frac{\delta A}{\delta q}|_{r}\frac{\delta B}{\delta p}|_{l}
-(-1)^{\eta_{A}\eta_{B}}\frac{\delta B}{\delta q}|_{r}\frac{\delta A} {%
\delta p}|_{l}
\eeq
where $\eta_{A}$ denotes the number of fermions called ghost number in $A$
and the subscript $r$ and $l$ right and left derivatives.

In the SU(2) Skyrmion model, the nilpotent BRST charge $Q$, the fermionic
gauge fixing function $\Psi$ and the BRST invariant minimal Hamiltonian $%
H_{m}$ are given by
\begin{eqnarray}
Q&=&{\cal C}^{i}\tilde{\Omega}_{i}+{\cal P}^{i}B_{i},~~~
\Psi=\bar{{\cal C}}_{i}\chi^{i}+\bar{{\cal P}}_{i}N^{i},  \nonumber \\
H_{m}&=&\tilde{H}^{\prime}-\frac{1}{2{\cal I}_{10}}{\cal C}^{1}
\bar{{\cal P}}_{2}
\end{eqnarray}
which satisfy the relations $\{Q,H_{m}\}=0$, $Q^{2}=\{Q,Q\}=0$, $\{\{\Psi,Q\},Q\}=0$.
The effective quantum Lagrangian is then described as
\begin{equation}
L_{eff}=\pi^{\mu}\dot{a}^{\mu}+\pi_{\theta}\dot{\theta} +B_{2}\dot{N}^{2}+%
\bar{{\cal P}}_{i}\dot{{\cal C}}^{i}+\bar{{\cal C}}_{2} \dot{{\cal P}}%
^{2}-H_{tot}
\end{equation}
with $H_{tot}=H_{m}-\{Q,\Psi\}$. Here $B_{1}\dot{N}^{1} +\bar{{\cal C}}_{1}%
\dot{{\cal P}}^{1}=\{Q,\bar{{\cal C}}_{1} \dot{N}^{1}\}$ terms are
suppressed by replacing $\chi^{1}$ with $\chi^{1} +\dot{N}^{1}$.

Now we choose the unitary gauge
\begin{equation}
\chi^{1}=\Omega_{1},~~~\chi^{2}=\Omega_{2}
\end{equation}
and perform the path integration over the fields $B_{1}$, $N^{1}$, $\bar{%
{\cal C}}_{1}$, ${\cal P}^{1}$, $\bar{{\cal P}}_{1}$ and ${\cal C}^{1}$, by
using the equations of motion, to yield the effective Lagrangian of the form
\begin{eqnarray}
L_{eff}&=&\pi^{\mu}\dot{a}^{\mu}+\pi_{\theta}\dot{\theta} +B\dot{N}+\bar{%
{\cal P}}\dot{{\cal C}}+\bar{{\cal C}}\dot{{\cal P}}  \nonumber \\
& &-M_{0}-\frac{1}{8{\cal I}_{10}}(\pi^{\mu}-a^{\mu}\pi_{\theta})(\pi^{\mu}
-a^{\mu}\pi_{\theta})\frac{a^{\sigma}a^{\sigma}}{a^{\sigma}a^{\sigma}+2\theta} 
-\frac{1}{4{\cal I}_{10}}\pi_{\theta}\tilde{\Omega}_{2}  \nonumber \\
& &+2a^{\mu}a^{\mu}\pi_{\theta}\bar{{\cal C}}{\cal C}+\tilde{\Omega}_{2}N
+B\Omega_{2}+\bar{{\cal P}}{\cal P}
\end{eqnarray}
with redefinitions: $N\equiv N^{2}$, $B\equiv B_{2}$, $\bar{{\cal C}}\equiv
\bar{{\cal C}}_{2}$, ${\cal C}\equiv {\cal C}^{2}$, $\bar{{\cal P}}\equiv
\bar{{\cal P}}_{2}$, ${\cal P}\equiv {\cal P}_{2}$.

Next, using the variations with respect to $\pi^{\mu}$, $\pi_{\theta}$, 
${\cal P}$ and $\bar{{\cal P}}$, one obtain the relations
\begin{eqnarray}
\dot{a}^{\mu}&=&\frac{1}{4{\cal I}_{10}}(\pi^{\mu}-a^{\mu}\pi_{\theta})
a^{\sigma}a^{\sigma} +a^{\mu}(\frac{1}{4{\cal I}_{10}}\pi_{\theta}-N-B)  
\nonumber\\
\dot{\theta}&=&-\frac{1}{4{\cal I}_{10}}a^{\mu}(\pi^{\mu}-a^{\mu}\pi_{\theta})
a^{\sigma}a^{\sigma} +a^{\mu}a^{\mu}(-\frac{1}{2{\cal I}_{10}}\pi_{\theta}
-2\bar{{\cal C}}{\cal C}+N) +\frac{1}{4{\cal I}_{10}}a^{\mu}\pi^{\mu}  
\nonumber \\
{\cal P}&=&-\dot{{\cal C}},~~~~~\bar{{\cal P}}=\dot{\bar{{\cal C}}}
\end{eqnarray}
to yield the effective Lagrangian
\begin{eqnarray}
L_{eff}&=&-M_{0}+\frac{2{\cal I}_{10}}{a^{\sigma}a^{\sigma}}\dot{a}^{\mu}\dot{a}^{\mu} 
-2{\cal I}_{10}\left[\frac{\dot{\theta}}{a^{\sigma}a^{\sigma}} +(B+2\bar{%
{\cal C}}{\cal C})a^{\sigma}a^{\sigma}\right]^{2}+B\dot{N}  \nonumber \\
& &+\frac{4{\cal I}_{10}}{a^{\sigma}a^{\sigma}}a^{\mu}\left[ \dot{a}%
^{\mu}+a^{\mu}(\frac{\dot{\theta}} {a^{\sigma}a^{\sigma}}+(B+2\bar{{\cal C}}%
{\cal C})a^{\sigma}a^{\sigma})\right] (B+N)+\dot{\bar{{\cal C}}}\dot{{\cal C}}.  
\nonumber 
\end{eqnarray}

Finally, with the identification 
$N=-B+\frac{\dot{\theta}}{a^{\sigma}a^{\sigma}}$, one can arrive at the 
BRST invariant Lagrangian \cite{hong001}
\begin{eqnarray}
L_{eff}&=&-M_{0}+\frac{2{\cal I}_{10}}{1-2\theta}\dot{a}^{\mu}\dot{a}^{\mu} 
-\frac{2{\cal I}_{10}}{(1-2\theta)^{2}}\dot{\theta}^{2} 
-2{\cal I}_{10}(1-2\theta)^{2}(B+2\bar{{\cal C}}{\cal C})^{2}  \nonumber \\
& &-\frac{\dot{\theta}\dot{B}}{1-2\theta} +\dot{\bar{{\cal C}}}\dot{{\cal C}},
\end{eqnarray}
which is invariant under the BRST transformation
\begin{eqnarray}
\delta_{B}a^{\mu}&=&\lambda a^{\mu}{\cal C},~~~ \delta_{B}\theta=-\lambda
(1-2\theta){\cal C},  \nonumber \\
\delta_{B}\bar{{\cal C}}&=&-\lambda B,~~~ \delta_{B}{\cal C}=\delta_{B}B=0.
\end{eqnarray}


\subsection{SU(3) Skyrmion with flavor symmetry breaking effects}


In this section, our starting SU(3) Skyrmion Lagrangian in Eq. 
(\ref{lagfsbapp}) is given by
\begin{eqnarray}
 {\cal L}&=&-\frac{1}{4}f_{\pi}^{2}{\rm tr}(l_{\mu}l^{\mu}) +\frac{1}{32e^{2}}%
 {\rm tr}[l_{\mu},l_{\nu}]^{2}+{\cal L}_{WZW}  \nonumber \\
 & &+\frac{1}{4}f_{\pi}^{2}{\rm tr}M(U+U^{\dag}-2)+{\cal L}_{FSB},  \nonumber
 \\
 {\cal L}_{FSB}&=&\frac{1}{6}(f_{K}^{2}m_{K}^{2}-f_{\pi}^{2}m_{\pi}^{2}) {\rm %
tr}((1-\sqrt{3}\lambda_{8})(U+U^{\dag}-2))  \nonumber \\
& &-\frac{1}{12}(f_{K}^{2}-f_{\pi}^{2}){\rm tr} ((1-\sqrt{3}%
\lambda_{8})(Ul_{\mu}l^{\mu} +l_{\mu}l^{\mu}U^{\dag})),
\end{eqnarray}
where the WZW action is given by Eq. (\ref{wzwsu3}). 

Now we consider only the rigid motions of the SU(3) Skyrmion
\[
U(\vec{x},t)={\cal A}(t)U_{0}(\vec{x}){\cal A}(t)^{\dag}. 
\]
Assuming maximal symmetry in the Skyrmion, we can use the hedgehog solution
$U_{0}$ given in Eq. (\ref{hedgehog}) embedded in the SU(2) isospin 
subgroup of SU(3) with the chiral angle $\theta (r)$ which is determined by 
minimizing the static mass $M_{0}$ in Eq. (\ref{skstaticmass}) and, for unit 
winding number, satisfies the boundary conditions 
$\lim_{r \rightarrow \infty} \theta (r)=0$ and $\theta (0)=\pi$.

Since ${\cal A}$ belongs to $SU(3)$, ${\cal A}^{\dag}\dot{{\cal A}}$ is
anti-Hermitian and traceless to be expressed as a linear combination of 
$\lambda_{a}$ as follows
\[
{\cal A}^{\dag}\dot{{\cal A}}=ief_{\pi}v^{a}\lambda_{a}=ief_{\pi}\left(
\begin{array}{cc}
\vec{v}\cdot\tau +\nu 1 & V \\
V^{\dag} & -2\nu
\end{array}
\right)
\]
where
\begin{equation}
\vec{v}=(v^{1},v^{2},v^{3}),~~V=\left(
\begin{array}{c}
v^{4}-iv^{5} \\
v^{6}-iv^{7}
\end{array}
\right),~~ \nu=\frac{v^{8}}{\sqrt{3}}.  \label{vs}
\end{equation}
After tedious algebraic manipulations, the FSB contribution to the Skyrmion
Lagrangian is then expressed as~\cite{hong01prd}
\begin{eqnarray}
{\cal L}_{FSB}&=&-(f_{K}^{2}m_{K}^{2}-f_{\pi}^{2}m_{\pi}^{2})(1-\cos\theta)\sin^{2}d  \nonumber \\
& &+\frac12 (f_{K}^{2}-f_{\pi}^{2})\sin^{2}d \left(\frac{8}{3}%
e^{2}f_{\pi}^{2}\vec{v}^{2}\sin^{2}\theta-\frac{2\sin^{2}\theta}{r^{2}} 
-\left(\frac{{\rm d}\theta}{{\rm d}r}\right)^{2}\right)\cos \theta  \nonumber \\
& &-(f_{K}^{2}-f_{\pi}^{2})e^{2}f_{\pi}^{2}\frac{\sin^{2}d}{d^{2}}
\left((1-\cos \theta)^{2}\| D^{\dag}V\|^{2}-\sin^{2}\theta\| D^{\dag} 
\tau\cdot\hat{r}V\|^{2}\right)  \nonumber \\
& &+\frac{i\sqrt{2}}{3}(f_{K}^{2}-f_{\pi}^{2})e^{2}f_{\pi}^{2}\frac{\sin 2d}{%
d} \sin^{2} \theta (D^{\dag}\vec{v}\cdot\tau V -(D^{\dag}\vec{v}
\cdot\tau V)^{*})
\nonumber \\
& &+(f_{K}^{2}-f_{\pi}^{2})e^{2}f_{\pi}^{2}\cos^{2}d~(1-\cos \theta)V^{\dag}V.
\label{fsblagsu3}
\end{eqnarray}

In order to separate the SU(2) rotations from the deviations into strange
directions, the time-dependent rotations can be written as~\cite{kleb90}
\begin{equation}
{\cal A}(t)=\left(
\begin{array}{cc}
A(t) & 0 \\
0 & 1
\end{array}
\right)S(t) 
\label{curlya}
\end{equation}
with $A(t) \in$ SU(2) and the small rigid oscillations $S(t)$ around the
SU(2) rotations.\footnote{Here one notes that fluctuations $\phi_{a}$ from 
collective rotations $A$ can be also separated by the other suitable 
parameterization~\cite{schw92}
$
U=A\sqrt{U_{0}}A^{\dagger}{\rm exp}(i\sum_{a=1}^{8}\phi_{a}\lambda_{a})
A\sqrt{U_{0}}A^{\dagger}.
$
}  Furthermore, we exploit the time-dependent angular velocity of the SU(2)
rotation through
\[
A^{\dagger}\dot{A}=\frac{i}{2}\dot{\alpha}\cdot\vec{\tau}.
\]
Note that one can use the Euler angles for the parameterization of the rotation
~\cite{schwei91}.  On the other hand the small rigid oscillations $S$, which were also 
used in Ref.~\cite{kleb94}, can be described as
\beq
S(t)={\rm exp}(i\sum_{a=4}^{7}d^{a}\lambda_{a})={\rm exp}(i{\cal D}),
\label{stexp}
\eeq
where
\beq
{\cal D}=\left(
\begin{array}{cc}
0 & \sqrt{2}D \\
\sqrt{2}D^{\dag} & 0
\end{array}
\right),~~ D=\frac{1}{\sqrt{2}}\left(
\begin{array}{c}
d^{4}-id^{5} \\
d^{6}-id^{7}
\end{array}
\right). 
\label{curlyd}
\eeq

Including the FSB correction terms in Eq. (\ref{fsblagsu3}), the Skyrmion
Lagrangian to order $1/N_{c}$ is then given in terms of the angular velocity 
$\alpha_{i}$ and the strange deviations $D$
\begin{eqnarray}
L&=&-M_{0}+\frac{1}{2}{\cal I}_{10}\dot{\alpha}\cdot\dot{\alpha}+(4{\cal I}_{20}
+\Gamma_{1})\dot{D}^{\dag}\dot{D}+\frac{i}{2}N_{c}(D^{\dag}\dot{D} 
-\dot{D}^{\dag}D) 
\nonumber \\
& &+i({\cal I}_{10}-2{\cal I}_{20}-\frac12 \Gamma_{1}+\Gamma_{2})
\left(D^{\dag}\dot{\alpha}\cdot\vec{\tau}\dot{D}-\dot{D}^{\dag}\dot{\alpha}\cdot\vec{\tau}D\right)
\nonumber\\
& &-\frac{1}{2}N_{c}D^{\dag}\dot{\alpha}\cdot\vec{\tau}D
+2\left({\cal I}_{10}-\frac{4}{3}{\cal I}_{20}-\frac{4}{3}\Gamma_{1}
+3\Gamma_{2}\right)(D^{\dag}D)(\dot{D}^{\dag}\dot{D})  \nonumber \\
& &-\frac{1}{2}\left({\cal I}_{10}-\frac{4}{3}{\cal I}_{20}-\frac{1}{3}
\Gamma_{1}+2\Gamma_{2}\right)(D^{\dag}\dot{D}+\dot{D} ^{\dag}D)^{2}
\nonumber \\
& &+\left(2{\cal I}_{20}+\frac12 \Gamma_{1}\right)(D^{\dag}\dot{D} -\dot{D}%
^{\dag}D)^{2} -\frac{i}{3}N_{c}(D^{\dag}\dot{D}-\dot{D}^{\dag}D)D^{\dag}D
\nonumber \\
& &-\frac12\Gamma_{0} m_{\pi}^{2}-\left(\Gamma_{0}(\chi^{2}m_{K}^{2}
-m_{\pi}^{2})+\Gamma_{3}\right)\left(D^{\dag}D -\frac{2}{3}%
(D^{\dag}D)^{2}\right)  \nonumber \\
& &-2(\Gamma_{1}-\Gamma_{2})(D^{\dag}\dot{D})(\dot{D}^{\dag}D),  
\label{lagsu3}
\end{eqnarray}
where $\chi=f_{K}/f_{\pi}$. Here the soliton energy $M_{0}$, the moment of
inertia ${\cal I}_{10}$ are given by Eqs. (\ref{skstaticmass}) and 
(\ref{skinertiamom}), and the other moment of inertia ${\cal I}_{20}$, the 
strength $\Gamma_{0}$ of the chiral symmetry breaking and the inertia 
parameters $\Gamma_{i}$ $(i=1,2,3)$
originated from the FSB term are respectively given by
\begin{eqnarray}
{\cal I}_{20}&=&\frac{2\pi}{e^{3}f_{\pi}}\int_{0}^{\infty}{\rm d}z~z^{2}
(1-\cos\theta) \left[1+\frac{1}{4}\left(\left(\frac{{\rm d}\theta}{{\rm d}z}\right)^{2}
+\frac{2\sin^{2}\theta}{z^{2}}\right)\right],  \nonumber \\
\Gamma_{0}&=&\frac{8\pi}{e^{3}f_{\pi}}\int_{0}^{\infty}{\rm d}z~z^{2}
(1-\cos\theta),
\nonumber \\
\Gamma_{1}&=&(\chi^{2}-1)\Gamma_{0},  \nonumber \\
\Gamma_{2}&=&(\chi^{2}-1)\frac{8\pi}{3e^{3}f_{\pi}}\int_{0}^{\infty}{\rm d}z
~z^{2}\sin^{2}\theta,  \nonumber \\
\Gamma_{3}&=&(\chi^{2}-1)\frac{4\pi f_{\pi}}{e}\int_{0}^{\infty}{\rm d}z~z^{2}
\left(\left(\frac{{\rm d}\theta}{{\rm d}z}\right)^{2} 
+\frac{2\sin^{2}\theta}{z^{2}}
\right)\cos\theta  
\label{enisu3}
\end{eqnarray}
with the dimensionless quantities $z=ef_{\pi}r$.

The momenta $\pi_{h}^{i}$ and $\pi_{s}^{\alpha}$, conjugate to the collective
coordinates $\alpha_{i}$ and the strange deviation $D_{\alpha}^{\dag}$ are
given by
\begin{eqnarray}
\vec{\pi}_{h}&=&{\cal I}_{10}\dot{\alpha}+i\left({\cal I}_{10} -2{\cal I}%
_{20}-\frac12\Gamma_{1}+\Gamma_{2}\right)\left(D^{\dag}\vec{\tau}\dot{D}
-\dot{D}^{\dag}\vec{\tau}\right)-\frac{1}{2}N_{c}D^{\dag}\vec{\tau}D,  
\nonumber\\
\pi_{s}&=&(4{\cal I}_{20}+\Gamma_{1})\dot{D}-\frac{i}{2}N_{c}D
-i\left({\cal I}_{10}-2{\cal I}_{20}-\frac12\Gamma_{1}+\Gamma_{2}\right)
\dot{\alpha}\cdot \vec{\tau}D  \nonumber \\
& &+2\left({\cal I}_{10}-\frac{4}{3}{\cal I}_{20}-\frac{4}{3}\Gamma_{1}
+3\Gamma_{2}\right)(D^{\dag}D)\dot{D}  \nonumber \\
& &-\left({\cal I}_{10}-\frac{4}{3}{\cal I}_{20}-\frac{1}{3}\Gamma_{1}
+2\Gamma_{2}\right)(D^{\dag}\dot{D}+\dot{D}^{\dag}D)D  \nonumber \\
& &-(4{\cal I}_{20}+\Gamma_{1})(D^{\dag}\dot{D}-\dot{D}^{\dag}D)D +\frac{i}{3}%
N_{c}(D^{\dag}D)D  \nonumber \\
& &-2(\Gamma_{1}-\Gamma_{2})(D^{\dag}\dot{D})D,  
\label{conjmomssu3}
\end{eqnarray}
which satisfy the Poisson brackets $\{\alpha_{i},\pi_{h}^{j}\}=\delta_{i}^{j}$, 
$\{D_{\alpha}^{\dag},\pi_{s}^{\beta}\}=\{D^{\beta},\pi_{s,\alpha}^{\dag}\}
=\delta_{\alpha}^{\beta}$.  Performing Legendre transformation, we obtain the 
Hamiltonian to order $1/N_{c}$ as follows
\begin{eqnarray}
H&=&M_{0}+\frac{1}{2}\Gamma _{0}m_{\pi }^{2}+\frac{1}{2{\cal I}_{10}}
\vec{\pi}_{h}^{2}
+\frac{1}{4{\cal I}_{20}^{\prime }}\pi _{s}^{\dag }\pi _{s}-i%
\frac{N_{c}}{8{\cal I}_{20}^{\prime }}(D^{\dag }\pi _{s}-\pi _{s}^{\dag }D)
\nonumber \\
& &+\left[\frac{N_{c}^{2}}{16{\cal I}_{20}^{\prime }}+\Gamma _{0}(\chi
^{2}m_{K}^{2}-m_{\pi }^{2})+\Gamma_{3}\right] D^{\dag }D
+i\left[ \frac{1}{2{\cal I}_{10}}-\frac{1}{4{\cal I}_{20}^{\prime}}
\left(1+\frac{\Gamma _{2}}{%
{\cal I}_{10}}\right) \right]  \nonumber \\
& &\cdot (D^{\dag}\vec{\pi}_{h}\cdot\vec{\tau}\pi _{s}-\pi _{s}^{\dag }\vec{\pi}_{h}\cdot\vec{\tau}D)
+\frac{N_{c}}{4{\cal I}_{20}^{\prime }}\left(1+\frac{\Gamma _{2}}
{{\cal I}_{10}}\right) D^{\dag }\vec{\pi}_{h}\cdot \vec{\tau}D  \nonumber \\
& &+\left[ \frac{1}{2{\cal I}_{10}}-\frac{1}{3{\cal I}_{20}^{\prime }}\left( 1+%
\frac{3}{2}\frac{\Gamma _{2}}{{\cal I}_{10}}\right) +\frac{\Gamma _{2}^{2}+%
{\cal I}_{10}(\Gamma _{1}-\Gamma _{2})}{8{\cal I}_{10}{\cal I}_{20}^{\prime 2}}%
\right] (D^{\dag }D)(\pi _{s}^{\dag }\pi _{s})  \nonumber \\
& &+\left[ \frac{1}{12{\cal I}_{20}^{\prime }}\left( 1+\frac{3}{2}\frac{%
\Gamma _{2}}{{\cal I}_{10}}\right) -\frac{1}{8{\cal I}_{10}}-\frac{\Gamma
_{2}^{2}-{\cal I}_{10}(\Gamma _{1}-\Gamma _{2})}{32{\cal I}_{10}{\cal I}%
_{20}^{\prime 2}}\right] (D^{\dag }\pi _{s}+\pi _{s}^{\dag }D)^{2}  \nonumber
\\
& &-\left( \frac{1}{8{\cal I}_{20}^{\prime }}+\frac{\Gamma _{1}-\Gamma _{2}}{%
32{\cal I}_{20}^{\prime 2}}\right) (D^{\dag }\pi _{s}-\pi _{s}^{\dag }D)^{2}
\nonumber \\
& &-i\frac{N_{c}}{8}\left[ \frac{1}{{\cal I}_{20}^{\prime }}
\left( 1-\frac{\Gamma_{2}}{{\cal I}_{10}}\right) +\frac{\Gamma _{2}^{2}
+2{\cal I}_{10}(\Gamma_{1}-\Gamma _{2})}{2{\cal I}_{10}
{\cal I}_{20}^{\prime 2}}\right] (D^{\dag}\pi_{s}-\pi _{s}^{\dag }D)(D^{\dag }D)  \nonumber \\
& &+\left[\frac{N_{c}^{2}}{12{\cal I}_{20}^{\prime}}-\frac{2}{3}
\Gamma _{0}(\chi^{2}m_{K}^{2}-m_{\pi }^{2})-\frac{2}{3}\Gamma _{3}\right.  
\nonumber \\
& &\left.+\frac{N_{c}^{2}}{32}\frac{\Gamma _{2}^{2}+2{\cal I}_{10} (\Gamma
_{1}-\Gamma_{2})}{{\cal I}_{10}{\cal I}_{20}^{\prime 2}}\right] (D^{\dag
}D)^{2},
\label{ham00su3}
\end{eqnarray}
where ${\cal I}_{20}^{\prime}={\cal I}_{20}+\frac{1}{4}\Gamma_{1}$.

Through the symmetrization procedure~\cite{oliveira97,hong991}, we can obtain 
the Hamiltonian of the form
\begin{eqnarray}
H &=&M_{0}+\frac{1}{2}\Gamma _{0}m_{\pi }^{2}+\frac{1}{2{\cal I}_{10}}(%
\vec{I}^{2}+\frac{1}{4})+\frac{1}{4{\cal I}_{20}^{\prime }}\pi _{s}^{\dag}
\pi _{s}-i\frac{N_{c}}{8{\cal I}_{20}^{\prime }}(D^{\dag }\pi _{s}
-\pi_{s}^{\dag }D)  \nonumber \\
& &+\left[\frac{N_{c}^{2}}{16{\cal I}_{20}^{\prime }}+\Gamma _{0}(\chi
^{2}m_{K}^{2}-m_{\pi }^{2})+\Gamma_{3}\right] D^{\dag }D  \nonumber \\
& &+i\left[ \frac{1}{2{\cal I}_{10}}-\frac{1}{4{\cal I}_{20}^{\prime }}\left(
1+\frac{\Gamma _{2}}{{\cal I}_{10}}\right) \right] (D^{\dag }\vec{I}\cdot
\vec{\tau}\pi _{s}-\pi _{s}^{\dag }\vec{I}\cdot \vec{\tau}D)   \nonumber\\
& &+\frac{N_{c}}{4{\cal I}_{20}^{\prime }}\left(1+\frac{\Gamma _{2}}
{{\cal I}_{10}}\right) D^{\dag }\vec{I}\cdot \vec{\tau}D+\cdots.
\label{nhtsu3}
\end{eqnarray}
where the isospin operator $\vec{I}$ is given by $\vec{I}=\vec{\pi}_{h}$ and 
the ellipsis stands for the strange-strange interaction terms of order 
$1/N_{c}$ which can be readily read off from Eq. (\ref{ham00su3}).  Here one 
notes that the overall energy shift $\frac{1}{8{\cal I}_{10}}$ originates from 
the Weyl ordering correction in the BFT Hamiltonian scheme as discussed 
before.  

\end{appendix}

\end{document}